\documentclass[showpacs,preprintnumbers,twocolumn,nofootinbib,eqsecnum,prd]{revtex4-1} 

\usepackage{graphicx}
\usepackage{latexsym}
\usepackage{amsmath}
\usepackage{amssymb}

\def\beq{\begin{equation}}
\def\eeq{\end{equation}}

\begin{document}

\title{Gravitational self-force corrections to two-body tidal  interactions\\
 and the effective one-body formalism}

\author{Donato \surname{Bini}$^1$}
\author{Thibault \surname{Damour}$^2$}

\affiliation{$^1$Istituto per le Applicazioni del Calcolo ``M. Picone,'' CNR, I-00185 Rome, Italy\\
$^2$Institut des Hautes Etudes Scientifiques, 91440 Bures-sur-Yvette, France}

\date{\today}

\begin{abstract}
Tidal interactions have a significant influence on the late dynamics of compact binary systems, which constitute the prime targets
of the upcoming network of gravitational-wave detectors.
We refine the theoretical description of tidal interactions (hitherto known only to the second post-Newtonian level) by extending our recently developed analytic self-force formalism, for extreme mass-ratio binary systems, to the computation of several tidal invariants.
Specifically, we compute, to linear order in the mass ratio and to the 7.5$^{\rm th}$ post-Newtonian order, the following tidal invariants: the square and the cube of the gravitoelectric quadrupolar tidal tensor, the square of the gravitomagnetic quadrupolar tidal tensor, and the square of the gravitoelectric octupolar tidal tensor.
Our high-accuracy analytic results are compared to recent numerical self-force tidal data by Dolan et al. \cite{Dolan:2014pja}, and, notably, provide an analytic understanding of the light ring asymptotic behavior found by them. We transcribe our kinematical tidal-invariant results in the more dynamically significant effective one-body description of the tidal interaction energy. By combining, in a synergetic manner, analytical and numerical results, we provide simple, accurate analytic representations of the global, strong-field behavior of the gravitoelectric quadrupolar tidal factor. A striking finding is that the linear-in-mass-ratio piece in the latter tidal factor changes sign in the strong-field domain, to become negative (while its previously known second post-Newtonian approximant was always positive).
We, however, argue that this will be more than compensated by a probable fast growth, in the strong-field domain, of the nonlinear-in-mass-ratio contributions in the tidal factor.
\end{abstract}
\maketitle

\section{Introduction}

The current development of gravitational wave detectors gives a new motivation for improving our theoretical understanding of the general relativistic dynamics of compact binary systems, i.e., systems comprising black holes and/or neutron stars. Recent work has shown that tidal interactions have a significant influence on the late dynamics of coalescing neutron star binaries 
\cite{Flanagan:2007ix,Read:2009yp,Baiotti:2010xh,Bernuzzi:2012ci,Damour:2012yf,Bernuzzi:2011aq,Read:2013zra,DelPozzo:2013ala,Hotokezaka:2013mm,Radice:2013hxh,Bernuzzi:2014kca}. It makes it urgent to refine the theoretical description of tidal effects in the strong-field regime reached in the last stages of the inspiralling phase of neutron star binaries.

In recent years, it has been understood that a useful strategy for studying the strong-field aspects of the dynamics of compact binaries is to combine, in a synergetic manner, information gathered from several different approximation methods, namely: the post-Newtonian (PN) formalism, the post-Minkowskian one, the gravitational self-force (SF) formalism, full numerical relativity simulations, and, the effective one-body (EOB) formalism. In particular, the EOB formalism
\cite{Buonanno:1998gg,Buonanno:2000ef,Damour:2000we,Damour:2001tu}  appears to define a useful framework which can combine, in an efficient and accurate manner, information coming from all the other approximation schemes, while also adding genuinely new information coming from EOB theory.
For recent examples of this synergetic use of EOB theory see Refs. 
\cite{Damour:2009sm,Barack:2010ny,Akcay:2012ea,Taracchini:2013rva,Damour:2014sva,Damour:2014afa,Bini:2014ica}.

In the present paper, we shall refine the theoretical description of tidal interactions by combining, within EOB theory, three  types of information: (i) the state-of-the-art PN knowledge of tidal interaction in comparable mass binary systems (which is limited to the 2PN level \cite{Bini:2012gu}); (ii) the extension to tidal effects of a recently developed high-accuracy analytic description of extreme-mass-ratio binary systems to linear order in the mass ratio $q=m_1/m_2$ \cite{Bini:2013zaa,Bini:2013rfa,Bini:2014nfa,Bini:2014ica}; and (iii)
recent numerical self-force computations of some tidal invariants to linear order in $q$ \cite{Dolan:2014pja}.

An important aspect of the present work will be to transcribe the purely kinematic knowledge of some tidal invariants (expressed as functions of the dimensionless frequency parameter $y=(Gm_2\Omega/c^3)^{2/3}$, where $\Omega$ is the orbital frequency) into the {\it dynamical} knowledge of the tidal interaction energy of binary systems.
This will be done within the EOB formulation of tidal effects proposed by Damour and Nagar \cite{Damour:2009wj}. 
Up to now this formulation has been developed only through PN theory, and was limited to the fractional second post-Newtonian (2PN) level, i.e., at the level where one includes relativistic corrections of order $(v/c)^4$ to the Newtonian tidal binding energy \cite{Bini:2012gu}. [For previous, 1PN accurate, tidal computations see 
\cite{Vines:2010ca,Damour:2009wj}.]

Here, thanks to the technology developed in our previous papers \cite{Bini:2013zaa,Bini:2013rfa,Bini:2014nfa,Bini:2014ica}, we shall be able to analytically compute several tidal invariants to a very high-order accuracy, namely 7.5PN, i.e. $(v/c)^{15}$ beyond the Newtonian level, but only to {\it linear} order in $q$.
We shall then transcribe this kinematic information into a more dynamically useful form. More precisely, we shall compute to 7.5PN accuracy the relativistic factors
$\hat A^{l^\epsilon}=1+O\left(\left(\frac{v}{c}\right)^2\right)$ that multiply the leading-order  EOB description of the tidal interaction energy in binary systems. As we shall see in detail later, the label $l$ denotes the multipolarity of the considered interaction, while $\epsilon$ denotes its parity: $+$ for even (or electric-like) parity, and $-$ for odd (or magnetic-like) parity. Our main focus will be the two dominant tidal interactions: quadrupolar-electric ($2^+$) and quadrupolar-magnetic ($2^-$).

Another important aspect of our synergetic study of tidal effects will be to compare our 7.5PN-accurate analytic results to the recent accurate numerical self-force results of Dolan et al. \cite{Dolan:2014pja} on some tidal invariants. In addition, we shall combine our analytic PN and EOB knowledge with the accurate data of 
\cite{Dolan:2014pja} to propose some simple, but numerically accurate, analytic representations of the EOB tidal relativistic factors $\hat A^{l^\epsilon}(u;X_1)$ valid in the strong field regime $u\sim O(1)$, relevant for describing the late stages of coalescing neutron star binaries. To guide the reader through our later developments, let us display here the notation we shall use for some important quantities throughout this paper.

The masses of the gravitationally interacting two bodies are $m_1$ and $m_2$, with the convention that $m_1\le m_2$.
We then define
\begin{eqnarray}
\label{eq:1.1}
M&\equiv& m_1+m_2\,,\quad \mu\equiv\frac{m_1 m_2}{M}\,,\nonumber\\
\nu &\equiv& \frac{\mu}{M}=\frac{m_1m_2}{(m_1+m_2)^2}
\end{eqnarray}
[Beware of the fact that in the SF literature the letter $M$ is often used to denote the large mass, i.e., $m_2$ in our notation, while the letter $\mu$ is often used to denote the small mass, i.e., $m_1$ in our notation.]

Besides the {\it symmetric} mass ratio $\nu$ just defined, we shall also use the other dimensionless mass ratios
\beq
\label{eq:1.2}
q\equiv \frac{m_1}{m_2}\le 1\,,\quad X_1\equiv \frac{m_1}{m_1+m_2}\le \frac12\,.
\eeq
Note the links
\beq
\label{eq:1.3}
\nu=\frac{q}{(1+q)^2}\,,\quad X_1=\frac{q}{1+q}\,,
\eeq
and the fact that, in the small mass-ratio case $q\ll 1$, we have $\nu \simeq X_1\simeq q\ll 1$.
[By contrast, $X_2\equiv \frac{m_2}{M}\equiv 1-X_1$ is equivalent to $1-q+O(q^2)$ in this limit.]

We will evaluate all SF quantities on the world line ${\mathcal L}_1$ of the smaller mass, $m_1$.
Tidal invariants will be expressed either in terms of the dimensionless frequency parameters, $x$ and $y$, or of the EOB dimensionless gravitational potential $u$.
Here, we define
\begin{eqnarray}
\label{eq:1.4}
x &=& \left( \frac{G(m_1+m_2)\Omega}{c^3} \right)^{2/3}\\
\label{y_def}
y &=& \left( \frac{G m_2 \Omega}{c^3} \right)^{2/3}\\
\label{u_def}
u&=& \frac{G(m_1+m_2) }{c^2 r_{\rm EOB}}\,.
\end{eqnarray}
In the Newtonian limit $u\simeq x \simeq (v/c)^2$ (while $x\equiv (1+q)^{2/3}y$).
In the following we shall often set $G=c=1$, except when it may be physically illuminating to re-establish the presence of $G$ or $c$ in some expression.

\section{Relativistic tidal effects in binary systems: recap of known results}

Before dealing with the new, high-PN order, tidal results that are the main aim of this work, let us recall the state of the art in the knowledge of relativistic tidal effects in binary systems.

Ref. \cite{Damour:1982wm} extended the concept of Love number (measuring the tidal polarizability of an extended body) to neutron stars, and showed that the corresponding  finite size effects start modifying the dynamics of compact binary systems at the (formal) 5PN level. In an effective field theory description of extended objects, finite size effects are treated by augmenting the leading-order skeletonized point-mass action describing gravitationally interacting compact objects \cite{Damour:1982wm},
\beq
\label{eq:2.1}
S_0=\int \frac{d^D x}{c}\, \frac{c^4}{16 \pi G}\sqrt{-g} R(g)-\sum_A \int m_A c^2 d\tau_A\,,
\eeq
where $d\tau_A\equiv c^{-1}(-g_{\mu\nu}(y_A)dy_A^\mu dy_A^\nu)^{1/2}$ is the (dimensionally) regularized proper time along the world line $y_A^\mu(\tau_A)$ of body $A$, by additional, nonminimal world line couplings involving higher-order derivatives of the field \cite{Damour:1995kt,Damour:1998jk,Goldberger:2004jt}. To classify the possible tidal-related nonminimal world line scalars, it is useful to appeal to the relativistic theory of tidal expansions \cite{Zhang86,Damour:1990pi,Damour:1991yw}. In the notation  of Refs. \cite{Damour:1990pi,Damour:1991yw}, the tidal expansion of the \lq\lq external metric\rq\rq felt by body $A$ (member of a $N$-body system) is expressed in terms of two types of tidal tensors: the gravitoelectric $G_L^A(\tau_A)\equiv G^A_{a_1\ldots a_l}(\tau_A)$, and gravitomagnetic $H_L^A(\tau_A)\equiv H^A_{a_1\ldots a_l}(\tau_A)$, symmetric trace-free (spatial) tensors, together with their proper time derivatives. (The spatial indices $a_i=1,2,3$ refer to   local-frame
coordinates, $X_A^0=c\tau_A, X_A^a$ attached to body $A$.) In terms of these tidal tensors, the most general nonminimal world line action has the form
\begin{eqnarray}
\label{eq:2.2}
S_{\rm nonminimal} 
&= &\sum_{A} \biggl\{ \frac{1}{4} \, \mu_A^{(2)} \int
d\tau_A \, G_{\alpha\beta}^A \, G_A^{\alpha\beta}\nonumber\\
&+ & \frac{1}{6 \, c^2} \, \sigma_A^{(2)} \int d\tau_A \, H_{\alpha\beta}^A \,
H_A^{\alpha\beta} \nonumber \\
&+ & \frac{1}{12} \, \mu_A^{(3)} \int d\tau_A \, G_{\alpha\beta\gamma}^A \,
G_A^{\alpha\beta\gamma} \nonumber \\
&+ & \frac{1}{4 \, c^2} \, \mu'^{(2)}_A \int d\tau_A \dot  
G_{\alpha\beta}^A\dot G_A^{\alpha\beta}  \nonumber \\
&+ & \ldots \biggl\} \, ,
\end{eqnarray}
where $\dot G_A^{\alpha\beta}\equiv  u_A^{\mu} \nabla_{\mu} 
G_{\alpha\beta}^A$ and
where the ellipsis refer to higher-order invariants involving, e.g., higher-than-quadratic tidal scalars, starting with a term cubic in the quadrupolar tidal tensor
$G_{ab}^A$:
\beq
\label{eq:2.3}
\int d\tau_A G^A_{ab}G^A_{bc}G^A_{ca}\,.
\eeq
In the text, we shall focus on the simplest invariants associated with the {\it quadrupolar} ($l=2$) electric-type and magnetic-type tidal tensors $G_{ab}$, $H_{ab}$ (see Appendix D for octupolar-level invariants). The latter are related as follows to the spatial components (in the local frame) of the \lq\lq electric" and \lq\lq magnetic" parts of the Riemann tensor (evaluated, with dimensional regularization, along the considered world line)
\begin{eqnarray}
\label{eq:2.4}
G^A_{\alpha\beta}&\equiv & -{\mathcal E}^A_{\alpha\beta}(U_A)\,,\\
\label{eq:2.5}
H^A_{\alpha\beta}&\equiv & +2\, c\, {\mathcal B}^A_{\alpha\beta}(U_A)\,,
\end{eqnarray} 
where ${\mathcal E}^A_{\alpha\beta}(U_A)$ and ${\mathcal B}^A_{\alpha\beta}(U_A)$ are defined as
\begin{eqnarray}
\label{riemann_em}
{\mathcal E}^A_{\alpha\beta}(U_A)&=& R_{\alpha\mu\beta\nu}U_A^\mu U_A^\nu\nonumber\\
{\mathcal B}^A_{\alpha\beta}(U_A)&=& R^*{}_{\alpha\mu\beta\nu}U_A^\mu U_A^\nu\,,
\end{eqnarray}
and where 
$U_A^\mu\equiv dy_A^\mu/d\tau_A$ denotes the $4$-velocity of body $A$.

We shall always assume that we are interested in the tidal invariant of the body labeled 1 (with mass $m_1$), member of a binary system.
For ease of notation, we shall henceforth often suppress the body label $A=1$.

The quadrupolar electric-like tidal tensor (\ref{eq:2.4}), in comparable mass binary systems,  has been computed to 1PN fractional accuracy in Refs. \cite{Damour:1992qi,Damour:1993zn} (see also Refs. \cite{Vines:2010ca,Taylor:2008xy}). Ref. \cite{JohnsonMcDaniel:2009dq} has also computed to 1PN accuracy the octupolar electric-like tidal tensor, $G_{abc}$, and the quadrupolar magnetic-like tidal tensor $H_{ab}\sim {\mathcal B}_{ab}$. The significantly more involved calculation of tidal effects, along general orbits, in binary systems at the 2PN fractional accuracy has been tackled by Bini, Damour and Faye \cite{Bini:2012gu}. 
For later comparison, let us quote the values of the 2PN-accurate tidal invariants computed in  \cite{Bini:2012gu} for the simple case of {\it circular} orbits.
There are two ways of expressing these results in a gauge-invariant way. First, one can express them in terms of the symmetric, dimensionless frequency parameter
$x=((m_1+m_2)\Omega)^{2/3}$, see Eq. (\ref{eq:1.4}).
Note that $x$ is related to the body-dissymmetric (SF motivated) frequency parameter $y$, Eq. (\ref{y_def}), via
\beq
\label{eq:2.7}
x\equiv \left(1+\frac{m_1}{m_2} \right)^{2/3} y =(1+q)^{2/3}y\,,
\eeq
where $q\equiv m_1/m_2.$

In terms of $x$ the 2PN-accurate results of \cite{Bini:2012gu} read (say, after using Eq. (4.12) there to replace the harmonic-coordinate radius $r^h_{12}$ in terms of $x$)
\begin{eqnarray}
\label{eq:2.8}
J_{e^2}\equiv [{\mathcal E}_{\alpha\beta}(U)]_1^2&=& \frac{6m_2^2}{M^6}x^6 \left[\frac{1-3x+3x^2}{(1-3x)^2}  \right.\nonumber\\
&+& (2X_1^2-X_1)x \nonumber\\  
&+& \left(\frac53 X_1^4-X_1^3+\frac{787}{84}X_1^2\right.\nonumber\\
&& \left.\left.+\frac14 X_1\right)x^2+O_{X_1}(x^3)\right]\,,\\
\label{eq:2.9}
J_{b^2}\equiv [{\mathcal B}_{\alpha\beta}(U)]_1^2&=& \frac{18m_2^2}{M^6}x^7 \left[\frac{1-2x}{(1-3x)^2}  \right.\nonumber\\
&+& \left(\frac{10}{3} X_1^2-2 X_1 \right)x\nonumber\\
&& \left. +O_{X_1}(x^2)\right]\,. 
\end{eqnarray}
Here, we recall that $X_1\equiv m_1/M\equiv q/(1+q)$, and we have included the exact results in the test-mass limit $X_1\to 0$. [The notation $O_a(x^n)$ denotes a term which vanishes with $a$ and which is $O(x^n)$.]

A second useful way of expressing these tidal scalars in a gauge-invariant way is to express them in terms of the EOB radial distance $r_{\rm EOB}$ (which has a gauge-invariant meaning). This can be done either by using Eqs. (5.28) and (5.29) in \cite{Bini:2012gu}, or, by using the exact relation between $u\equiv GM/(c^2 r_{\rm EOB})$ and $x$ predicted by EOB theory \cite{Damour:2009sm}, viz
\beq
\label{eq:2.10}
x=u \left( \frac{-\frac12 A'(u)}{h^2(u)} \right)^{1/3}\,,
\eeq
where
\beq
\label{eq:2.11}
h^2(u)=1+2\nu \left(\frac{A(u)}{\sqrt{\tilde A(u)}}-1 \right)\,,
\eeq
\beq
\label{eq:2.12}
\tilde A(u)\equiv  A(u)+\frac12 u A'(u)\,.
\eeq
Here, $A(u; \nu)$ (making also its dependence on $\nu$ explicit) is the basic EOB radial potential, which genera\-lizes the fa\-mous Schwarz\-schild po\-ten\-tial $1-2GM/(c^2r)$ to the two-body case.
Thanks to many studies over the last years, a lot is known about the EOB radial potential $A(u;\nu)$, both for what concerns its PN expansion (i.e., its expansion in powers of $u=GM/(c^2r_{\rm EOB})$), and its self-force (SF) expansion (i.e., its expansion in powers of $\nu=m_1m_2/M^2=q/(1+q)^2$).
The full PN expansion of $A(u;\nu)$ has been recently determined to the 4PN level \cite{Bini:2013zaa}. For our present purpose, we only need to use the (remarkably simple) 2PN-accurate value of $A(u;\nu)$, namely \cite{Buonanno:1998gg}
\beq
\label{eq:2.13}
A^{\rm 2PN}(u;\nu)=1-2u+2\nu u^3+O_\nu(u^4)\,.
\eeq
Inserting this result in Eq. (\ref{eq:2.10}) yields the links
\begin{eqnarray}
\label{eq:2.14}
x&=&u +\frac13 \nu u^2 +\left(-\frac54+\frac29\nu \right)\nu u^3 +O_\nu(u^4)\,,\\  
\label{eq:2.15}
u&=&x -\frac13 \nu x^2 +\frac54 \nu x^3 +O_\nu(x^4) \,.  
\end{eqnarray}
Inserting (\ref{eq:2.14}) in Eqs. (\ref{eq:2.8}) and (\ref{eq:2.9}) yields
\begin{eqnarray}
\label{eq:2.16}
J_{e^2}&=& \frac{6m_2^2}{r_{\rm EOB}^6}  \left[\frac{1-3u+3u^2}{(1-3u)^2}  \right.\nonumber\\
&+& \left.  X_1 u +\frac{1}{28}(295 X_1^2-7 X_1)u^2+O_{X_1}(u^3)\right]\,,\nonumber\\
\label{eq:2.17}
J_{b^2}&=& \frac{18m_2^2}{r_{\rm EOB}^7} \left[\frac{1-2u}{(1-3u)^2}  \right.\nonumber\\
&+& \left.  \frac13 (3X_1^2+X_1)u+O_{X_1}(u^2)\right]\,.
\end{eqnarray}
It should be noted that the $X_1$-dependence of the invariants $J_{e^2}$ and $J_{b^2}$ is simpler when these scalars are expressed in terms of the EOB radial distance $r_{\rm EOB}$ (with $u\equiv GM/(c^2r_{\rm EOB})$). In particular, the 1PN correction to $J_{e^2}$ is linear in $X_1$, and the 2PN correction is quadratic in $X_1$.
The same holds when expressing $J_{e^2}$ in terms of the harmonic (or ADM) radial distance. By contrast, when expressing $J_{e^2}$ in terms of the frequency parameter $x$, the 1PN correction is quadratic in $X_1$, while the 2PN one is already quartic in $X_1$, see Eq. (\ref{eq:2.8})
[A similar increase in mass-ratio complexity was noticed in Ref. \cite{Bini:2013rfa}, when expressing the binding energy in terms of $x$ instead of $u$.]

We shall come back below to the importance of the nonlinear dependence of the quadrupolar electric tidal invariant $J_{e^2}$ on the mass fraction $X_1=m_1/M$.
To conclude this recap section, let us quote the values taken by $J_{e^2}$ and $J_{b^2}$ when taking the small-mass-ratio limit $X_1=q/(1+q)\to 0$, and expressing them in terms of the SF-friendly (but body-dissymmetric) frequency parameter $y$, Eq. (\ref{y_def}).
For finite values of $q$, the link (\ref{eq:2.7}) between $x$ and $y$ involves nasty powers of $1+q$. However, the link simplifies when considered to first order in $q$, namely
\beq
\label{eq:2.18}
x=\left(1+\frac23 q +O(q^2)  \right)y\,.
\eeq 
Similarly, the 2PN-accurate links (\ref{eq:2.14}) and (\ref{eq:2.15}) yield, to first order in $q$:
\begin{eqnarray}
\label{2.19}
y&=&\left(1-\frac23 q\right)u+\frac13 qu^2  - \frac54 q u^3 +O(u^4)\nonumber\\
&& +O(q^2)\,,\\
\label{2.20}
u&=&\left(1+\frac23 q\right)y -\frac13 qy^2  + \frac54 q y^3 +O(y^4)\nonumber\\
&& +O(q^2) \,.\
\end{eqnarray}
Using these, and working with (SF-motivated) body-dissymmetric dimensionless tidal invariants involving extra powers of the large mass $m_2$, we have
\begin{eqnarray}
\label{eq:2.21}
m_2^4 J_{e^2}&=& 6y^6 \frac{1-3y+3y^2}{(1-3y)^2}+q\left( -12 y^6  \right.\nonumber\\
&& \left. -30 y^7 -\frac{93}{2}y^8 +O(y^9) \right)+O(q^2)\,,\\
\label{eq:2.22}
m_2^4 J_{b^2}&=& 18 y^7 \frac{1-2y}{(1-3y)^2}+q\left( -24 y^7  \right.\nonumber\\
&& \left. -84 y^8  +O(y^9) \right)+O(q^2)\,.
\end{eqnarray}
Let us finally quote the form of the corresponding results for the redshift-rescaled scalars
\begin{eqnarray}
\label{eq:2.23}
{\mathcal J}_{e^2}&=&[{\mathcal E}_{\alpha\beta}(k)]^2=\Gamma_1^{-4}[{\mathcal E}_{\alpha\beta}(U)]^2 \,,\\
\label{eq:2.24}
{\mathcal J}_{b^2}&=&[{\mathcal B}_{\alpha\beta}(k)]^2=\Gamma_1^{-4}[{\mathcal B}_{\alpha\beta}(U)]^2  
\,.
\end{eqnarray}
Here $\Gamma_1\equiv (dt/d\tau)_1\equiv U_1^t\equiv 1/z_1$  is the (inverse) redshift factor along the world line of $m_1$ while $k=\Gamma_1^{-1}U_1=\partial_t +\Omega \partial_\phi$ denotes the Killing vector associated with the helical symmetry of the circular binary system. Both the PN expansion, and the SF expansion, of $\Gamma_1$ have been the focus of many studies in recent years \cite{Detweiler:2008ft,Blanchet:2009sd,Blanchet:2010zd,LeTiec:2011ab,Blanchet:2014bz,Blanchet:2012at,Akcay:2012ea}. For the purpose of this section, we only need the expansion of $\Gamma_1$ up to 2PN-accuracy in $y$, and to first order in $q$
\begin{eqnarray}
\label{eq:2.25}
\Gamma_1 &=& (1-3y)^{-1/2}+q \left( -y \right.\nonumber\\
&& \left. -2y^2-5y^3 +O(y^4)\right)+O(q^2)\,,
\end{eqnarray}
or, equivalently
\begin{eqnarray}
\label{eq:2.26}
\Gamma_1^{-2} &=&  1-3y +q \left( 2y \right.\nonumber\\
&& \left. -5y^2-\frac54y^3 +O(y^4)\right)+O(q^2)\,.
\end{eqnarray}
Inserting this result in Eqs. (\ref{eq:2.23}) and (\ref{eq:2.14}) yields
\begin{eqnarray}
\label{eq:2.27}
m_2^4 {\mathcal J}_{e^2}&=& 6y^6 (1-3y+3y^2)+q\left( -12 y^6  \right.\nonumber\\
&& \left. +66 y^7 -\frac{69}{2}y^8 +O(y^9) \right)+O(q^2)\,,\\
\label{eq:2.28}
m_2^4 {\mathcal J}_{b^2}&=& 18 y^7 (1-2y)+q\left(-24 y^7  \right.\nonumber\\
&& \left. +132 y^8  +O(y^9) \right)+O(q^2)\,.
\end{eqnarray}
In some sections of this paper, the behavior of various SF-expanded quantities as the considered circular orbit approaches the light ring (LR) [i.e., as $x\to 3+O(q)$, $u\to 3+O(q)$ or $y\to 3+O(q)$] will play an important role. Note that this behavior depends very much on the considered quantity. E.g., the $O(q^0)$ pieces in 
$J_{e^2}$ and $ J_{b^2}$ blow up as $\sim (1-3x)^{-2}\sim (1-3u)^{-2}\sim (1-3y)^{-2}$, while their redshifted-rescaled counterparts have finite limits at the LR. [The $O(q^0)$ piece in $\Gamma_1$ goes to infinity as $(1-3y)^{-1/2}$ near the LR.] We shall discuss below the LR behavior of the $O(q^1)$ contributions in these scalars.

\section{Tidal invariants to the fractional 7.5PN level, up to the first order in the mass ratio}

The previous section has recalled the current knowledge (up to the fractional 2PN level) of tidal invariants in {\it comparable-mass} circular binaries. We shall henceforth consider {\it first-order}  gravitational self-force contributions to tidal invariants in {\it small mass ratio} circular binaries.
More precisely, we shall show how to analytically compute $J_{e^2}$, $J_{b^2}$ and several other tidal scalars when working to first order in $q=m_1/m_2\ll 1$. The technique we shall use is a rather straightforward generalization of the approach we used in several recent works \cite{Bini:2013zaa,Bini:2013rfa,Bini:2014nfa,Bini:2014ica}. Let us briefly recall the main features of our technique.

The first feature (which generalizes an idea introduced by Detweiler \cite{Detweiler:2008ft}) is to consider a gauge-invariant function. Here, we regularize and evaluate several scalars $J_{e^2}$, $J_{b^2}$, etc., along the world line ${\mathcal L}_1$ of the small mass $m_1$ in a circular binary. One then considers the functional dependence of these gauge-invariant scalars on the (gauge-invariant) frequency parameter $y$, Eq. (\ref{y_def}).

The second feature is to express the above invariants in terms of the mass-ratio rescaled first-order self-force (1SF) perturbation $h_{\mu\nu}$ of the background metric 
$g_{\mu\nu}^{(0)}$:
\beq
\label{eq:3.1}
g_{\alpha\beta}(x^\mu)=g^{(0)}_{\alpha\beta}(x^\mu,m_2)
+q\, h_{\alpha\beta}(x^\mu)+O(q^2)\,. 
\eeq
Here, $g^{(0)}_{\alpha\beta}$ is taken to be a Schwarzschild metric of mass $m_2$ and we recall that $q=m_1/m_2$.

The third feature is to compute $h_{\alpha\beta}(x^\mu)$ near the world line of $m_1$ by combining several analytical approaches to Regge-Wheeler-Zerilli theory.
The analytical approach used depends on the multipolar order $l$ in the tensor multipolar expansion of $h_{\alpha\beta}(x^\mu)$. The nonradiative multipoles $0\le l \le 1$ are treated analytically, by transforming the results of \cite{Zerilli:1970se,Zerilli:1971wd} to an asymptotically flat gauge.
The radiative multipoles $2\le l \le 5$ are treated by computing the near-zone re-expansion of the hypergeometric-expansion form of Regge-Wheeler-Zerilli theory introduced by Mano, Suzuki and Tagasugi \cite{Mano:1996vt,Mano:1996mf,Mano:1996gn}. The generic, higher-$l$ radiative multipoles $l\ge 6$ are obtained by solving an inhomogeneous Regge-Wheeler equation by a straightforward PN expansion. As discussed in our previous paper \cite{Bini:2014ica} the hypergeometric-expansion treatment of the $l^{\rm th}$ multipole allows one to correctly include the near zone tail effects up to the $(l+2)^{\rm th}$ PN level. The fact that we use such a treatment up to $l=5$ therefore means that our near zone metric starts being inaccurate (because of the use of a straightforward PN expansion) only at the 8PN level. In other words, our near zone metric will be accurate up to the 7.5PN level included (which is the same accuracy that we used in our recent spin-orbit computation \cite{Bini:2014ica}).  This accuracy will allow us to compute the  quadrupolar electric and magnetic tidal invariants $J_{e^2}$, $J_{b^2}$, Eq. (\ref{eq:2.8}) to the fractional 7.5PN accuracy, i.e. to the order $(Gm_2/(c^2r))^{7.5}\sim y^{7.5}$ beyond the Newtonian level result, $J_{e^2}^{\rm Newton}=6m_2/r^6$. [We have the same fractional PN accuracy in $J_{e^2}$ and $J_{b^2}$ because our error in the electric (magnetic) curvature comes from neglecting some tail terms in the corresponding electric (magnetic) $l=6$ multipoles.]

To relieve the tedium, and because many aspects of our present study are similar to our previous works \cite{Bini:2013zaa,Bini:2013rfa,Bini:2014nfa,Bini:2014ica} we shall relegate the technical details of our computation to some Appendices. Let us only stress here the features of our calculations that are conceptually different from those of Ref. \cite{Bini:2014ica}. The first such feature is that we are now evaluating a combination of terms involving up to the {\it second} derivatives of the metric perturbation $h_{\mu\nu}$. 

To be more precise, let $k=\partial_t+\Omega \partial_\phi$ be the helical Killing vector of the spacetime around a binary system (of masses $m_1$ and $m_2$) on circular orbits. We recall that the $4$-velocity vector of body 1, say $U_1^\mu$, is parallel to the value of the Killing vector $k^\mu$ along the world line ${\mathcal L}_1$ of body 1, i.e.,
\beq
\label{eq:3.2}
U_1^\mu =\Gamma_1 k^\mu\,.
\eeq
The proportionality factor $\Gamma_1$ (equal to Detweiler's redshift variable $U_1^t=dt/d\tau_1$) is obtained from the normalization condition
$U_1\cdot U_1=-1$, i.e.,
\beq
\label{eq:3.3}
-\Gamma_1^{-2}=[g_{\mu\nu}k^\mu k^\nu]_1=[g_{tt}+2\Omega g_{t\phi}+\Omega^2 g_{\phi\phi}]_1\,.
\eeq
Here, and below, the brackets $[\ldots ]_1$ indicate that one evaluates (after regularization) a spacetime-varying quantity on the world line ${\mathcal L}_1$.
When the context makes clear what is intended, we shall omit to indicate those evaluation brackets (as we did in Eqs.  (\ref{eq:2.4}), (\ref{riemann_em} above). In addition, as, in the following, we shall always evaluate quantities on the world line of body 1, rather than body 2, we shall often simplify the notation by omitting the body label 1 on quantities such as $U=U_1$ or 
$\Gamma=\Gamma_1$.
Let us  also recall that all the invariants are initially computed as functions of the coordinate radius $r_0$ of particle 1. One then needs to re-express $r_0$ in terms of the gauge-invariant frequency parameter $y$. This is done by using the following relation~\cite{Detweiler:2008ft} 
\beq
\Omega=\sqrt{\frac{m_2}{r_0^3}}
\left( 1-q\frac{r_0^2}{4m_2}[\partial_r h_{kk}]_1  +O(q^2) \right)\,.
\eeq
This relation follows from the geodesic character of ${\mathcal L}_1$ (which also implies the condition $[\partial_{\phi} h_{kk}]_1= 0$). Here, 
 $h_{kk}\equiv  h_{\mu\nu}k^\mu k^\nu$ denotes  the double $k-$contraction of the metric perturbation. 
 
Instead of working with the invariants made of the usual gravitoelectric $U$-projection of the curvature tensor, ${\mathcal E}_{\alpha\beta}(U)=[R_{\alpha\mu \beta \nu} U^\mu U^\nu]_1$, we found convenient to work with the gravitoelectric $k$-projection of the curvature tensor, i.e., 
\beq
\label{eq:3.4}
{\mathcal E}_{\alpha\beta}(k)\equiv [R_{\alpha\mu \beta \nu} k^\mu k^\nu]_1= \Gamma^{-2}{\mathcal E}_{\alpha\beta}(U)\,.
\eeq
The simplest invariant associated with ${\mathcal E}_{\alpha\beta}(k)$ is
\beq
\label{eq:3.5}
{\mathcal J}_{e^2}\equiv [g^{\alpha\alpha'} g^{\beta \beta'}{\mathcal E}_{\alpha\beta}(k){\mathcal E}_{\alpha'\beta'}(k)]_1\equiv {\rm Tr} {\mathcal E}^2(k)  \, .
\eeq
When considering the $m_2$-adimensionalized \footnote{Using $G=c=1$.} version, $\tilde {\mathcal J}_{e^2}\equiv m_2^4 {\mathcal J}_{e^2}$ of ${\mathcal J}_{e^2}$, and inserting in Eq. (\ref{eq:3.5}) the perturbed metric (\ref{eq:3.1}), we get, to first order in the mass ratio $q=m_1/m_2$
\beq
\label{eq:3.6}
\tilde {\mathcal J}_{e^2}\equiv m_2^4 {\mathcal J}_{e^2}=6y^6(1-3y+3y^2)+q\, \delta_{e^2}(y)+O(q^2)\, ,
\eeq
where the first term on the right-hand-side (r.h.s.) is the unperturbed ($m_2$-Schwarzschild background) scalar ${\rm Tr} [{\mathcal E}^2(k)]$ expressed in terms of the $m_2$-scaled frequency parameter $y\equiv (m_2\Omega)^{2/3}$, and where the $O(q)$ perturbation is given by the following combination of derivatives of $h_{\mu\nu}$
\begin{widetext}
\begin{eqnarray}
\label{eq:3.7}
\delta_{e^2}(y) &=& - y^5 \partial_{\theta\theta} h_{kk} -\frac{(3y-1)^2 y^5}{ 1-2 y}\partial_{\bar\phi\bar\phi} h_{kk} + (1-2y)(2-3y)y^3 m_2^2 \partial_{rr} h_{kk}\nonumber\\
&&+2
 (1-3y)y^{11/2} \partial_{\bar \phi}h_{tr}+2\frac{(1-3y)y^7}{m_2 }\partial_{\bar \phi} h_{r\phi}
+2 (1-3y)(2-3y)y^4m_2 \partial_r h_{kk}\nonumber\\
&& -2\frac{(1-3y)y^7}{m_2 }\partial_{r}h_{\phi\phi}-2 (1-3y)y^{11/2} \partial_{r}h_{t\phi} \nonumber\\
&& -2 (1-2y)(18y^2-18y+5)y^6 h_{rr} +2\frac{(1-3y)(1-y)y^8}{(1-2y)m_2^2}h_{\phi\phi}+4\frac{(1-3y)y^{15/2}}{(1-2y)m_2}h_{t\phi}\nonumber\\
&& -2 \frac{y^8}{m_2^2}h_{\theta\theta}+  \frac{2y^7}{ (1-2 y)} h_{kk}\,.
\end{eqnarray}
\end{widetext}
The r.h.s. of Eq. (\ref{eq:3.7})  (here written without making use of Einstein's equations) is meant to be regularized and evaluated at the location of particle 1. We have already used the fact that particle 1 moves along an equatorial $(\theta=\pi/2)$ circular orbit located at the radius $r_0=m_2/y+O(q)$.
The evaluation, and regularization, of $\delta$ is done along the same lines as our previous work  \cite{Bini:2014ica}. The metric perturbation $h_{\mu\nu}$ is decomposed into tensor spherical harmonics $h_{\mu\nu}^{lm}$ (of even and odd types), and is computed in Regge-Wheeler gauge. Each $lm$ multipolar contribution to $\delta$ is finite. As already mentioned, the explicit computation of $h_{\mu\nu}^{lm}$, and the corresponding $\delta_{lm}\equiv \delta[h_{\mu\nu}^{lm, \rm even}]+\delta[h_{\mu\nu}^{lm, \rm odd}]$, depends on the value of $l$. The non radiative multipoles $0\le l \le 1$ are exactly known. The low radiative multipoles $2\le l \le 5$ are computed as hypergeometric-function expansions, which are then re-expanded in powers of $\omega r$ and $m_2/r$. The higher radiative multipoles $l\ge 6$ are directly computed as a PN expansion, i.e., as a near-zone expansion in powers of $\omega r$ and $m_2/r$.
For each value of $l$, after the near-zone expansion, the dependence on the \lq\lq magnetic multipolar number" $m$ is sufficiently explicit to allow one to perform the summation of $\delta_{lm}$ over $m$ (from $-l$ to $+l$), thanks to the existence of standard summation rules \cite{Nakano:2003he}.

The singular nature of $\delta$ as the location of particle 1 is approached shows up in the fact that the value of the r.h.s. of Eq. (\ref{eq:3.7})  depends on whether the radial coordinate $r$ of the field point $x^\lambda$, where $h_{\mu\nu}^{lm}(x^\lambda)$ and its derivatives are evaluated, approaches the radial location $r_0$
of particle 1 from above or from below.

Let $\delta_{lm}^+$ denote the result obtained when $r\to r_0^+$, and $\delta_{lm}^-$ the result obtained when $r\to r_0^-$. The corresponding results after summation over $m$ are denoted $\delta_l^\pm=\sum_{m=-l}^l \delta_{lm}^\pm$.
As in Ref. \cite{Bini:2014ica}, it is convenient to focus on the average between the two limits, say
\beq
\label{eq:3.8}
\delta_l^0\equiv \frac12 (\delta_l^+ + \delta_l^-)\,.
\eeq
Indeed, this radial-limit average eliminates some singular terms (namely, those that are odd under reflection around the particle location, such as singular terms of the type $\partial_\mu \rho^{-1}$, where $\rho$ denotes the distance between the field point and the world line).

When considering gauge-invariant perturbed quantities $\delta$ that depended only on $h_{\mu\nu}$ and, possibly, its first derivatives \cite{Bini:2013zaa,Bini:2013rfa,Bini:2014nfa,Bini:2014ica}, the average $\delta_l^0$, Eq. (\ref{eq:3.8}) was found to have a limit as $l\to \infty$. Here, the presence of second derivatives in $\delta$, Eq. (\ref{eq:3.7}), corresponds to a more singular spacetime behavior around ${\mathcal L}_1$ (involving $\partial_{\mu\nu}\rho^{-1}$). As a priori expected, we found that this implied a quadratic growth of $\delta_l^0$ as $l\to \infty$. More precisely, we found (from our generic-$l$ PN-expanded analytic solution) that the large $l$ behavior of $\delta_l^0$ has the form 
\beq
\label{eq:3.9}
\delta_l^0 = b_0(y) l(l+1)+b_1(y)+O\left(\frac{1}{l^2}\right)\,.  
\eeq

A convenient technical feature of our approach is that we can (by using our PN-expanded solution) analytically compute, to any preassigned order, the PN expansion of the two coefficients $b_0(y)$ and $b_1(y)$. For instance, in the case of $\tilde {\mathcal J}_{2e}$  we found
\begin{eqnarray}
\label{eq:3.10}
b_0(y)&=& 6 y^6-\frac{63}{2} y^7+\frac{1251}{32} y^8+\frac{105}{128} y^9+\frac{15435}{8192} y^{10}\nonumber\\
&& +\frac{143073}{32768} y^{11}+\frac{5353803}{524288} y^{12}+\frac{50560281}{2097152} y^{13}\nonumber\\
&& +O(y^{14})\\
\label{eq:3.11}
b_1(y)&=&
-\frac{183}{8}y^7 +\frac{4335}{64}y^8-\frac{43437}{512}y^9+\frac{105447}{8192}y^{10}\nonumber\\
&& +\frac{4328493}{131072}y^{11}+\frac{89808549}{1048576} y^{12}+\frac{1882340487}{8388608} y^{13}\nonumber\\
&& +O(y^{14})\,.
\end{eqnarray}

This allows us to compute the \lq\lq subtraction term"
\beq
B(y; l)\equiv b_0(y) l(l+1)+b_1(y)
\eeq
to any preassigned PN order.
Finally, the regularized value of $\delta$ is given by the convergent series (see Eq. (\ref{eq:3.9}))
\beq
\label{eq:3.12}
\delta^{\rm reg}=\sum_{l=0}^\infty \left( \delta_l^0-B(y;l)\right)\,.
\eeq
With this technique we were able to compute the PN expansion of the function $\delta^{\rm reg}(y)$ up to the fractional 7.5PN accuracy, i.e., modulo a fractional error term $O_{\ln{}}(y^8)$, or an absolute error term $y^6\, O_{\ln{}}(y^8)=O_{\ln{}}(y^{14})$. [Here, $O_{\ln{}}(y^n)$ denotes a term of order $y^n$ modulo logarithmic corrections.] The technical details of our computation are given in Appendix A. Our final result for the  $O(q)$ (1SF) term in Eq. (\ref{eq:3.6}) reads
\begin{widetext}
\begin{eqnarray}
\label{eq:3.13}
\delta^{\rm reg}_{e^2}(y)&=& -12 y^6+66 y^7 -\frac{69}{2} y^8 +\left(-\frac{2407}{4} +\frac{1779}{128}\pi^2\right) y^9\nonumber\\
&+&\left(\frac{2339879}{800} -\frac{38949}{512}\pi^2-\frac{9216}{5}\ln(2)-\frac{4608}{5}\gamma-\frac{2304}{5}\ln(y)\right) y^{10}\nonumber\\
&+& \left(\frac{206336}{35}\gamma-\frac{232180789}{33600} +\frac{103168}{35}\ln y +\frac{501504}{35}\ln 2 +\frac{1393795}{4096}\pi^2 -\frac{17496}{7}\ln 3 \right) y^{11}\nonumber\\
&-&\frac{164352}{175}\pi y^{23/2}\nonumber\\
&+&\left(\frac{9913288243}{1209600}-\frac{4725416287}{1179648}\pi^2-\frac{6580119}{524288}\pi^4-\frac{199504}{189}\gamma+30618\ln(3)\right.\nonumber\\
&& \left. -\frac{37677392}{945}\ln(2)-\frac{99752}{189}\ln(y)\right) y^{12}\nonumber\\
&+&\frac{7660504}{1225}\pi y^{25/2}\nonumber\\
&+&\left(\frac{7299159446817431}{32598720000}-\frac{11139849}{80}\ln(3)-\frac{203838659456}{1819125}\gamma+\frac{70709473888}{1819125}\ln(2)\right.\nonumber\\
&-&\frac{101919329728}{1819125}\ln(y)-\frac{9765625}{528}\ln(5)+\frac{5259264}{175}\ln(2)\gamma+\frac{2629632}{175}\ln(y)\ln(2)\nonumber\\
&+&\frac{1314816}{175}\ln(y)\gamma-\frac{42271455505841}{3303014400}\pi^2+\frac{328704}{175}\ln^2(y)+\frac{1314816}{175}\gamma^2\nonumber\\
&+&\left.\frac{5259264}{175}\ln^2(2)-\frac{73728}{5}\zeta(3)+\frac{16267066167}{33554432}\pi^4\right) y^{13}\nonumber\\
&-& \frac{675068098}{218295}\pi y^{27/2}
 +O_{\ln{}}(y^{14})\,.
\end{eqnarray}
\end{widetext}
 
Using the same technique we computed several other tidal invariants as function of $y$; see Appendices B, C, D for details.

First, besides the {\it quadratic} tidal electric invariant ${\rm Tr}[{\mathcal E}^2(k)]$, we also computed the trace of the {\it cube} of the tidal electric matrix ${\mathcal E}^\mu{}_\nu(k)$.
Writing the 1SF contribution to the adimensionalized version of this cubic invariant in {\it factorized} form,
\beq
\label{eq:3.14}
m_2^6 {\rm Tr}[{\mathcal E}^3(k)]=-3(1-3y)(2-3y)y^9 \left(1+q \widehat \delta_{e^3}(y) \right) +O(q^2)\,,
\eeq
we found the 7.5PN-accurate result (see Appendix B for details):
\begin{widetext}
\begin{eqnarray}
\label{eq:3.15}
\widehat \delta^{\rm reg}_{e^3}(y)&=&   -3+\frac{15}{2}y+\frac{147}{8}y^2+\left(\frac{1779}{512}\pi^2-\frac{1561}{16}\right)y^3\nonumber\\
&& +\left(\frac{1336679}{3200}-\frac{2304}{5}\ln(2)-\frac{576}{5}\ln(y)-\frac{1152}{5}\gamma-\frac{2271}{256}\pi^2\right)y^4\nonumber\\
&& +\left(-\frac{11479819}{134400}+\frac{907147}{16384}\pi^2+\frac{2336}{7}\ln(y)+\frac{68928}{35}\ln(2)+\frac{4672}{7}\gamma-\frac{4374}{7}\ln(3)\right) y^5\nonumber\\
&& -\frac{41088}{175} \pi y^{11/2} \nonumber\\
&& +\left(-\frac{21915684437}{4838400}-\frac{6580119}{2097152}\pi^4-\frac{900450163}{4718592}\pi^2-\frac{386860}{189}\ln(2)\right. \nonumber\\
&& \left. +\frac{1186538}{945}\ln(y)+\frac{2373076}{945}\gamma+\frac{37179}{7}\ln(3)\right) y^6\nonumber\\
&& +\frac{181694}{245}\pi y^{13/2} \nonumber\\
&&+\left(\frac{170773056511481}{130394880000}-\frac{17943532507}{1819125}\ln(y)-\frac{29165103}{2240}\ln(3)+\frac{328704}{175}\ln(y)\gamma\right. \nonumber\\
&&+\frac{657408}{175}\ln(y)\ln(2)+\frac{1314816}{175}\ln(2)\gamma-\frac{35887065014}{1819125}\gamma+\frac{82176}{175}\ln^2(y)-\frac{9765625}{2112}\ln(5)\nonumber\\
&&-\frac{4286862278}{1819125}\ln(2)+\frac{1314816}{175}\ln^2(2)+\frac{7573535048959}{13212057600}\pi^2+\frac{328704}{175}\gamma^2-\frac{18432}{5}\zeta(3)\nonumber\\
&& \left.+\frac{5955078711}{134217728}\pi^4\right) y^7 +\frac{497879621}{218295} \pi y^{15/2}+O_{\ln{}}(y^8)\,.
\end{eqnarray}
\end{widetext}
In addition we considered the quadratic tidal-magnetic invariant
\beq
\label{eq:3.16}
{\rm Tr}[{\mathcal B}^2(k)]=[{\mathcal B}^\mu{}_\nu(k){\mathcal B}^\nu{}_\mu(k)]_1
\eeq
where ${\mathcal B}_{\mu\nu}(k)=R^*_{\mu\alpha\nu\beta}k^\alpha k^\beta\equiv \Gamma^{-2}{\mathcal B}_{\mu\nu}(U)$. The 1SF accurate expansion (in nonfactorized form) of its $m_2$-adimensionalized version reads
\beq
\label{eq:3.17}
m_2^4{\rm Tr}[{\mathcal B}^2(k)]=18 (1-2y) y^7 +q \delta_{b^2}(y)+O(q^2)\,,
\eeq
where the 7.5PN accurate value of the 1SF correction reads (see Appendix C for details)
\begin{widetext}
\begin{eqnarray}
\label{eq:3.18}
\delta^{\rm reg}_{b^2}(y)  
&=&-24 y^7+132 y^8-201 y^9+\left(-\frac{1591}{2} +\frac{123}{4}\pi^2\right) y^{10}\nonumber\\
&& +\left(\frac{56441}{240} +\frac{57815}{256}\pi^2-\frac{3616}{5}\ln(y)-2880\ln(2)-\frac{7232}{5}\gamma  \right) y^{11}\nonumber\\
&& +\left(-\frac{54794167}{5600} +\frac{191973}{256}\pi^2+\frac{127992}{35}\ln(y)+\frac{255984}{35}\gamma+\frac{622064}{35}\ln(2)-\frac{21870}{7}\ln(3)\right) y^{12}\nonumber\\
&& -\frac{10272}{7}\pi y^{25/2}\nonumber\\
&& +\left(-\frac{7934674343}{294912}\pi^2+\frac{657864577393}{4233600} +\frac{80074047}{131072}\pi^4+\frac{2981848}{945}\ln(y)+\frac{5963696}{945}\gamma-\frac{25619152}{945}\ln(2)\right.\nonumber\\
&& \left. +\frac{161109}{5}\ln(3)\right) y^{13} \nonumber\\
&& +\frac{5722586}{735} \pi y^{27/2}\nonumber\\
&& +\left(\frac{324324616007067631}{146694240000} -\frac{21379755466}{202125}\ln(2)-\frac{172041813}{1540}\ln(3)-\frac{9765625}{594}\ln(5)\right.\nonumber\\
&& -\frac{11088032174}{67375}\gamma-\frac{5544016087}{67375}\ln(y)+\frac{24556928}{525}\ln^2(2)-\frac{114944}{5}\zeta(3)-\frac{14269241969123}{206438400}\pi^2\nonumber\\
&&  +\frac{1537376}{525}\ln^2(y)+\frac{6149504}{525}\gamma^2 -\frac{116483667391}{8388608}\pi^4+\frac{4095104}{175}\ln(2)\ln(y)\nonumber\\
&& \left.+\frac{8190208}{175}\ln(2)\gamma+\frac{6149504}{525}\ln(y)\gamma\right) y^{14}\nonumber\\
&& +\frac{296749969}{72765}\pi y^{29/2} +O_{\ln {}}(y^{15})\,.
\end{eqnarray}
\end{widetext}

\section{Comparison with numerical tidal self-force results of Dolan et al.}

Dolan et al. \cite{Dolan:2014pja}
 have recently numerically evaluated the 1SF contribution to the eigenvalues of the tidal-electric, and tidal-magnetic, quadrupolar tensors $m_2^2{\mathcal E}^\mu{}_\nu(U)$,  $m_2^2{\mathcal B}^\mu{}_\nu(U)$. These eigenvalues are such that 
\begin{eqnarray}
\label{eq:4.1}
m_2^2 {\mathcal E}(U)&=& {\rm diag} [\lambda_1^{\rm (E)},\lambda_2^{\rm (E)},-(\lambda_1^{\rm (E)}+\lambda_2^{\rm (E)})]\nonumber\\
m_2^2 {\mathcal B}(U)&=& {\rm diag} [\lambda^{\rm (B)},-\lambda^{\rm (B)},0]\,,
\end{eqnarray}
where we used their tracelessness, and the existence of a zero eigenvalue of ${\mathcal B}(U)$ \cite{Dolan:2014pja}. Let us introduce a notation for the eigenvalues of the corresponding Killing-scaled tidal tensors
\begin{eqnarray}
\label{eq:4.2}
m_2^2 {\mathcal E}(k)&=& {\rm diag} [\sigma_1^{\rm (E)},\sigma_2^{\rm (E)},-(\sigma_1^{\rm (E)}+\sigma_2^{\rm (E)})]\nonumber\\
m_2^2 {\mathcal B}(k)&=& {\rm diag} [\sigma^{\rm (B)},-\sigma^{\rm (B)},0]\,.
\end{eqnarray} 
The unperturbed (0SF) values of these eigenvalues, as functions of the background frequency parameter $y=(m_2\Omega)^{2/3}$ are
\begin{eqnarray}
\label{eq:4.3}
\lambda_1^{{\rm (E)}0}&=& -y^3 \frac{2-3y}{1-3y}\nonumber\\
\lambda_2^{{\rm (E)}0}&=& y^3 \frac{1}{1-3y}\nonumber\\
\lambda_3^{{\rm (E)}0}&\equiv&-(\lambda_1^{{\rm (E)}0}+\lambda_2^{{\rm (E)}0})=y^3 \nonumber\\
\lambda^{{\rm (B)}0}&=& 3y^{7/2}\frac{\sqrt{1-2y}}{1-3y} \,,
\end{eqnarray}
\begin{eqnarray}
\label{eq:4.4}
\sigma_1^{{\rm (E)}0}&=& -y^3(2-3y)\nonumber\\
\sigma_2^{{\rm (E)}0}&=& y^3\nonumber\\
\sigma_3^{{\rm (E)}0}&\equiv&-(\sigma_1^{{\rm (E)}0}+\sigma_2^{{\rm (E)}0})=y^3(1-3y) \nonumber\\
\sigma^{{\rm (B)}0}&=& 3y^{7/2}\sqrt{1-2y}\,.
\end{eqnarray}

Let us write the SF expansion of any $m_2$-adimensionalized (gauge-invariant) function of $y$ as
\beq
\label{eq:4.4bis}
f(y)=f^0(y)+qf^{\rm 1SF}(y)+O(q^2)\,.
\eeq
Dolan et al. \cite{Dolan:2014pja} have numerically computed $\lambda_1^{\rm (E)1SF}$, $\lambda_2^{\rm (E)1SF}$ and $\lambda^{\rm (B)1SF}$. To compare our high-order analytic results to their numerical estimates we have used our three invariants ${\rm Tr}[{\mathcal E}^2(k)]$, ${\rm Tr}[{\mathcal E}^3(k)]$, ${\rm Tr}[{\mathcal B}^2(k)]$ to analytically compute $\sigma_1^{{\rm (E)}}$, $\sigma_2^{{\rm (E)}}$ and $\sigma^{{\rm (B)}}$, and then used the exact link
\begin{eqnarray}
\label{eq:4.5}
\lambda_a^{\rm (E)}=\Gamma^2 \sigma_a^{\rm (E)}\,,\qquad
\lambda_a^{\rm (B)}=\Gamma^2 \sigma_a^{\rm (B)}\,,
\end{eqnarray}
together with the 8.5PN accurate 1SF expansion of $\Gamma$ derived in our previous work \cite{Bini:2014nfa}, to analytically compute high-order PN expansions of 
$\lambda_1^{\rm (E)1SF}$, $\lambda_2^{\rm (E)1SF}$, and $\lambda^{\rm (B)1SF}$.
More precisely, if we introduce the notation 
\begin{eqnarray}
\label{eq:4.6}
\alpha_{\rm 1SF}&=&\frac12 \delta_{e^2}(y) \nonumber\\
\beta_{\rm 1SF}&=& -(1-3y)(2-3y)y^9  \, \widehat \delta_{e^3}(y)\,,
\end{eqnarray}
so that
\begin{eqnarray}
\label{eq:4.7}
\frac12 m_2^4 {\rm Tr}[{\mathcal E}^2(k)]&=& 3y^6(1-3y+3y^2)+q\alpha_{\rm 1SF}+O(q^2) \nonumber\\
\frac13 m_2^6 {\rm Tr}[{\mathcal E}^3(k)]&=& -(1-3y)(2-3y)y^9+q \beta_{\rm 1SF}\nonumber\\
&& +O(q^2) \,,
\end{eqnarray}
the 1SF perturbation of the exact equations
\begin{eqnarray}
\label{eq:4.8}
\frac12 m_2^4 {\rm Tr}[{\mathcal E}^2(k)]&=&\sigma_1^{\rm (E)}{}^2+\sigma_2^{\rm (E)}{}^2+\sigma_1^{\rm (E)}\sigma_2^{\rm (E)}\nonumber\\
\frac13 m_2^6  {\rm Tr}[{\mathcal E}^3(k)]&=& -\sigma_1^{\rm (E)} \sigma_2^{\rm (E)}(\sigma_1^{\rm (E)}+\sigma_2^{\rm (E)})\,,
\end{eqnarray} 
yields a linear system of two equations for the two unknowns $\sigma_1^{\rm (E)1SF}$, $\sigma_2^{\rm (E)1SF}$ with $\alpha_{\rm 1SF}$ and $\beta_{\rm 1SF}$ as r.h.s.'s.
The (unique) solution of this system reads
\begin{eqnarray}
\label{eq:4.9}
 \sigma_1^{\rm (E) 1SF}  &=& \frac{\alpha_{\rm 1SF}  \sigma_1^{\rm (E)0} +\beta_{\rm 1SF} }{ ( \sigma_1^{\rm (E)0}-  \sigma_2^{\rm (E)0})
(2 \sigma_1^{\rm (E)0}+   \sigma_2^{\rm (E)0})} \nonumber\\
&=&\frac{-y^3(2-3y)  \alpha_{\rm 1SF} +\beta_{\rm 1SF} }{ 9y^6(1-y)(1-2y))} \nonumber\\
\sigma_2^{\rm (E) 1SF}  &=& \frac{\alpha_{\rm 1SF}  \sigma_2^{\rm (E)0} +\beta_{\rm 1SF} }{ ( \sigma_2^{\rm (E)0}-  \sigma_1^{\rm (E)0})
(2  \sigma_2^{\rm (E)0}+   \sigma_1^{\rm (E)0})}\nonumber\\
&=&  \frac{y^3\alpha_{\rm 1SF}  +\beta_{\rm 1SF} }{9y^7(1-y)}\,.
\end{eqnarray}

Note that the denominators $(2 \sigma_1^{\rm (E)0}+   \sigma_2^{\rm (E)0})$ and $(2  \sigma_2^{\rm (E)0}+   \sigma_1^{\rm (E)0})$ have different PN orders. Indeed, in the Newtonian limit ($y\to 0$) $\sigma_1^{\rm (E)0}\simeq -2y^3$, $\sigma_2^{\rm (E)0}\simeq +y^3$, so that $(2 \sigma_1^{\rm (E)0}+   \sigma_2^{\rm (E)0})\simeq -3y^3$, while $(2  \sigma_2^{\rm (E)0}+   \sigma_1^{\rm (E)0})=O(y^4)$ is of 1PN fractional magnitude.

As a consequence, when inserting in Eqs. (\ref{eq:4.9}) our 7.5PN accurate results for  $\alpha_{\rm 1SF}$ and $\beta_{\rm 1SF}$ (using Eqs. (\ref{eq:3.13}) and (\ref{eq:3.15}) above), we were able to determine $\sigma_1^{\rm (E)0}$ to the fractional 7.5PN accuracy while we lost one 1PN level in the analytic accuracy of $\sigma_2^{\rm (E)0}$. Using then the exact link (\ref{eq:4.5}), together with our 8.5PN accurate result for $\Gamma^2(y)$ \cite{Bini:2014nfa}, we also computed the corresponding (7.5PN and 6.5PN accurate) expressions of $\lambda_1^{\rm (E)1SF}$ and  $\lambda_2^{\rm (E)1SF}$.
For brevity, let us only quote here our results for the $U$-normalized eigenvalues
\begin{widetext}
\begin{eqnarray}
\label{eq:4.10a}
 \lambda_1^{\rm (E) 1SF}  &=& 2 y^3+2 y^4-\frac{19}{4}y^5+\left(\frac{227}{3}-\frac{593}{256}\pi^2\right) y^6\nonumber\\
&&+\left(-\frac{71779}{4800}-\frac{719}{256}\pi^2+\frac{1536}{5}\ln(2)+\frac{384}{5}\ln(y)+\frac{768}{5}\gamma\right)y^7\nonumber\\
&&+\left(\frac{35629703}{100800}-\frac{1008787}{24576}\pi^2-\frac{8576}{105}\ln(y)-\frac{5248}{7}\ln(2)-\frac{17152}{105}\gamma+\frac{2916}{7}\ln(3)\right)y^8\nonumber\\
&& +\frac{27392}{175}\pi y^{17/2}\nonumber\\
&& +\left(-\frac{5435624}{2835} \gamma+\frac{4692901483}{7077888} \pi ^2-\frac{2717812}{2835} \ln(y)+\frac{877432}{2835} \ln(2)-\frac{20898}{7} \ln(3)+\frac{2193373}{1048576} \pi ^4 \right. \nonumber\\
&& \left.
-\frac{6746904013}{7257600}  \right) y^9\nonumber\\
&& -\frac{254116}{1225} \pi  y^{19/2}\nonumber\\
&& +\left(\frac{58241403128}{5457375} \gamma-\frac{876544}{175} \gamma \ln(2)-\frac{219136}{175} \ln(y) \gamma
-\frac{438272}{175} \ln(y) \ln(2)\right. \nonumber\\
&& +\frac{113134518813241}{19818086400} \pi ^2+\frac{29120701564}{5457375} \ln(y)+\frac{6396680456}{5457375} \ln(2)
+\frac{6028101}{1120} \ln(3)\nonumber\\
&& -\frac{6653357405}{67108864} \pi ^4+\frac{12288}{5} \zeta(3)+\frac{9765625}{3168} \ln(5)-\frac{876544}{175} \ln^2(2)
-\frac{219136}{175} \gamma^2\nonumber\\
&& \left. -\frac{54784}{175} \ln^2(y)-\frac{1964481413350639}{48898080000}  \right) y^{10}\nonumber\\
&& -\frac{5977039346}{3274425} \pi  y^{21/2}+O_{\ln{}}(y^{11}) \,,
\end{eqnarray}
\begin{eqnarray}
\label{eq:4.10b}
 \lambda_2^{\rm (E) 1SF}  &=&  -y^3-\frac32 y^4-\frac{23}{8}y^5+\left(-\frac{2593}{48}+\frac{1249}{1024}\pi^2\right) y^6\nonumber\\
&& +\left(-\frac{362051}{3200}-\frac{128}{5}\ln(y)+\frac{1737}{1024}\pi^2-\frac{256}{5}\gamma-\frac{512}{5}\ln(2)\right) y^7\nonumber\\
&& +\left(\frac{917879}{1280}-\frac{7637151}{65536} \pi ^2+\frac{16592}{105} \ln(2)+\frac{88}{7} \ln(y)+\frac{176}{7} \gamma-\frac{729}{7} \ln(3)\right) y^8 \nonumber\\
&& -\frac{27392}{525} \pi  y^{17/2}\nonumber\\
&& +\left(\frac{1193824}{2835} \gamma-\frac{24327985735}{14155776} \pi ^2+\frac{596912}{2835} \ln(y)
+\frac{2368}{405} \ln(2)+\frac{1215}{2} \ln(3)+\frac{29225393}{2097152} \pi ^4\right. \nonumber\\
&& \left. +\frac{35725395527}{2903040}  \right) y^9\nonumber\\
&&+\frac{58087}{1575} \pi  y^{19/2}+O_{\ln{}}(y^{10}) \,,
\end{eqnarray}
\begin{eqnarray}
\label{eq:4.10c}
 \lambda_3^{\rm (E) 1SF}  &=& -y^3-\frac12 y^4+\frac{61}{8}y^5+\left(-\frac{1039}{48}+\frac{1123}{1024}\pi^2\right) y^6\nonumber\\
&& +\left(\frac{1229711}{9600}+\frac{1139}{1024}\pi^2-\frac{256}{5}\ln(y)-\frac{512}{5}\gamma-\frac{1024}{5}\ln(2)\right)y^7\nonumber\\
&& +\left(-\frac{431650697}{403200}+\frac{30981749}{196608} \pi ^2+\frac{7256}{105} \ln(y)+\frac{62128}{105} \ln(2)+\frac{14512}{105} \gamma-\frac{2187}{7} \ln(3)\right) y^8\nonumber\\
&& -\frac{54784}{525} \pi  y^{17/2}\nonumber\\
&& +\left(\frac{848360}{567} \gamma+\frac{14942182769}{14155776} \pi ^2+\frac{424180}{567} \ln(u)
-\frac{894008}{2835} \ln(2)+\frac{33291}{14} \ln(3)-\frac{33612139}{2097152} \pi ^4  \right. \nonumber\\
&& \left.-\frac{165133169609}{14515200}  \right) y^9\nonumber\\
&&+\frac{376087}{2205} \pi  y^{19/2}+O_{\ln{}}(y^{10})
 \,.
\end{eqnarray}
\end{widetext}

Concerning the quadrupolar tidal-magnetic eigenvalue $\lambda^{\rm (B)}$, Eq. (\ref{eq:4.1}), it was enough to use our computation of 
\beq
\label{eq:4.11}
\frac12 m_2^4 {\rm Tr}[{\mathcal B}^2(k)]=\left(\sigma^{\rm (B)}\right)^2
\eeq
to determine the 7.5PN accurate expansion of $\sigma^{\rm (B)}=\sigma^{{\rm (B)}0}+q\sigma^{(B)1SF}+O(q^2)$.
Here we use the positive sign for the $0$SF  magnetic eigenvalue $
\sigma^{{\rm (B)}0}=3y^{7/2}\sqrt{1-2y}$. [Note that Ref. \cite{Dolan:2014pja}  works instead with the opposite-sign eigenvalue.]
 Using again the link $\lambda^{\rm (B)}=\Gamma^2 \sigma^{\rm (B)}$ and the results of Ref.  \cite{Bini:2014nfa} we got (note the minus sign on the left-hand side)
\begin{widetext}
\begin{eqnarray}
\label{eq:4.12}
 -\lambda^{\rm (B) 1SF}  &=&  
2y^{7/2}+3y^{9/2}+\frac{59}{4}y^{11/2}-\left(\frac{41}{16}\pi^2-\frac{2761}{24} \right)y^{13/2}\nonumber\\
&&-\left(\frac{112919}{3072}\pi^2-\frac{1808}{15}\gamma-240\ln(2)-\frac{1618039}{2880}-\frac{904}{15}\ln(y)  \right)y^{15/2}\nonumber\\
&&-\left(-\frac{2756}{105}\gamma-\frac{3645}{14}\ln(3)-\frac{1378}{105}\ln(y)-\frac{491047651}{201600}+\frac{565685}{3072}\pi^2+\frac{4492}{21}\ln(2)  \right)y^{17/2}\nonumber\\
&& +\frac{856}{7}\pi y^9 \nonumber\\
&& -\left(\frac{200961}{140}\ln(3)+\frac{1940698}{2835}\ln(y)+\frac{3881396}{2835}\gamma+\frac{1992212}{2835}\ln(2)+\frac{26691349}{524288}\pi^4 \right. \nonumber\\
&& \left. 
+\frac{454873888681}{50803200}-\frac{7377893735}{3538944}\pi^2  \right)y^{19/2} \nonumber\\
&&-\frac{69473}{22050}\pi y^{10} \nonumber\\
&& -\left(\frac{6139232}{1575}\ln^2(2)-\frac{89531499967}{100663296}\pi^4
-\frac{42496203125923}{2477260800}\pi^2-\frac{83360241649}{10914750}\gamma-\frac{1412559}{1232}\ln(3)  \right. \nonumber\\
&& -\frac{9765625}{7128}\ln(5)-\frac{82889847697}{10914750}\ln(2)
-\frac{83360241649}{21829500}\ln(y)+\frac{1537376}{1575}\gamma^2-\frac{28736}{15}\zeta(3)+\frac{2047552}{525}\gamma\ln(2) \nonumber\\
&& \left. +\frac{384344}{1575}\ln^2(y)
+\frac{423235951437871681}{1760330880000}+\frac{1023776}{525}\ln(y)\ln(2)
+\frac{1537376}{1575}\ln(y)\gamma  \right)y^{21/2} \nonumber\\
&&-\frac{5843221973}{4365900}\pi y^{11} 
+O_{\ln{}}(y^{23/2})
 \,.
\end{eqnarray}
\end{widetext}
The analytical values of the first three coefficients (i.e. up to the fractional 2PN level) in the expansions above for
$ \lambda_1^{\rm (E) 1SF}$, $ \lambda_2^{\rm (E) 1SF}$, $ \lambda_3^{\rm (E) 1SF}$ and $ \lambda^{\rm (B) 1SF}$
agree with the values inferred in  Ref. \cite{Dolan:2014pja} from their accurate numerical data.
We then compared our full 7.5PN-accurate analytical result (\ref{eq:4.10a}) for $ \lambda_1^{\rm (E) 1SF}$ to the numerical results displayed in Table I of the August 2014 Arxive preprint of Dolan et al. \cite{Dolan:2014pja}. In doing this comparison it is useful to work with the following Newtonian-rescaled version of $ \lambda_1^{\rm (E) 1SF}$:
\beq
\label{eq:4.13}
\widehat \lambda_1^{\rm (E) 1SF} \equiv \frac{\lambda_1^{\rm (E) 1SF} }{y^3}=2+2y-\frac{19}{4}y^2+\ldots +O_{\ln{}}(y^8)
\eeq
To gauge the quality of the agreement between the analytical and the numerical results it is useful to derive a plausible upper bound on the analytical error term $O_{\ln{}}(y^8)$ in Eq. (\ref{eq:4.13}). Using the analytical knowledge (see next section) that $\lambda_1^{\rm (E) 1SF}$ has a singular behavior $\propto (1-3y)^{-5/2}$ at the light-ring $y\to \frac13$, one can analytically expect that the PN error term should be roughly of the type
\beq
\label{eq:4.14}
\frac{(3y)^8}{(1-3y)^{5/2}}(c_0+c_1 \ln(3y) +c_2 \ln^2 (3y))\,.
\eeq
Here, we factorized out the numerical coefficient $3^8$ (linked to the convergence radius associated with the existence of a singularity at $y=\frac13$), so that 
we a priori expect the remaining numerical coefficients $c_0$, $c_{1,2}$ to be (roughly) of order unity. For simplicity, we shall neglect the (PN) expected logarithmic running of this term. We found that replacing the parenthesis $c_0+c_1 \ln(3y) +c_2 \ln^2 (3y)$ in Eq. (\ref{eq:4.14}) by a constant, say $c$, led to reasonable results, when doing comparisons with numerical data. More precisely, we found that the value $c=10$, i.e., a final PN error estimate
\beq
\label{eq:4.15}
10\frac{(3y)^8}{(1-3y)^{5/2}}\,,
\eeq
seemed to constitute an acceptable upper bound on the (absolute value of) the analytical error $O_{\ln{}}(y^8)$ in Eq. (\ref{eq:4.13}).

\begin{figure}
\label{fig:FIG1}
\includegraphics[scale=0.35]{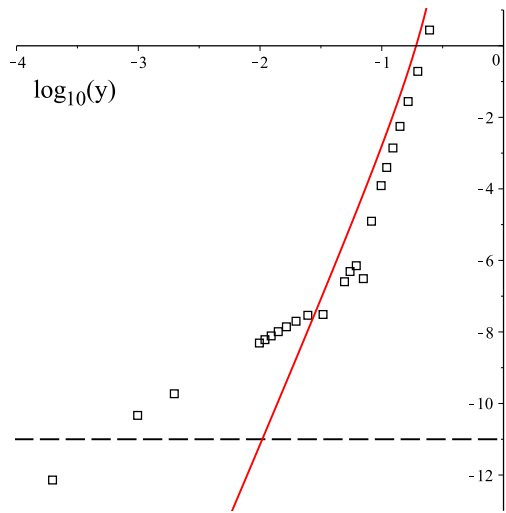}
\caption{The base-10 logarithm of the Newtonian-rescaled numerical-minus-analytical difference $
[\widehat \lambda_1^{\rm (E) 1SF}]^{\rm num}-[\widehat \lambda_1^{\rm (E) 1SF}]^{\rm 7.5PN}$
versus the base-10 logarithm of $y$. The slanting  (red online) solid line indicates the analytical error estimate while the dashed horizontal line (located at $-11$) indicates the (rough) numerical error level as in Ref. 
\cite{Dolan:2014pja}.
}
\end{figure}
In Fig. 1, we plot the base-10 logarithm of the Newtonian-rescaled numerical-minus-analytical difference
\beq
\label{eq:4.16}
[\widehat \lambda_1^{\rm (E) 1SF}]^{\rm num}-[\widehat \lambda_1^{\rm (E) 1SF}]^{\rm 7.5PN}
\eeq
versus the base-10 logarithm of $y$. The slanting  (red online) solid line indicates the analytical error estimate. The dashed horizontal line (located at $-11$) indicates the (rough) numerical error level  that would correspond to the number of significant digits (namely eleven) displayed in the first column of Table I in Ref. 
\cite{Dolan:2014pja}. If both error estimates were correct, the results for the differences (\ref{eq:4.16}) (displayed as boxes) should all lie {\it below} at least one of the two error lines. We see that this is the case for all the data points corresponding to $y\le \frac{1}{20}$ (i.e., $r_\Omega/M\le 20$ in the notation of \cite{Dolan:2014pja}). This is a nice confirmation of both the validity of our analytical results, and the validity of the numerical results of \cite{Dolan:2014pja} in the most important strong-field domain $\frac{1}{20}\le y \le \frac14$ (i.e., $4\le r_\Omega/M\le 20$).

On the other hand, the data points on the left (except the leftmost one) corresponding to $\frac{1}{1000}\le y \le \frac1{30}$ (i.e., $30\le r_\Omega/M\le 1000$) lie {\it above} both the expected PN error level, and the 11-digits horizontal line. This suggests that they have been affected by some small systematic error, kicking in at large radii. Note, however, that {\it all} data points are below either the estimated PN error curve, or a horizontal line at $-7.5$. In other words, at the remarkably good 8-digits level (which is enough for all practical purposes), there is agreement between analytical and numerical results in the very extended domain $\frac{1}{5000}\le y \le \frac14$ within the estimated PN error level (\ref{eq:4.15}). As a further test of our high PN-accuracy analytical results, it would be interesting to recompute some of the weak field numerical data to see if the agreement can be improved down to the PN error line (especially when $\frac{1}{100}\le y \le \frac{1}{30}$).

Finally, let us mention in passing that we have also compared the numerical results of \cite{Dolan:2014pja} for Detweiler's 1SF redshift quantity $\Delta U=\Gamma^{\rm 1SF}(y)$ to our 8.5PN accurate  analytical results \cite{Bini:2014nfa}. When computing the difference between $h_{kk}^{\rm num}=2(1-3y)^{3/2}\Delta U$ and $h_{kk}^{\rm 8.5PN}$, we have found that they are {\it all} lied {\it below} either our estimated 9PN error $(3y)^9$ or the $10^{-20}$ level corresponding to the number of digits displayed in Table III of Ref. \cite{Dolan:2014pja}.

These results show again the interest of comparing analytical and numerical results. In particular, we think that our Fig. 1 is the first such result  where high-accuracy analytical results inform numerical studies in suggesting the hidden presence of (small) systematic numerical errors kicking in at large radii.

\section{Light ring behavior of tidal invariants}

Several previous self-force studies have pointed out the existence of a singular behavior of some 1SF corrections as one approaches the LR: $y\to \frac13$.  This was first pointed out in Ref. \cite{Akcay:2012ea} for the redshift-related quantity $h_{kk}=h_{\mu\nu}^F k^\mu k^\nu=2(1-3y)^{3/2}\Gamma^{\rm 1SF}$, and the associated EOB potential $a=a^{\rm 1SF}$. It was found there that, as $y\to \frac13$,
\beq
\label{eq:5.1}
a^{\rm 1SF}(y)\sim \frac{\zeta}{4}(1-3y)^{-1/2}\,,
\eeq
and correspondingly (with the superscript F indicating that $h_{\mu\nu}$  is evaluated in an asymptotically {\it flat} gauge, rather than the Lorentz gauge used in 
\cite{Akcay:2012ea})
\beq
\label{eq:5.2}
h_{kk}^{\rm F}\sim -\frac12 \left( \zeta -\frac49 \right) (1-3y)^{-1/2}\,.
\eeq
The numerical value of the so introduced parameter $\zeta$ is close to one. A summary of the numerical estimates of $\zeta$ given in 
\cite{Akcay:2012ea} is (as noted in \cite{Bini:2014ica})
\beq
\label{eq:5.3}
\zeta =1.006(3)\,.
\eeq
As explained in \cite{Akcay:2012ea}, the origin of the LR singularities (\ref{eq:5.1}), (\ref{eq:5.2}) is simply the fact that, as one approaches the LR, the components of the stress-energy tensor of the perturbing source $\sqrt{-g}\, T_1^{\mu\nu}=m_1 \int U_1^\mu U_1^\nu \delta (x^\lambda -y_1^\lambda) d\tau_1$ are proportional to $m_1\Gamma_1$ rather than simply to $m_1$, with $\Gamma_1=dt/d\tau_1=(1-3y)^{-1/2}+O(q)$. By using results of EOB theory, it was then pointed out in Ref. 
\cite{Bini:2014ica}  that the LR singularities (\ref{eq:5.1}), (\ref{eq:5.2}) propagate into a corresponding singular behavior of the 1SF spin-orbit
function $\psi^{\rm 1SF}=-\sqrt{1-3y}\, \delta_{\rm spinorbit}^R(y)$ of the type
\beq
\label{eq:5.4}
\psi^{\rm 1SF}\sim -\frac14 \left( \zeta -\frac49 \right) (1-3y)^{-1}\,.
\eeq
The latter predicted behavior [which numerically reads $-0.1404(1)(1-3y)^{-1}$] has been recently confirmed in Ref.  \cite{Dolan:2014pja}.
In addition, Ref.  \cite{Dolan:2014pja} has numerically found that the 1SF contribution to the electric tidal eigenvalue $\lambda_1^{\rm (E)}$ diverges, near the LR, as
\beq
\label{eq:5.5}
\lambda_1^{\rm 1SF}\sim +0.01039\,  (1-3y)^{-5/2}\,.
\eeq
We wish to point out here that, similarly to what happened for the 1SF spin-orbit correction $\psi^{\rm 1SF}$, the behavior (\ref{eq:5.5}) is actually  an analytical consequence of the behavior (\ref{eq:5.1}), (\ref{eq:5.2}). Indeed, we have shown above that $\lambda_1^{\rm (E)}$ is related to the $k$-scaled eigenvalue $\sigma_1^{\rm (E)}$ by $\lambda_1^{\rm (E)}=\Gamma^2 \sigma_1^{\rm (E)}$. Let us show that, in the SF expansion of this relation, the singular behavior of $(\Gamma^2)^{\rm 1SF}$ dominates over that of $\sigma_1^{\rm (E)1SF}$. Indeed, the  $k$-normalization of ${\mathcal E}^\mu{}_\nu(k)=R^\mu{}_{\alpha\nu\beta}k^\alpha k^\beta$ (where the components of $k=\partial_t +\Omega \partial_\phi$ are regular at the LR), implies, when computing the curvature of $g_{\mu\nu}=g_{\mu\nu}^{(0)}+qh_{\mu\nu}+O(q^2)$, with $h_{\mu\nu}\propto \Gamma$ as explained above, that, sketchily, ${\mathcal E}(k)\sim {\mathcal E}^{(0)}(k)+q \Gamma+O(q^2)$.
Therefore, the LR singularity in the $\sigma_a^{\rm (E)}$'s (as well as in $\sigma^{\rm (B)}$) is $\sigma_a^{\rm (E)}=\sigma_a^{\rm (E)(0)}+q\Gamma+O(q^2)$, where the $0$SF contribution $\sigma_a^{{\rm (E)}0}$ are {\it regular}, and of order unity  at the LR (except for $\sigma_3^{{\rm (E)}0}$ which vanishes there).

In conclusion, we have, near the LR, $\sigma_1^{{\rm (E)}}=\sigma_1^{{\rm (E)}0}(1+O(q\Gamma))$, with a fractional 1SF correction of order $q\Gamma\sim q(1-3y)^{-1/2}$. By contrast, as $\Gamma^{-2}=-g_{\mu\nu}k^\mu k^\nu$, we have the relation
\beq
\label{eq:5.6}
\Gamma^{-2}=1-3y-qh_{kk}+O(q^2)\,,
\eeq
so that
\beq
\label{eq:5.7}
\Gamma= \frac{1}{\sqrt{1-3y}}\left(1+q\frac{h_{kk}}{2(1-3y)}+O(q^2)  \right)  \,.
\eeq
The LR behavior $h_{kk}\sim \Gamma$, then implies that the {\it fractional} 1SF correction to $\Gamma$ is of order $q\Gamma^3$, i.e., much stronger than the $O(q\Gamma)$ fractional correction to $\sigma_1^{\rm (E)}$. As a consequence, the dominant LR singularity in $\lambda_1^{\rm (E)}=\Gamma^2 \sigma_1^{\rm (E)}$  comes, as announced, from the 1SF correction in $\Gamma$, Eq. (\ref{eq:5.7}). Using Eq. (\ref{eq:5.2}) this yields the analytical prediction
\beq
\label{eq:5.8}
\lambda_1^{\rm (E)} \simeq \frac{\sigma_1^{{\rm (E)}0}}{1-3y}\left(1+q\frac{h_{kk}}{1-3y}  \right)\,,
\eeq
so that (using $\sigma_1^{{\rm (E)}0}=-y^3(2-3y)=-(\frac13)^3$ near the LR)
\beq
\label{eq:5.9}
\lambda_1^{\rm (E)1SF} \simeq \frac{\sigma_1^{{\rm (E)}0}}{(1-3y)^2}h_{kk}\simeq +\frac{1}{54}\left( \zeta-\frac49 \right)(1-3y)^{-5/2}\,.
\eeq
In addition, our reasoning yields analogous predictions for $\lambda_2^{\rm (E)}$, as well as for the magnetic eigenvalue $\lambda^{\rm (B)}$. These predictions are simply obtained by replacing $\sigma_1^{{\rm (E)}0}$ in Eq. (\ref{eq:5.9}) respectively by  $\sigma_2^{{\rm (E)}0}$ and $\lambda^{{\rm (B)}0}$. As 
$\sigma_2^{{\rm (E)}0}=y^3$ and $\sigma^{{\rm (B)}0}=3y^{7/2}\sqrt{1-2y}$ happen to take the same numerical value (namely ($\frac13)^3$) as $-\lambda_1^{{\rm (E)}0}$ at the LR, their predicted LR behavior is simply
\beq
\label{eq:5.9bis}
\lambda_2^{\rm (E)1SF} \simeq \sigma^{{\rm (B)}1SF} \simeq -\frac{1}{54}\left( \zeta-\frac49 \right)(1-3y)^{-5/2}\,.
\eeq
These analytical predictions agree with the behaviors found numerically in Ref. \cite{Dolan:2014pja}. Moreover, the coefficients of $(1-3y)^{-5/2}$ on the r.h.s.'s of Eqs. (\ref{eq:5.9}) and (\ref{eq:5.9bis}) are analytically predicted to be 
\beq
\label{eq:5.10}
C_\lambda\equiv C_{\lambda_1}=-C_{\lambda_2}=-C_{\sigma_B}=\frac{1}{54}\left( \zeta-\frac49 \right)\,.
\eeq
When using the numerical estimate of $\zeta$ \cite{Akcay:2012ea} summarized in Eq. (\ref{eq:5.3}) this yields
\beq
\label{eq:5.11}
C_{\lambda}^{\rm  Akcay\, et \, al.}=0.0104(1)\,.
\eeq
This nicely agrees with the recent result of Dolan et al. \cite{Dolan:2014pja}
\beq
\label{eq:5.12}
C_{\lambda}^{\rm  Dolan\, et \, al.}=0.01039(1)\,.
\eeq
Here, we quoted the smaller error bar obtained for $C_{\sigma_B}$ in \cite{Dolan:2014pja}. 
In turn, if we combine the more accurate numerical result (\ref{eq:5.12}) with the analytical prediction (\ref{eq:5.10}), we can derive a more accurate estimate of the parameter $\zeta$, namely
\beq
\label{eq:5.13}
\zeta=54\,  C_{\lambda} +\frac49=1.0055(5)\,.
\eeq

On the other hand, the fact that $\sigma_3^{{\rm (E)}0}=y^3(1-3y)$ vanishes near the LR renders our reasoning inconclusive.
Indeed, in that case, the expected {\it fractional} 1SF correction to $\sigma_3^{{\rm (E)}0}$ will be $\sim q\Gamma/\sigma_3^{{\rm (E)}0}\sim q\Gamma^3$, i.e., comparable to the fractional correction to $\Gamma$.

\section{Dynamical transcription of tidal results within EOB theory}

EOB theory  \cite{Buonanno:1998gg,Buonanno:2000ef,Damour:2000we,Damour:2001tu} is an analytical formalism that reformulates the dynamics of binary systems, with masses $m_1$ and $m_2$, in terms of the dynamics of an \lq\lq effective one body\rq\rq problem, where a particle of mass $\mu=m_1m_2/(m_1+m_2)$ is gravitationally coupled to an effective external metric
$g_{\mu\nu}^{\rm eff}(x^\lambda; M,\nu)$ which depends both on the total mass $M=m_1+m_2$ and on the symmetric mass ratio $\nu=\mu/M=m_1m_2/(m_1+m_2)^2$.

The effective external metric is parametrized (for non spinning bodies) as
\begin{eqnarray}
\label{eq:6.1}
ds^2=g_{\mu\nu}^{\rm eff} dx^\mu dx^\nu&=& -A(u;\nu) dt^2+B(u;\nu) dr_{\rm EOB}^2\nonumber\\
&& +r_{\rm EOB}^2(d\theta^2+\sin^2\theta d\phi^2)\,.
\end{eqnarray}
[In addition, the EOB dynamics includes higher-order-in-momenta terms that we will not need to discuss here.]
Here, $u=M/r_{\rm EOB}$ and the two metric functions $A(u;\nu)$ and $B(u;\nu)$ are $\nu-$deformed versions of the well known Schwarzschild result
$A^{\rm S}(u)=1-2u=1/B^{\rm S}(u)$. In other words, when $\nu \to 0$, we have $A(u;0)=1-2u=1/B(u;0)$. In the following, we shall refer to the crucially important function
$A(u;\nu)=-g_{00}^{\rm eff}$ as the main EOB radial potential.

It was proposed in \cite{Damour:2009wj} to represent, within the EOB formalism, the tidal interactions in binary systems, comprising tidally deformable compact bodies (namely neutron stars) by adding to the point-mass (or binary black hole) radial potential $A^{\rm BBH}(u;\nu)$ extra tidal contributions:
\beq
\label{eq:6.2}
A^{\rm total}(u;\nu)=A^{\rm BBH}(u;\nu)+A_1^{\rm tidal}(u;\nu)+A_2^{\rm tidal}(u;\nu)\,.
\eeq
The nontidal contribution $A^{\rm BBH}(u;\nu)$ is fully known up to the $4$PN level \cite{Bini:2013zaa}, while its 1SF contribution $a_{\rm 1SF}(u)$, such that
\beq
\label{eq:6.3}
A^{\rm BBH}(u;\nu)=1-2u+\nu a_{\rm 1SF}(u) +O(\nu^2)
\eeq
has been analytically determined up to the 8.5PN order \cite{ Bini:2013rfa,Bini:2014nfa} (see also \cite{Shah:2013uya}  for analytical-numerical results up to 10.5PN) and its global strong-field shape has been numerically determined (and analytically fitted) in Ref. \cite{Akcay:2012ea}.
The two tidal contributions $A_1^{\rm tidal}$ and $A_2^{\rm tidal}$ to $A^{\rm total}$ in Eq. (\ref{eq:6.2}) are additional radial potentials, associated, respectively, with the tidal deformations of body 1 and body 2. They can be decomposed into various multipolar contributions, labeled by multipole order $l$ and parity  $\epsilon$ ($+$ or $-$, i.e., electric- or magnetic-like); e.g., for body 1
\begin{eqnarray}
\label{eq:6.4}
A_1^{\rm tidal}(u)&=&\left(\sum_{l\ge 2}\sum_{\epsilon=\pm}A_{1}^{(l^\epsilon)LO}(u)\hat A_{1}^{(l^\epsilon)}(u)\right)+\ldots \nonumber\\
&=& \sum_{l\ge 2}\left( A_{1}^{(l^+)LO}(u)\hat A_{1}^{(l^+)}(u)\right.\\
&& \left. +A_{1}^{(l^-)LO}(u)\hat A_{1}^{(l^-)}(u)\right)+\ldots \,.\nonumber
\end{eqnarray}
Following Ref. \cite{Damour:2009wj} we have factorized each multipolar contribution in a leading order (LO), or \lq\lq Newtonian" piece,
$A_{1}^{(l^\epsilon)LO}(u)$ and a relativistic correcting factor $\hat A_{1}^{(l^\epsilon)}=1+O((v/c)^2)$. The LO piece is proportional to an $l$-dependent power of the interbody distance (in EOB coordinates), e.g.,
\beq
\label{eq:6.5}
 A_{1}^{(l^+)LO}(u)=-\kappa_1^{(l)}u^{2l+2}\,,  
\eeq
where $\kappa_1^{(l)}$ is a dimensionless combination involving the $l^{\rm th}$ electric tidal Love number $k_1^{(l)}$ of body 1, its radius $R_1$ and the two masses $m_1$ and $m_2$ (see either Eq. (25) in Ref. \cite{Damour:2009wj} or Eq. (11.7) in \cite{Bini:2012gu}). On the other hand, the correcting factor $\hat A_{1}^{(l)}$ (of electric or magnetic type) measures the distance-dependent effect of higher-PN relativistic tidal interactions (i.e., all effects beyond the Newtonian level interaction energy). The main aim of the present work is to use analytical (and numerical) gravitational self-force theory to improve our knowledge of several such relativistic 
tidal correcting factors. Before doing so, let us recall the current knowledge of these relativistic tidal correcting factors.

The currently most accurately known relativistic factor is the (physically most important) electric quadrupolar one $\hat A_{1}^{(l^+)}(u)$. Ref. \cite{Damour:2009wj} computed it at the 1PN accuracy, while \cite{Bini:2012gu} computed both its 2PN contribution and its exact test-mass value.
Combining these pieces of information yields
\begin{eqnarray}
\label{eq:6.6}
\hat A_{1}^{(2^+)}(u; X_1)&=& 1+\frac{3u^2}{1-3u}+\frac52 X_1 u \\
&& +\left(\frac18 X_1+\frac{337}{28}X_1^2  \right)u^2 +O_{X_1}(u^3)\,.\nonumber
\end{eqnarray}
Let us quote the known results for the two most important sub-leading tidal interactions: the magnetic-quadrupolar (with $ A_{1}^{(2^-)LO}\propto u^7$) and
electric-octupolar (with $ A_{1}^{(3^+)LO}\propto u^8$) ones. Their relativistic tidal factors read \cite{Bini:2012gu}
\begin{eqnarray}
\label{eq:6.7}
\hat A_{1}^{(2^-)}(u; X_1)&=& \frac{1-2u}{1-3u}+\left(\frac{11}{6} X_1+X_1^2 \right) u \nonumber\\
&& +O_{X_1}(u^3)\,. \\
\label{eq:6.8}
\hat A_{1}^{(3^+)}(u; X_1)&=& 1-2u+\frac{8}{3}\frac{ u^2(1-2u)}{1-3u}+\frac{15}2 X_1 u \nonumber\\ 
&& +\left(-\frac{311}{24} X_1+\frac{110}{3}X_1^2  \right)u^2\nonumber\\
&&  +O_{X_1}(u^3)\,. 
\end{eqnarray}
The relativistic tidal factors of body 1, Eqs. (\ref{eq:6.6})-(\ref{eq:6.8}), have been expressed in terms of the EOB dimensionless gravitational potential $u=GM/(c^2r_{\rm EOB})$ (with $M=m_1+m_2$), and of the mass fraction $X_1=m_1/M$. Note that the corresponding results for the tidal contribution of body 2 is simply obtained by replacing $X_1$ by  $X_2=m_2/M(=1-X_1)$.  As already used above, the notation $O_{X_1}(u^n)$ denotes a term which vanishes (at least) proportionally to $X_1$ and which is of order $u^n$ (i.e., of the $n^{th}$ PN order). Let us also note that the $X_1$ dependence of the above {\it electric}  relativistic factors (expressed in terms of the EOB variable $u$) is simpler than what it would be if expressed either in terms of the frequency parameters $x$, Eq. (\ref{eq:1.4}) or $y$, Eq. (\ref{y_def}).
Indeed, when re-expressed in terms of $x$, $\hat A_{1}^{(2^+)}(x; X_1)$ would involve $X_1^2$ already at the linear order in $x$ (1PN order), and would involve $X_1^3$ and $X_1^4$ at order $x^2$ (2PN order).

In the following we shall focus on the SF expansion of the relativistic  tidal factors $\hat A_{1}^{(l^\epsilon)}$, i.e., their expansion in powers of $X_1$ (rather than their PN expansion in powers of $u$). We shall then use the notation
\begin{eqnarray}
\label{eq:6.9}
\hat A_{1}^{(l^\epsilon)}(u; X_1)&=&\hat A_{1}^{(l^\epsilon)\rm 0SF}(u)+X_1 \hat A_{1}^{(l^\epsilon)\rm 1SF}(u)\nonumber\\
&& +X_1^2 \hat A_{1}^{(l^\epsilon)\rm 2SF}(u) + O(X_1^3) \,. 
\end{eqnarray}
The structure of the known PN results (\ref{eq:6.6})-(\ref{eq:6.8}) suggests that the 3SF, and higher, contributions $O(X_1^3)$ to the electric relativistic factors
start at order $u^3$, i.e., at the 3PN order. The zero SF contributions $\hat A_{1}^{(l^\epsilon)\rm 0SF}(u)$ are the (exactly known) test-mass results displayed as the first terms at the r.h.s.'s of Eqs. (\ref{eq:6.6})-(\ref{eq:6.8}).
For instance,
\beq
\label{eq:6.10}
\hat A_{1}^{(2^+)\rm 0SF}(u)=1+\frac{3u^2}{1-3u}\,.
\eeq
Current SF technology (both numerical and analytical) only allows one to access the 1SF corrections $\hat A_{1}^{(l^\epsilon)\rm 1SF}(u)$. On the other hand, the combination of PN theory and EOB theory exhibited in Eqs. (\ref{eq:6.6})-(\ref{eq:6.8}) shows that we already have some knowledge of the 2SF contributions, as will be further discussed below.

The general relation between the (dynamically significant) EOB relativistic tidal factors $\hat A_{1}^{(l^\epsilon)}(u)$ and the (kinematically invariant) $U-$normalized tidal scalars, such as $J_{e^2}={\rm Tr}[{\mathcal E}^2(U)]$, has been derived in \cite{Bini:2012gu}. Let us recall the final result of \cite{Bini:2012gu} in the simple case of circular orbits. With each irreducible $U-$normalized multipolar tidal invariant $J$ (evaluated along the world line of body 1) is associated a corresponding contribution, say $A_1^J=A_1^{J\rm LO}\hat A_1^J$, in the tidal piece (linked with body 1) of the total EOB radial potential $A^{\rm total}$, Eq. (\ref{eq:6.2}).

Using Eq. (5.19) in \cite{Bini:2012gu}, this general link implies that
\beq
\label{eq:6.11}
\hat A_1^J(u; X_1)=\sqrt{F(u)}\Gamma_1^{-1}\frac{J}{J^{\rm Newt}(r_{\rm EOB})}\,.
\eeq
Here, the first factor (which is a specific prediction of EOB theory) is given in terms of the EOB $A-$potential via the definitions
(consistently with Eqs. (\ref{eq:2.10})--(\ref{eq:2.12}) above) 
\begin{eqnarray}
\label{eq:6.12}
\tilde A(u) &\equiv & A(u)+\frac12 u A'(u) \\
\label{eq:6.13}
h^2(u)&\equiv & 1+2\nu \left( \frac{A(u)}{\sqrt{\tilde A(u)}}-1 \right)\\
\label{eq:6.14}
F(u) & \equiv & \tilde A(u) h^2(u)\,. 
\end{eqnarray}
The second factor is the redshift factor $\frac{d\tau_1}{dt}=\Gamma_1^{-1}$ along the world line of body 1.
The last factor is the ratio between the two-body value of the tidal invariant $J$ (for body 1), and the value that $J$ would take at Newtonian order, when expressed in terms of the EOB distance $r_{\rm EOB}$. For instance, we see from Eq. (\ref{eq:2.14}) and from Eqs. (6.15) in \cite{Bini:2012gu}, that
\begin{eqnarray}
\label{eq:6.15}
J_{e^2}^{\rm Newt}(r_{\rm EOB}) &= &\frac{6m_2^2}{r_{\rm EOB}^6}\\
\label{eq:6.16}
J_{b^2}^{\rm Newt}(r_{\rm EOB}) &= &\frac{18m_2^2M}{r_{\rm EOB}^7}\\
\label{eq:6.17}
J_{(l=3^+)}^{\rm Newt}(r_{\rm EOB}) &= &\frac{90 m_2^2}{r_{\rm EOB}^8}\,.
\end{eqnarray}

The r.h.s. of Eq. (\ref{eq:6.11}) must be expressed  in terms of the EOB gravitational potential $u=M/r_{\rm EOB}$ and of $X_1=m_1/M$.
The EOB-related factors $\sqrt{F(u)}$ and $J^{\rm Newt}(r_{\rm EOB})$ are already expressed in terms of $u$ or $r_{\rm EOB}=M/u$. [$F(u)$ defined by Eqs. (\ref{eq:6.12})-(\ref{eq:6.13}) is an explicit function of $u$ and $\nu=X_1X_2=X_1(1-X_1)$.] One must, however, express both $d\tau_1/dt=\Gamma_1^{-1}$ and $J$ in terms of $u$ and $X_1$.

Let us now focus on the first order SF expansion of $\hat A_1^J$, i.e., on the first two terms of Eq. (\ref{eq:6.9}). At this linear order in $X_1$, all small mass ratios are equivalent to $q=m_1/m_2$: $X_1=q/(1+q)=q+O(q^2)$, $\nu=q/(1+q)^2=q+O(q^2)$.
The 1SF accurate expansion of the redshift function $\Gamma(y)=\Gamma_1(y)$ is related to the metric perturbation $h_{kk}=h_{\mu\nu}k^\mu k^\nu$ by Eq. (\ref{eq:5.7}).
In Ref. \cite{Bini:2014nfa} we have computed the 8.5PN expansion of $h_{kk}$ in powers of $y$ (see Eqs. (21)-(24) in \cite{Bini:2014nfa}). 
On the other hand, we have analytically computed above the 1SF contribution to the function ${\mathcal J}_{e^2}(y)=\Gamma^{-4}J_{e^2}$ at the 7.5PN accuracy. Combining these two results yields the 7.5PN-accurate expansion of the factor
$$
\Gamma^{-1}J_{e^2}=\Gamma^3 {\mathcal J}_{e^2}
$$
in Eq. (\ref{eq:6.11}) to first order in $q$, and as a function of $y$. In order to re-express this result as a function of the EOB variable $u$ we need the transformation linking $u$ and $y$ to (at least) 7.5PN-accuracy. This transformation follows from basic results in EOB theory that we have recalled in Sec. II above. Namely, the frequency parameter $x=(M\Omega)^{2/3}$ is related to $u$ via Eq. (\ref{eq:2.10}), while $y=(m_2\Omega)^{2/3}$  is related to $x$ via Eq. (\ref{eq:2.7}), i.e., Eq. (\ref{eq:2.15}) at linear order in $q$.
The 1SF-accurate version of the $u-y$ link is then obtained from the 1SF-accurate expansion of the basic EOB potential $A(u;\nu)$, i.e.,
\beq
A(u;\nu)=1-2u+\nu a_{\rm 1SF}(u)+O(\nu^2)\,.
\eeq
This yields \cite{Damour:2009sm} 
\beq
x=u \left[1-\frac16 \nu a'_{\rm 1SF}(u) -\frac23 \nu \left(\frac{1-2u}{\sqrt{1-3u}}-1  \right)+O(\nu^2) \right]
\eeq
and
\beq
\label{y_expanded}
y=u \left[1-\frac16 q a'_{\rm 1SF}(u) -\frac23 q \frac{1-2u}{\sqrt{1-3u}}+O(q^2) \right]
\eeq
Finally, by inserting Eqs. (\ref{eq:5.7}) and (\ref{y_expanded}) in Eqs. (\ref{eq:6.11})  
and by taking care of the extra mass-ratio effects linked to the adimensionalization of 
the various tidal invariants by suitable powers of $m_2$ \footnote{Note for instance that the $m_2-$adimensionalized version of $J_{e^2}^{\rm Newt}=6m_2^2/r_{\rm EOB}^6$ contains six powers of $m_2/M=1/(1+q)$: $m_2^4 J_{e^2}^{\rm Newt}=6(m_2/M)^6u^6$.}, we obtain the following analytic results for the 1SF contributions $\hat A_1^{(l^\epsilon)\rm 1SF}$, Eq. (\ref{eq:6.9}, to the quadrupolar-electric and quadrupolar-magnetic relativistic tidal corrections (respectively linked to $J_{e^2}$ and $J_{b^2}$).
\begin{widetext} 
\begin{eqnarray}
\label{eq:6.20}
\hat A_{1}^{(2^+)1\rm SF}(u)&=&  \frac{5}{2} u + \frac18 u^2+\left(-\frac{3487}{16} +\frac{1905}{256} \pi^2\right)u^3\nonumber\\
&+& \left(-\frac{2816}{5} \ln(2)-\frac{1408}{5} \gamma+\frac{4114603}{9600} -\frac{25337}{1024} \pi^2 -\frac{704}{5} \ln u \right)u^4\nonumber\\
&+& \left( \frac{1146014221}{403200} -\frac{10387277}{24576}\pi^2 +\frac{66056}{105} \ln u+\frac{132112}{105} \gamma+ 3152 \ln 2 -\frac{4374}{7} \ln 3 \right)u^5\nonumber\\
&-&\frac{6848}{21}\pi u^{11/2}\nonumber\\
&+& \left(\frac{30246655583}{2903040}+\frac{34840}{63}\ln(u)+\frac{69680}{63}\gamma-\frac{5150464}{945}\ln(2)+6075\ln(3)\right. \nonumber\\
&& \left. -\frac{2494654027}{786432}\pi^2+\frac{76231071}{1048576}\pi^4\right) u^6\nonumber\\
&+&\frac{11417267}{7350}\pi u^{13/2}\nonumber\\
&+&\left(\frac{9943210070208659}{55883520000}-\frac{19182623242}{779625)}\ln(2)-\frac{13278187073}{779625}\ln(u)-\frac{37249370460407}{2831155200}\pi^2 \right.\nonumber\\
&-&\frac{26556374146}{779625}\gamma-\frac{49369095}{2464}\ln(3)-\frac{19782291875}{67108864}\pi^4+\frac{438272}{75}\ln(2)\ln(u)+\frac{219136}{75}\ln(u)\gamma \nonumber\\
&-&\left. \frac{28672}{5}\zeta(3)-\frac{37109375}{9504}\ln(5)+\frac{876544}{75}\ln^2(2)+\frac{219136}{75}\gamma^2+\frac{876544}{75}\gamma\ln(2)+\frac{54784}{75}\ln^2(u)\right) u^7 \nonumber\\
&+& \frac{283918559}{485100}\pi u^{15/2}+O_{\ln{}}(u^{8})\,.
\end{eqnarray}

\begin{eqnarray}
\label{eq:6.21}
\hat A_{1}^{(2^-)1\rm SF}(u)&=&   
\frac{11}{6} u 
 -\frac{123}{8}  u^2
+  \left( -\frac{11219}{48}+\frac{123}{16}\pi^2 \right)  u^3\nonumber\\
&&+\left( -\frac{2086369}{17280}+\frac{16483}{9216}\pi^2-\frac{5168}{45}\ln(u)-\frac{10336}{45}\gamma-\frac{1376}{3}\ln(2) \right)  u^4\nonumber\\
&& +\left( \frac{1267671359}{403200}-\frac{724321}{1536}\pi^2+\frac{15128}{35}\ln(u)+\frac{30256}{35}\gamma+\frac{678448}{315}\ln(2)-\frac{2916}{7}\ln(3) \right)u^5\nonumber\\
&& -\frac{439984}{1575}\pi u^{11/2}\nonumber\\
&& +\left(\frac{6946843011179}{304819200}-\frac{17275719071}{3538944}\pi^2+\frac{2150179}{2835}\ln(u)+\frac{4300358}{2835}\gamma-\frac{4116562}{2835}\ln(2)\right. \nonumber\\
&& \left. +\frac{503739}{140}\ln(3)+\frac{63541825}{524288}\pi^4\right) u^6\nonumber\\
&& +\frac{147766331}{132300}\pi u^{13/2}\nonumber\\
&& \left(-\frac{478755043703}{16372125}\gamma+\frac{3273344}{315}\ln(2)\gamma-\frac{26635748466577}{1238630400}\pi^2-\frac{122281237411}{3274425}\ln(2)\right. \nonumber\\
&&-\frac{11398887}{1232}\ln(3)-\frac{151780905715}{150994944}\pi^4-\frac{80078125}{42768}\ln(5)+\frac{12278464}{4725}\gamma^2+\frac{49093312}{4725}\ln^2(2)\nonumber\\
&&-\frac{229504}{45}\zeta(3)+\frac{1636672}{315}\ln(u)\ln(2)+\frac{12278464}{4725}\ln(u)\gamma+\frac{3346345862649884593}{10561985280000}\nonumber\\
&&\left. -\frac{478755043703}{32744250}\ln(u)+\frac{3069616}{4725}\ln^2(u) \right) u^7\nonumber\\
&& +\frac{110208790429}{78586200}\pi u^{15/2}+ O_{\ln {}}(u^{8})\,.
\end{eqnarray}

\end{widetext}

Note that both PN expansions are fractionally 7.5PN accurate.
When working within the usual PN formalism (where $g_{00}$ is more accurately determined than $g_{0i}$, and $g_{0i}$ more accurately that $g_{ij}$) one looses one order of PN accuracy when computing a \lq\lq magnetic" quantity.
However, here we have used an expansion of the Regge-Wheeler metric based on correctly including both the $l=5$ electric quadrupole and  the $l=5$  magnetic one. As a consequence, all our results are accurate up to, and including, the appearance of the first logarithm of $5$ (which is linked to near zone tail effects associated with $l=5$ \cite{Bini:2013zaa, Bini:2013rfa, Bini:2014nfa}), as well as the next half-PN order.

\section{Global, strong-field behavior of the relativistic tidal factors}

We have obtained above high-order expansions in powers of $u$ for the $O(X_1)$ contribution to the quadrupolar electric and quadrupolar-magnetic relativistic tidal factors $\hat A_1^{(l)}(u;X_1)$ that describe dynamical tidal effects within the EOB formalism, see Eq. (\ref{eq:6.4}). In this section we shall combine these analytic results with several other sources of information (test-mass limit, numerical SF data, EOB theory, LR behavior) in order to come up with a plausible, global description of the behavior of the functions $\hat A_1^{(l)}(u;X_1)$  in the entire, physically relevant domain of variation of the two variables $u$ and $X_1$.

As was stressed in Refs. \cite{Damour:2009wj,Bini:2012gu}, in order to accurately describe the dynamical influence of tidal effects in coalescing binary neutron stars, one needs to know the functions $\hat A_1^{(l^\epsilon)}(u;X_1)$ for a mass fraction $X_1\simeq \frac12$ and for values of $u$ up to contact, i.e., up to
\beq
\label{eq:7.1}
r_{\rm EOB}^{\rm contact}=R_1+R_2\equiv \frac{m_1}{{\mathcal C}_1}+\frac{m_2}{{\mathcal C}_2}\,,
\eeq 
corresponding to
\beq
\label{eq:7.2}
u^{\rm contact}= \left( \frac{X_1}{{\mathcal C}_1}+\frac{X_2}{{\mathcal C}_2}\right)^{-1}\,.
\eeq 
Here, we have defined the \lq\lq compactness" of each neutron star as ${\mathcal C}_1\equiv Gm_1/(c^2R_1)=m_1/R_1$. If we consider a neutron star of mass $m_1\simeq 1.35 M_\odot\simeq 2\, $km and radius $R_1$ between $10\, $km and $12\, $km, its compactness ${\mathcal C}_1$ will range between $1/6\approx 0.1667$ and $0.2$.
In the equal-mass case, Eq. (\ref{eq:7.2}) yields $u^{\rm contact}={\mathcal C}_1$. It is therefore desirable to know the behavior of $\hat A_1^{(l^\epsilon)}(u;X_1)$ up to
$u^{\rm contact}=0.2$. In view of the recent discovery of higher mass neutron stars, it is possible that even larger values of $u^{\rm contact}$ might become physically important. In the following, we shall combine strong-field SF data \cite{Akcay:2012ea,Dolan:2014pja} with analytic information to describe the behavior of the tidal factors up to $u=\frac14=0.25$ (and even beyond), and (hopefully) for values of $X_1$ of order unity.

\subsection{Strong-field behavior of the 1SF quadrupolar-electric tidal factors}

In this subsection, we only consider the $X_1-$linear piece $\hat A_{1}^{(2^+)1\rm SF}(u)$ of  $A_{1}^{(2^+)} (u; X_1)$.
Let us first consider the successive PN approximants to $\hat A_{1}^{(2^+)1\rm SF}(u)$, i.e., the successive terms in its expansion, Eq. (\ref{eq:6.20}), in powers of $u$, modulo $O_{\ln{}}(u^8)$.

\begin{figure}
\label{fig:FIG2}
\includegraphics[scale=0.35]{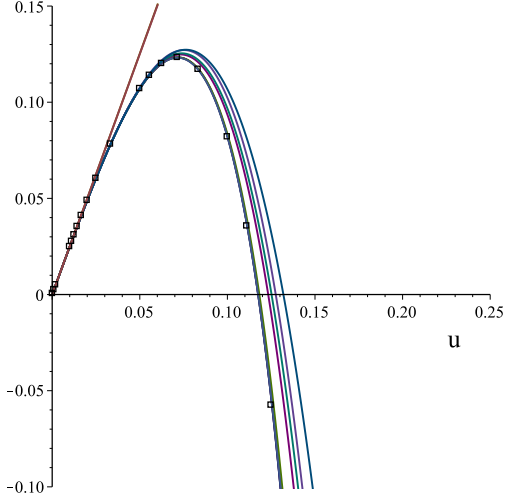}
\caption{The successive PN approximants to the $X_1-$linear piece $\hat A_{1}^{(2^+)1\rm SF}(u)$ in the quadrupolar-electric tidal factor are plotted as functions of $u$, starting from the 1PN approximant (straight line) up to the 7.5 PN approximant. The 2PN approximant is a parabola  (with upward concavity) close to the 1PN straight line. The higher approximants can be identified by looking at the position of the zero (close to $u=0.12$) as given in Table I below. The boxes indicate the numerical SF data obtained by combining the results of \cite{Dolan:2014pja} and \cite{Akcay:2012ea}. 
}
\end{figure}
Fig. 2 displays the successive PN approximants from 1PN (i.e.,  $\hat A_{1}^{(2^+)1\rm SF}(u)=\frac52 u$) up to 7.5 PN. A remarkable result is that, apart from the 1PN ($\frac52 u$) and the 2PN ($\frac52 u +\frac18 u^2$) approximants (which stay positive and monotonically increasing) all the higher-PN-approximants, after increasing away from zero, reach a positive maximum, after which they decrease, cross the zero level around $u=0.12$, and then continue dipping down towards more and more negative values as $u$ enters the strong field domain. We list in Table I the values of $u$ where the successive PN approximants to $\hat A_{1}^{(2^+)1\rm SF}(u)$ vanish.
Note that, as the PN order increases, the values of these zeros exhibit an approximate convergence towards $u\approx 0.1175$. As last line in the table we have displayed, for comparison,
the estimate of this zero coming from our best fit to the numerical SF data ($f_{23}$, see Eq. (\ref{eq:7.26}) below).

\begin{table}[h]
  \caption{\label{tab:zeros_pnorder} Zeros of the successive PN approximants to $\hat A_{1}^{(2^+)1\rm SF}(u)$}
  \begin{center}
      \begin{tabular}{|c|c|}
\hline
PN order &	$u$ of the zeros of $\hat A_{1}^{(2^+)1\rm SF}$ \\
\hline
3&  0.1319695\\ \hline 
4& 0.1275991\\ \hline 
5& 0.1252101\\ \hline 
5.5& 0.1232954\\ \hline 
6& 0.1175261\\ \hline 
6.5& 0.1182420\\ \hline 
7& 0.1175496\\ \hline 
7.5& 0.1175803\\ \hline
num & 0.1171848\\ 
\hline
\end{tabular}
\end{center}
\end{table}

Let us show that this analytic prediction is in agreement with numerical SF data. To do that, we need to be able to compute the EOB function $\hat A_{1}^{(2^+)1\rm SF}(u)$ from SF data. We can do this by combining two sets of SF data: those of Ref. \cite{Akcay:2012ea} on the EOB $a_{\rm 1SF}(u)$ function, and the recent data of Ref. \cite{Dolan:2014pja} on the tidal eigenvalues $\lambda_1^{\rm (E)}$, $\lambda_2^{\rm (E)}$. Indeed, the quantity we are interested in can be explicitly written as
\beq
\label{eq:7.3}
\hat A_{1}^{(2^+)}(u; X_1)=\frac{\sqrt{F(u;\nu)}}{(1-X_1)^6}\Gamma^{-1}(y(u))\frac{m_2^4J_{e^2}(y(u))}{6u^6}\,.
\eeq 
Here, $F(u;\nu)$ is given, to 1SF order, by inserting Eq. (\ref{eq:6.3}) in the definitions (\ref{eq:6.12})-(\ref{eq:6.14}), i.e.,
\begin{eqnarray}
\label{eq:7.4}
F(u;\nu) &=& \left[1-3u +\nu \left(a_{\rm 1SF}(u)+\frac12 u a'_{\rm 1SF}(u)  \right)   + O(\nu^2) \right]\times \nonumber\\
&& \left[ 1+2\nu \left( \frac{1-2u}{\sqrt{1-3u} }-1  \right) + O(\nu^2) \right ] \,,
\end{eqnarray}
while the functions of $y$, $\Gamma^{-1}(y)$ and $m_2^4{J_{e^2}(y)}$ are given, to 1SF order, by
\begin{eqnarray}
\label{eq:7.5}
\Gamma^{-1}(y) &=&\sqrt{1-3y}\left(1-\frac12 q \frac{h_{kk}}{1-3y}  +O(q^2)\right)  \\
\label{eq:7.6}
\frac12 m_2^4 J_{e^2}(y)&=& (\lambda_1^{\rm (E)})^2+(\lambda_2^{\rm (E)})^2 +\lambda_1^{\rm (E)}\lambda_2^{\rm (E)}\nonumber\\
&=& 3y^6 (1-3y+3y^2)\nonumber\\
&& +q \left( \frac{y^3(-3+6y)}{1-3y}\lambda_1^{\rm (E)1SF}\right.\nonumber\\
&& \left. +\frac{3y^4}{1-3y}\lambda_2^{\rm (E)1SF}\right) +O(q^2)\,.
\end{eqnarray}
In the latter expressions, we need to insert the 1SF-accurate expression of $y$ in terms of $u$, i.e.,
\begin{eqnarray}
\label{eq:7.7}
y &=& u\left( 1-\frac16 q a'_{\rm 1SF}(u) -\frac23 q \frac{1-2u}{\sqrt{1-3u}}+O(q^2)\right)\,.
\end{eqnarray}
The final 1SF-accurate result for $\hat A_{1}^{(2^+)}(u; X_1)$ yields for the coefficient of $X_1$ (or $q$) a linear combination (with $u-$dependent)
coefficients of the quantities:
$a_{\rm 1SF}(u)$, $a'_{\rm 1SF}(u)$, $h_{kk}$, $\lambda_1^{\rm (E)1SF}$ and $\lambda_2^{\rm (E)1SF}$. In addition, let us recall that the 1SF EOB potential
$a_{\rm 1SF}(u)$ is itself given in  terms of $h_{kk}$ \cite{LeTiec:2011dp,Bini:2014nfa} by
\beq
\label{eq:7.8}
a_{\rm 1SF}(u)=-\frac12 h_{kk}(u)-\frac{u(1-4u)}{\sqrt{1-3u}}\,,
\eeq
and that the $k$-normalized metric perturbation $h_{kk}$ is related to the 1SF contribution to the function $\Gamma(y)$ via
\beq
\label{eq:7.9}
\Gamma^{\rm 1SF}(y)=+\frac12 \frac{h_{kk}}{(1-3y)^{3/2}}\,.
\eeq
The recent work of Dolan et al. \cite{Dolan:2014pja} computed accurate numerical values for $\lambda_1^{\rm (E)1SF}$, $\lambda_2^{\rm (E)1SF}$ and $\Gamma^{\rm 1SF}(y)$ for a sample of values of $y$ ranging from $\frac{1}{5000}$ up to $\frac14$. By inserting their results in Eqs. (\ref{eq:7.3})-(\ref{eq:7.8}) above we can directly compute from numerical SF data all the 1SF contributions to $\hat A_{1}^{(2^+)1\rm SF}(u; X_1)$, {\it except} for the terms linear in the derivative $a'_{\rm 1SF}(u)$ (which come either from $F(u)$, Eq. (\ref{eq:7.4}), or from the $y\to u$ transformation (\ref{eq:7.7})). To compute these terms, we have used the work of Akcay et al. \cite{Akcay:2012ea}, which derived accurate global analytical representations of the function $a_{\rm 1SF}(u)$ from numerical SF data on $h_{kk}$. More precisely, we used the \lq\lq model \# 14" in Ref.  \cite{Akcay:2012ea} to analytically represent $a_{\rm 1SF}(u)$, and thereby analytically compute its derivative 
$a'_{\rm 1SF}(u)$. [In passing, we have checked that there was an excellent agreement, between the latter analytic model for $a_{\rm 1SF}(u)$ and the numerical data for $\Gamma^{\rm 1SF}$ given in \cite{Dolan:2014pja}, with differences smaller than about $10^{-11}$ over the full range $\frac{1}{5000}\le u \le \frac14$; in keeping with Fig. 4 in \cite{Akcay:2012ea}.] The result of this numerical computation is represented by small squares in Fig. 2. As shown by this Figure there is a rather good \lq\lq convergence" of the successive PN approximants towards the numerical SF result in the semi-strong-field domain $0\le u \lesssim \frac16$.  
On the other hand, as will be exhibited in Fig. 3 below, in the domain $\frac 16 \lesssim u \le \frac14$, even the 7.5 accurate expansion exhibits visible differences with the numerical result.
It is to be noted that the analytical prediction of a nontrivial strong-field behavior of the function $\hat A_{1}^{(2^+)1\rm SF}(u)$ (with a maximum and a change of sign) is fully confirmed by the numerical data. [A similar analytic and numerical agreement was found for spin-orbit effects in Refs. \cite{Bini:2014ica,Dolan:2013roa}.]

\subsection{Analytically expected light ring behavior of the SF-expanded quadrupolar-electric tidal factor}

Analogously to what was argued for the spin-orbit function $\psi(y)$ \cite{Bini:2014ica}, the plunging behavior towards negative values, for $u \gtrsim 0.2$ of the 1SF function $\hat A_{1}^{(2^+)1\rm SF}(u)$, can be understood from an analytically expected behavior of the latter function as $u\to \frac13$, i.e., as $u$ approaches the LR.

Indeed, if we rewrite the expression (\ref{eq:7.3}) in terms of the $k$-normalized quadrupolar electric invariant ${\mathcal J}_{e^2}=\sum_i(\sigma_i^{\rm (E)})^2=\Gamma^{-4}J_{e^2}$ we get
\beq
\label{eq:7.10}
\hat A_{1}^{(2^+)}(u)= \frac{\sqrt{F(u;\nu)}}{(1-X_1)^6} \Gamma^3(y(u))\frac{m_2^4{\mathcal J}_{e^2}(y(u))}{6u^6}\,,
\eeq
 where the factor $\Gamma^{-1}(y(u))$ in Eq. (\ref{eq:7.3}) has been replaced by $\Gamma^3(y(u))$.

As already argued above, near the LR ($y\to \frac13$), the function $m_2^4{\mathcal J}_{e^2}(y(u))$ will behave as
\beq
\label{eq:7.11}
m_2^4{\mathcal J}_{e^2}(y)\sim  6y^6 (1-3y+3y^2)+q\Gamma +O(q^2)\,,
\eeq
where the term $q\Gamma$ (which is to be interpreted in a rough sense, i.e., modulo some $y-$dependent coefficient which has a finite limit at the LR)
 denotes the effect of the metric perturbation $h_{\mu\nu}$, which blows up as $O(\Gamma)$ near the LR.

It is important to note in Eq. (\ref{eq:7.11}) that the $0$SF contribution (namely $6y^6(1-3y+3y^2)$) remains bounded and nonzero as $y \to \frac13$.
Sketchily, we have therefore ${\mathcal J}_{e^2}(y)\sim 1+q\Gamma$ as $y \to \frac13$. On the other hand, when considering the $q$-expansion of the function ${\mathcal J}_{e^2}(y(u))$ we shall have a stronger fractional blow up near the LR. Indeed, let us consider a general function of $y$ and $q$, say
\beq
\label{eq:7.12}
f(y;q)=f_0(y)+qf_1(y)+O(q^2)\,,
\eeq
and let us effect the change of variable $y=y(u,q)$ in the function $f$. The latter change of variables is given in $q-$expanded form in Eq. (\ref{y_expanded}). 
Among the two contributions of order $q$ in this equation, the second one blows up as $\Gamma$ near the LR, while the first one blows up as $a'_{\rm 1SF}(u)$. At this stage, we need to recall that $a_{\rm 1SF}(u)$ blows up as $\frac{\zeta}{4}\Gamma(y)$ near the LR, see Eq. (\ref{eq:5.1}).
By differentiating the latter equation, we conclude that $a'_{\rm 1SF}(u)$ blows up as
\beq
\label{eq:7.13}
a'_{\rm 1SF}(u)\simeq \frac{\zeta}{4}\Gamma'(y)\simeq \frac38 \zeta\,  \Gamma^3 (y)\,.
\eeq
Therefore, the most singular term, near the LR, in the 1SF transformation $y=u+qy^{1\rm SF}(u)+O(q^2)$ will be induced by
\beq
\label{eq:7.14}
y\simeq u -\frac16 \nu u  a'_{\rm 1SF}(u) \,.
\eeq

Inserting this result in Eq. (\ref{eq:7.12}) yields 
\begin{eqnarray}
\label{eq:7.14bis}
f(y(u;q);q)&\simeq & f_0(u)+q\left(f_1(u)-\frac16 u  a'_{\rm 1SF}(u) f_0'(u) \right)\nonumber\\
&&+O(q^2)\,,
\end{eqnarray}
corresponding to a fractional $O(q)$ change in $f(y(u))=f_0(y(u))+q f_1(y(u))+O(q^2)$ equal to
\begin{eqnarray}
\label{eq:7.15}
\frac{f(y(u;q); q)}{f_0(u)}&\simeq & 1+q\left(\frac{f_1(u)}{f_0(u)}-\frac16 a'_{\rm 1SF}(u) \frac{u f_0'(u)}{f_0(u)}  \right)\nonumber\\
&& +O(q^2)\,.
\end{eqnarray}
When applying this general result to ${\mathcal J}_{e^2}(y(u))$ (for which both $f_1/f_0$ and $f_0'/f_0$ are of order unity near the LR)
we conclude that ${\mathcal J}_{e^2}(y(u))/{\mathcal J}^{0\rm SF}_{e^2}(u)\sim 1+q\Gamma^3$. By contrast, we shall have a faster fractional LR blow up when applying the result (\ref{eq:7.15}) to the factor $\Gamma^3(y(u))$ in Eq. (\ref{eq:7.10}). 
Indeed, in that case though the term $f_1(u)/f_0(u)$, i.e., $3\Gamma^{1\rm SF}(y(u))/\Gamma^{0\rm SF}(u)$, is of order $\Gamma^3$ (see Eq. (\ref{eq:7.9})), the last term on the r.h.s. of Eq. (\ref{eq:7.15}) blows up even faster near the LR, namely as $\Gamma^5$. We conclude (using $\Gamma'_{0\rm SF}/\Gamma_{0\rm SF}=\frac32 (1-3u)^{-1}$) that the LR dominant term   in $\Gamma^3(y(u))/\Gamma^{0\rm SF}(u)$ is
\beq
\label{eq:7.16}
\frac{\Gamma^3(y(u))}{\Gamma^{0\rm SF}(u)}\simeq 1-\frac34 q \frac{u a'_{\rm 1SF}(u)}{1-3u}+O(q^2)\,.
\eeq
In addition, the factor $\sqrt{F(u;\nu)}$ in Eq. (\ref{eq:7.10}) contributes a term having a similar LR blow up. Indeed, near the LR we have
\beq
\label{eq:7.17}
\tilde A(u; \nu) =1-3u +\frac12 \nu  a'_{\rm 1SF}(u)+O(\nu^2)
\eeq
so that 
\beq
\label{eq:7.18}
\sqrt{\frac{F(u;\nu)}{F(u;0)}}\simeq \sqrt{\frac{\tilde A(u; \nu)}{1-3u}}\simeq 1+\frac14 \nu \frac{u a'_{\rm 1SF}(u)}{1-3u} +O(\nu^2)\,.
\eeq
Combining the factors (\ref{eq:7.16}) and (\ref{eq:7.18}) yields a {\it fractional} blow up behavior for $\hat A_{1}^{(2^+)}(u)$ given by
\beq
\label{eq:7.19}
\frac{\hat A_{1}^{(2^+)}(u;X_1)}{\hat A_{1}^{(2^+)}(u;0)}\simeq 1-\frac12 q \frac{u a'_{\rm 1SF}(u)}{1-3u} +O(q^2)\,.
\eeq
Inserting the $0$SF value of $\hat A_{1}^{(2^+)}(u)$, namely $1+3u^2(1-3u)^{-1}$ yields an {\it absolute} blow up behavior of the $O(q)$ term in $\hat A_{1}^{(2^+)}(u)$ given by
\beq
\label{eq:7.20}
\hat A_{1}^{(2^+)1\rm SF}(u;X_1)\simeq -\frac32 \frac{u^3 a'_{\rm 1SF}(u)}{(1-3u)^2}\simeq -\frac{\zeta}{16} \frac{9 \, u^3}{(1-3u)^{7/2}}\,,
\eeq
where we used the asymptotic behavior (\ref{eq:7.13}) for $a'_{\rm 1SF}(u)$.

In view of the positive value $\zeta \simeq 1$, we see that Eq. (\ref{eq:7.20}) predicts that $\hat A_{1}^{(2^+)1\rm SF}(u)$ must plunge, rather fast, towards $-\infty$ as $u\to \frac13$. This \lq\lq explains" the shape of the higher PN approximants, and of the numerical results. Note, finally, that Eq. (\ref{eq:7.20}) suggests to consider the following LR-rescaled version of $\hat A_{1}^{(2^+)1\rm SF}(u)$,
\beq
\label{eq:7.21}
\widetilde A_{1}^{(2)1\rm SF}(u):= (1-3u)^{7/2} \hat A_{1}^{(2^+)1\rm SF}(u)\,.
\eeq
This LR-rescaled function  should remain finite as $u\to \frac13$, and is predicted by Eq. (\ref{eq:7.20}) to take the limiting value
\beq
\label{eq:7.22}
\lim_{u\to \frac13}\, \widetilde A_{1}^{(2)1\rm SF}(u)=-\frac{\zeta}{48}=-0.02095(1)\,.
\eeq
In the numerical estimate we have used the value (\ref{eq:5.13}) for $\zeta$.

\subsection{Global analytic representations of the $X_1-$linear quadrupolar-electric tidal factor}

Let us now combine the various pieces of information (PN, numerical, LR behavior) we have acquired about the quadrupolar-electric tidal factor to find convenient analytic representations of the function $\hat A_{1}^{(2^+)1\rm SF}(u)$, that are valid in the strong-field domain.

Let us first mention that if one plots (as we shall do in Fig. 3 below) the discrete sample of numerically determined values of the LR-rescaled function $\widetilde A_{1}^{(2)1\rm SF}(u)$, Eq. (\ref{eq:7.21}), one obtains a set of points which 
approximately lie on a {\it cubic curve} of the form
\beq
\label{eq:7.23}
\widetilde A^{ \rm cubic}(u; a_1,a_2)=\frac52 u (1-a_1u)(1-a_2u)\,,
\eeq 
with $a_1\simeq 8.5$ and $a_2\simeq 3$. Here, the factor $\frac52 u$ is the 1PN approximant, the parameter $a_1$ parametrizes the first zero (outside the origin) located around $1/a_1\approx 0.12$  (see Table I), and $a_2$ parametrizes a second zero (located in the neighborhood of $\frac13$) of $\widetilde A_{1}^{(2)1\rm SF}(u)$. One can anticipate the existence of such a second zero from the result (\ref{eq:7.22}). Indeed, the numerical value of $\widetilde A_{1}^{(2)1\rm SF}(u)$ at the last strong field point in the data of Ref. \cite{Dolan:2014pja}, i.e., at $u=\frac14$ is $\widetilde A_{1}^{(2)1\rm SF}(\frac14)\simeq -0.09727$.
This value is about five times larger (in absolute value) than the LR value (\ref{eq:7.22}) analytically estimated above. Moreover, the numerical value
for the data point closest to $\frac14$ suggests that $\widetilde A_{1}^{(2)1\rm SF}(u)$ approximately reaches a minimum near $\frac14$. All this suggests that  $\widetilde A_{1}^{(2)1\rm SF}(u)$ will cross again the horizontal axis a little bit {\it after} the LR, i.e., for some $u=\frac13 (1+\varepsilon)$ with $ \varepsilon$ small and positive (corresponding to an $a_2=\frac{3}{1+\varepsilon}$ in Eq. (\ref{eq:7.23})). As we do not have in hand numerical data for $\frac14 < u < \frac13$ we cannot confirm this prediction about the value of $a_2$.

Anyway, we found that one could fit the numerical data points rather accurately by means of fitting functions of the type
\beq
\label{eq:7.24}
\widetilde A^{\rm fit}(u; a_1,a_2;f_0)=\widetilde A^{ \rm cubic}(u; a_1,a_2) f_0(u)
\eeq
where the extra factor remains close to one in the entire fitting domain.

First, a not very accurate fit (but the analytically simplest we could get) is obtained by least-squares fitting the 23 numerical data points ssociated with the sample  of Ref. \cite{Dolan:2014pja} (completed by model \# 14 of Ref. \cite{Akcay:2012ea}) to a simple cubic, i.e., by taking $f_0(u)=1$ in Eq. (\ref{eq:7.24}). The best-fit parameters for this cubic fit were found to be $a_1^{\rm cubic}=8.34925$, $a_2^{\rm cubic}=3.45232$, and the standard deviation of the residuals was $\sigma^{\rm cubic}_{\rm res}=2.78 \times 10^{-3}$. [All our fits use, for simplicity, equal weights for the data points.]

By contrast we obtained a much more accurate fit (with $\sigma_{\rm res}=6.01 \times 10^{-5}$) by including a fudge factor $f_0(u)$ in Eq. (\ref{eq:7.24}) of the form
\beq
\label{eq:7.25}
f_0(u)=\frac{1+n_1u }{1+d_2 u^2}
\eeq
and by fitting the four parameters $a_1$, $a_2$, $n_1$ and $d_2$. The best-fit parameters were then found to be
\begin{eqnarray}
\label{eq:7.26}
a_1&=&8.53353\,,\quad a_2=3.04309\,,\nonumber\\
n_1&=&0.840058\,,\quad d_2=17.73239\,.
\end{eqnarray}
We shall refer to this fit as $f_{23}$.
We have also fitted the template (\ref{eq:7.24}), (\ref{eq:7.25}) to an extended data set obtained by adding to the previous 23 numerical data points a $24^{\rm th}$ point given by the analytically predicted LR value (\ref{eq:7.22}). This gave 
residuals such that $\sigma_{\rm res}=7.8 \times 10^{-4}$ and best-fit parameters $a_1=8.56877$, $a_2=2.89147$, $n_1=1.243446$, $d_2=25.73912$. We shall refer to this fit as $f_{24}$. [Note that the latter value of $a_2$ is slightly smaller than 3, as analytically expected, while the former value of $a_2$ was slightly larger than 3.]
Let us also mention the result of a fit obtained by constraining the values of $a_1$, $a_2$ and $n_1$ by the relation
\beq
\label{eq:7.27}
n_1=a_1+a_2-\frac{209}{20}
\eeq 
predicted by the 2PN-accurate expansion of $\hat A_{1}^{(2^+)1\rm SF}(u)$. The corresponding 3-parameter fit (say $f_{23}^{\rm 2PN}$) to the unextended numerical data set led to 
$\sigma_{\rm res}=2.35 \times 10^{-4}$ for $a_1=8.54583$, $a_2=3.01805$, $d_2=20.63317$. Finally, if we impose the 2PN constraint and fit to the extended 24 data points one gets ($f_{24}^{\rm 2PN}$ fit) $\sigma_{\rm res}=7.9 \times 10^{-4}$ for $a_1=8.56185$, $a_2=2.89201$, $d_2=23.66920$.

\begin{figure}
\label{fig:PN_and_fits}
\[
\begin{array}{c}
\includegraphics[scale=0.35]{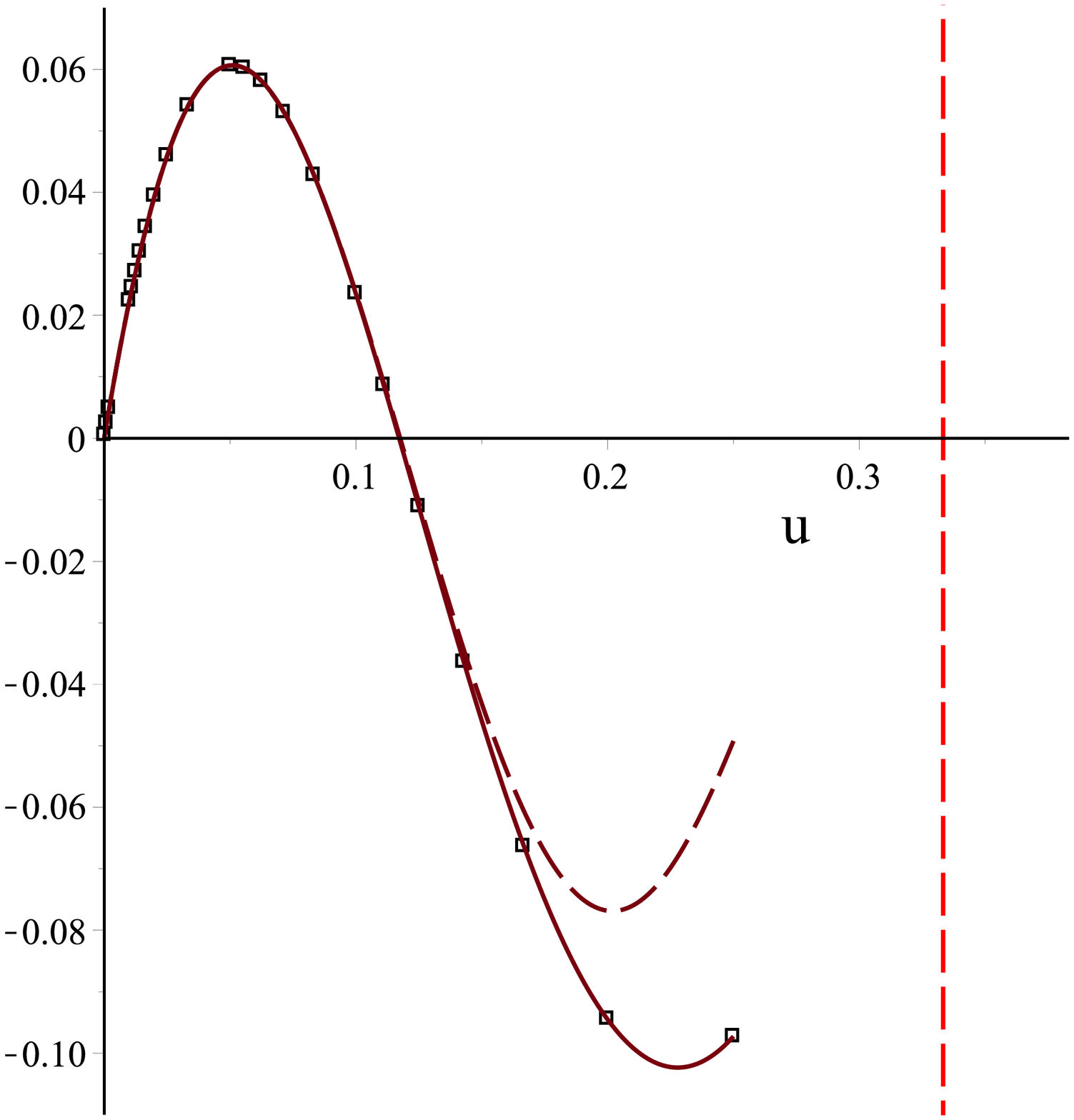}\\
(a)\\
\includegraphics[scale=0.35]{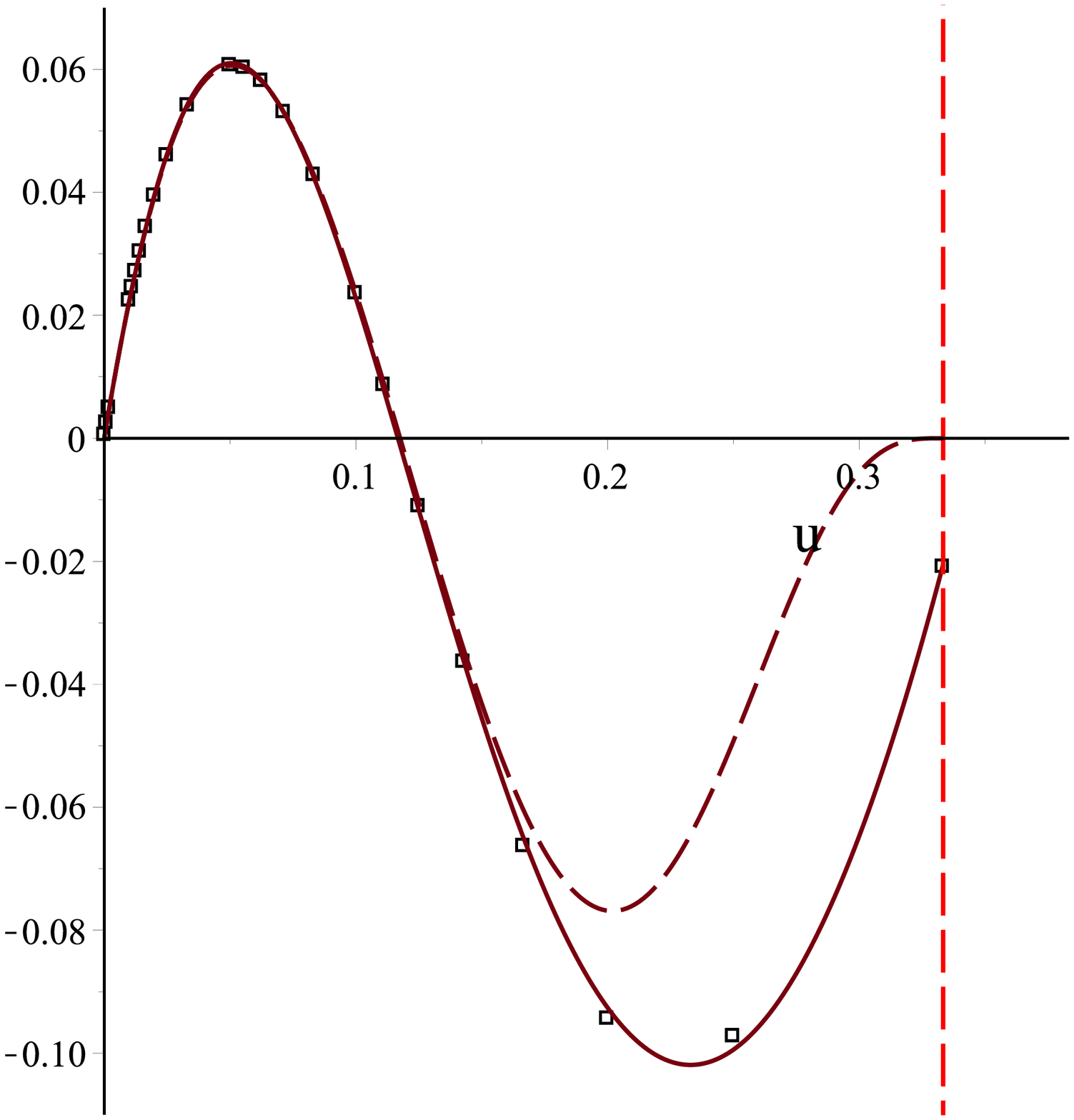}\\
(b)
\end{array}
\]
\caption{In panel (a)  we  compare the plots of three different estimates of  the light-ring-rescaled $X_1-$linear piece  $\widetilde  A_1^{(2^+)1\rm SF}(u)$ in the quadrupolar-electric tidal factor, Eq. (\ref{eq:7.20}): i) the numerical relativity data points \cite{Dolan:2014pja}, \cite{Akcay:2012ea} (boxes); ii) the fitting function $f_{23}(u)$, Eqs. (\ref{eq:7.24})-(\ref{eq:7.27}) (solid line); and iii) the product of the 7.5PN series for $\hat A_1^{(2^+)1\rm SF}(u)$ by $(1-3u)^{7/2}$ (dashed curve).
Panel (b) shows (now up to $u=\frac13$) the same estimates as in panel (a), except that the fitting function $f_{23}(u)$ is replaced by $f_{24}(u)$.
The vertical dashed line indicates the position of  the light ring $u=\frac13$.     
}
\end{figure}

In the two panels of Fig. 3  we  compare four different global estimates  of  the light-ring-rescaled 1SF function $\widetilde  A_1^{(2^+)1\rm SF}(u)$, Eq. (\ref{eq:7.21}): i) the numerical SF data points obtained by combining the results of Refs. \cite{Dolan:2014pja} and \cite{Akcay:2012ea}; 
ii) the 7.5PN-accurate analytic estimate defined as (note that the LR-prefactor on the r.h.s. is not re-expanded in powers of $u$) 
\beq
\widetilde A_{1\,\rm  7.5PN}^{(2)1\rm SF}(u):= (1-3u)^{7/2} \hat A_{1\,\rm  7.5PN}^{(2^+)1\rm SF}(u)\,;
\eeq
iii)  in panel (a), the fitting function $f_{23}(u)$, Eqs. (\ref{eq:7.24})-(\ref{eq:7.27});
 and iv)  in panel (b), the fitting function $f_{24}(u)$, defined above.
 The plots of the associated fitting functions  $f_{23}^{\rm 2PN}(u)$ and   $f_{24}^{\rm 2PN}(u)$ are omitted as they are almost indistinguishable from the corresponding   $f_{23}(u)$ and   $f_{24}(u)$ plots.

\subsection{Global analytic representations of the full, $X_1-$nonlinear quadrupolar-electric tidal factor}

A surprising result of the previous subsections has been the fact that the $X_1-$linear piece in the relativistic quadrupolar-electric tidal factor, Eqs. (\ref{eq:6.9}) and (\ref{eq:6.10}), was becoming more and more negative as $u$ enters the strong-field domain. This is surprising, because some of the comparisons between fully numerical 3D simulations of binary neutron stars and the EOB description of  tidally-interacting binary systems have suggested the need to include relativistic tidal factors
$\hat A_{1}^{(2^+)}(u;X_1)$, $\hat A_{2}^{(2^+)}(u;X_2)$ somewhat larger (near coalescence) than their 2PN approximants
$\hat A_{1}^{(2^+)1\rm SF}(u;X_1)=1+\alpha_1 u+\alpha_2 u^2$  \cite{Damour:2009wj,Baiotti:2010xh,Bernuzzi:2012ci}. We think that this might be explained by the behavior of the terms nonlinear in $X_1$ in Eq. (\ref{eq:6.9}). Indeed, let us recall that, for binary neutron star systems, we expect to have $X_1\simeq \frac12$ so that we cannot rely on the sole knowledge of the $X_1-$linear (1SF) contribution that we discussed in detail above.
We already know from 2PN results \cite{Bini:2012gu}  that we have, at least, a term quadratic in $X_1$, namely
\beq
\label{eq:7.28}
X_1^2 \hat A_{1}^{(2^+)2\rm SF}(u)=\frac{337}{28}X_1^2 u^2\left(1+O(u)\right)\,.
\eeq
The numerical coefficient entering this term is rather large ($\frac{337}{28}\simeq 12.0357$). When $X_1=\frac12$, the 2PN term (\ref{eq:7.28}) numerically dominates the corresponding $O(X_1)$ 2PN term, viz $\frac18 X_1 u^2$. It is therefore reasonable to expect that the higher PN corrections $1+O(u)$ in Eq. (\ref{eq:7.28}) might play an important role, and may compensate the change of sign of the $X_1-$linear contribution when $u\gtrsim 0.12$.
Actually, the original, defining expression for the $\hat A_1^{(l^\epsilon)}(u;X_1)$'s, Eq. (\ref{eq:6.11}) is a product of {\it positive} quantities. [Note that any quadratic irreducible tidal invariant $J$ is necessarily positive as it is the square of a tensor having only spatial components in the local frame of $U_1$: $J^{(2^+)}=G_{ab}G^{ab}$, $J^{(2^-)}=H_{ab}H^{ab}$, $J^{(3^+)}=G_{abc}G^{abc}$, \ldots .] This proves that, if,  when doing a SF expansion in powers of $X_1$, the 1SF term formally tends to $-\infty$ as $-c_1X_1(1-3u)^{-p_1}$ (with $c_1>0$) when $u\to \frac13^-$, the 2SF piece must tend to $+\infty$ as 
$+c_2X_1^2(1-3u)^{-p_2}$, with $c_2>0$ and $p_2>p_1$. [This is ssen by considering the class of formal limits where $X_1$ tends to zero as some powers of $(1-3u)$.] More precisely, as the SF expansion proceeds, near the LR, in powers of $\Gamma\sim X_1 (1-3u)^{-1/2}$, we must have $p_2\ge p_1+\frac12$. In the electric ($l=2$) case, this shows that, near the LR, $X_1^2 A_1^{(2)2\rm SF}$ must blow up as
\beq
\label{eq:7.29}
+\frac{c_2X_1^2}{(1-3u)^p}\,,\qquad p\ge 4 \,.
\eeq
Actually, the dominating power $p$ in the LR behavior (\ref{eq:7.29}) is likely to be strictly larger than 4. By extending the reasoning made above for the LR behavior of $\hat A_{1}^{(2)1\rm SF}(u)$ one anticipates that the highest possible power of $\Gamma$ will come from the $O(q^2)$ term in the $q-$expansion of fractional corrections such as 
\beq
\label{eq:7.29bis}
(1+cq\Gamma^5)^\alpha=1+c\alpha q \Gamma^5 +\frac{\alpha (\alpha -1)}{2}c^2 q^2 \Gamma^{10}+O(q^3)\,.
\eeq
In other words, this suggests that the difference between $p_2$ and $p_1$ will be given by the ratio $q^2\Gamma^{10}/(q\Gamma^5)\sim q \Gamma^5\sim q(1-3u)^{-5/2}$, so that $p_2=p_1+\frac52$. When $p_1=\frac72$, this yields $p_2=6$.

Combining this information with the 2PN knowledge (\ref{eq:7.28}) suggests that a plausible representation of the full $X_1-$nonlinear quadrupolar-electric  tidal factor reads
\begin{eqnarray}
\label{eq:7.30}
\hat A_{1}^{(2^+)}(u; X_1)&=&1+\frac{3u^2}{1-3u}+X_1 \frac{\widetilde A^{(2^+)1\rm SF}_{1}(u)}{(1-3u)^{7/2}}\nonumber\\
&&+X_1^2 \frac{\widetilde A^{(2^+)2\rm SF}_{1}(u)}{(1-3u)^{p}}+O(X_1^3u^3)\,,
\end{eqnarray}
with $4\le p \le 6$ and
\beq
\label{eq:7.31}
\widetilde A^{(2^+)2\rm SF}_{1}(u)=\frac{337}{28}u^2+\ldots
\eeq
With our current, incomplete knowledge, we suggest as best-guess estimate of $\hat A_{1}^{(2^+)}(u; X_1)$: i) to replace $\widetilde A^{(2^+)1\rm SF}_{1}$ by our best strong-field fit $f_{23}$ discussed above; ii)  
to approximate the LR-rescaled 2SF contribution simply by using its 2PN accurate value $\widetilde A^{(2^+)2\rm SF}_{1}(u)=(337/28)u^2$; and iii)
to neglect the unknown, 3PN level term $O(X_1^3u^3)$ in Eq. (\ref{eq:7.30}). One can then  test different values of $p$, within the range $4\le p \le 6$, against full numerical relativity simulations.

\begin{figure}
\label{fig:4}
\includegraphics[scale=0.35]{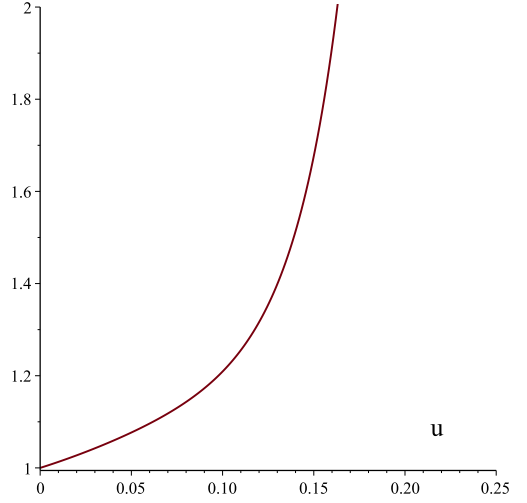}
\caption{The full quadrupolar-electric tidal factor  $\hat A_{1}^{(2^+)}(u; X_1)$ is plotted as a function of the EOB variable $u$, Eq. (\ref{eq:7.30})  for the choice of parameters $p=4$ and $X_1=\frac12$.   
}
\end{figure} 

For illustrative purposes, we display in Fig. 4 the full factor $\hat A_{1}^{(2^+)}(u; X_1)$ so defined, when using $p=4$ and $X_1=\frac12$. As we see in Fig. 4 this relativistic tidal factor stays always larger than 1. Let us note that, if contact occurs at $u_{\rm contact}={\mathcal C}_1=\frac16$ (taken as a typical neutron star compactness), the successive contributions, evaluated at contact, to the relativistic correction factor (\ref{eq:7.30}) read
\begin{eqnarray}
\label{eq:7.32}
\hat A_{1}^{(2^+)}\left(\frac16; X_1\right)&=& 1.16667-0.37463(2X_1)\nonumber\\
&& +1.333730 (2X_1)^2\,,
\end{eqnarray}
where the coefficient of $(2X_1)^2$ is analytically $337/252$. This expression suggests that the 2SF (and higher) contributions to
$\hat A_{1}^{(2^+)}$ largely dominate the 1SF one, and might even dominate the 0SF one, when one approaches contact. This shows the potential importance
of extending the presently available 1SF numerical results to the 2SF level.
It also shows the need to extend the 2PN-accurate results of \cite{Bini:2012gu} to higher PN levels. Eq. (\ref{eq:7.30}) suggests that the coefficient of the 3PN term
$X_1^2u^3$ will be rather large, namely of order $3p \frac{337}{28}$, which varies between $\approx +144$ and $\approx +217$ as $p$ ranges between $4$ and $6$.

Let us finally note that the appearance of tricky LR-singular terms, when fully expanding $\hat A_{1}^{(2^+)}(u; X_1)$ in powers of $X_1$, suggests that a more accurate representation (for practical uses) of this function might be obtained by keeping in non-SF-expanded form all the factors in the exact EOB expression (\ref{eq:6.11}) that can be explicitly expressed in terms of the full (non-SF-expanded) EOB potential $A(u;\nu)$. [This is notably the case of the first factor $\sqrt{F(u;\nu)}$ in Eq. (\ref{eq:6.11}).] Indeed, one way to understand the origin of the LR behavior 
\beq
\hat A_{1}^{(2^+)} \sim c_0 (1-3u)^{-1}-c_1X_1 (1-3u)^{-p_1} + c_2X_1^2 (1-3u)^{-p_2}\,,
\eeq
with alternating signs and $p_2>p_1>1$, is that it essentially results from expanding in powers of $\nu$ the inverse of the  (positive) modified EOB radial potential $\tilde A (u;\nu)$  according to 
\beq
\left(\tilde A (u;\nu)\right)^{-1}\sim [1-3u+c\nu (1-3u)^{-3/2}  ]^{-1}\,.
\eeq
Over the last years the many comparisons between EOB theory and numerical relativity simulations 
\cite{Buonanno:2007pf,Damour:2009kr,Buonanno:2009qa,Damour:2012ky,Taracchini:2013rva,Damour:2014sva} have led to a good knowledge of the function $A(u;\nu)$ in the comparable mass case, i.e., up to $\nu=\frac14$. There is therefore no need to worsen the the numerical accuracy of Eq. (\ref{eq:6.11}) by expanding $A(u;\nu)$ and $F(u;\nu)$ in powers of $\nu$.

\section{Global, strong-field behavior of the quadrupolar magnetic tidal factor}

In the previous section we have considered the quadrupolar-{\it electric} tidal factor, here we shall consider by contrast the 
quadrupolar-{\it magnetic} one. Let us start by writing down the magnetic analog of the EOB result Eq. (\ref{eq:7.3}).
It reads
\begin{eqnarray}
\label{eq:8.1}
\hat A_{1}^{(2^-)}(u;X_1)&=&\frac{\sqrt{F(u;\nu)}}{(1-X_1)^6}\Gamma^{-1}(y(u)) \frac{m_2^4 J_{b^2}(y(u))}{18 u^7}\nonumber\\
&=&\frac{\sqrt{F(u;\nu)}}{(1-X_1)^6}\Gamma^{3}(y(u)) \frac{m_2^4 {\mathcal J}_{b^2}(y(u))}{18 u^7}\,.
\end{eqnarray}
The reasoning applied above to the electric case shows again that, near the LR, the dominant singular behavior will come from the same factors as in the electric case, namely the SF expansion of $\sqrt{F(u;\nu)} \Gamma^{3}(y(u))$. The magnetic LR behavior is then obtained by multiplying Eq. (\ref{eq:7.20}) by the $0$SF value of 
$\hat A_{1}^{(2^-)}$ which is
\beq
\label{eq:8.2}
\hat A_{1}^{(2^-)0 \rm SF}=\frac{1-2u}{1-3u}\,,
\eeq
instead of
\beq
\label{eq:8.3}
\hat A_{1}^{(2^+)0 \rm SF}=1+\frac{3u^2}{1-3u}=\frac{1-3u+3u^2}{1-3u}\,.
\eeq
However, it is easily seen that, near the LR, $\hat A_{1}^{(2^-)0 \rm SF}$ behaves exactly as $\hat A_{1}^{(2^+)0 \rm SF}$, namely as $\frac13 (1-3u)^{-1}$. This shows that the $O(q)$ LR blow up of $\hat A_{1}^{(2^-)1 \rm SF}$ will be given by the same equation as its electric counterpart, Eq. (\ref{eq:7.21}), i.e.,
\beq
\label{eq:8.4}
\hat A_{1}^{(2^-)1\rm SF}(u)\simeq - \frac{\zeta}{48}\frac{(3u)^3}{(1-3u)^{7/2}}\,.
\eeq
As in the electric case, we therefore expect $\hat A_{1}^{(2^-)1\rm SF}(u)$, which starts, near $u=0$, as \cite{Bini:2012gu}
\beq
\label{eq:8.5}
\hat A_{1}^{(2^-)1\rm SF}(u)=\frac{11}{6}u+O(u^2)\,,
\eeq
to have a maximum, and then to decrease, to cross zero and to plunge towards large negative values in the strong-field domain.
This expected behavior is confirmed both by our analytic computation of the 7.5PN-accurate expansion of $\hat A_{1}^{(2^-)1\rm SF}(u)$, and by the recent numerical SF data of \cite{Dolan:2014pja}. We have already written down above, in Eq. (\ref{eq:6.21}), the 7.5PN-accurate  expansion of $\hat A_{1}^{(2^-)1\rm SF}(u)$.

Contrary to the electric case, here, even the 2PN contribution is negative (with value $-123/8=-15.375$, whose absolute value is about 8 times larger than the 1PN coefficient).

\begin{figure}
\label{fig:5}
\includegraphics[scale=0.35]{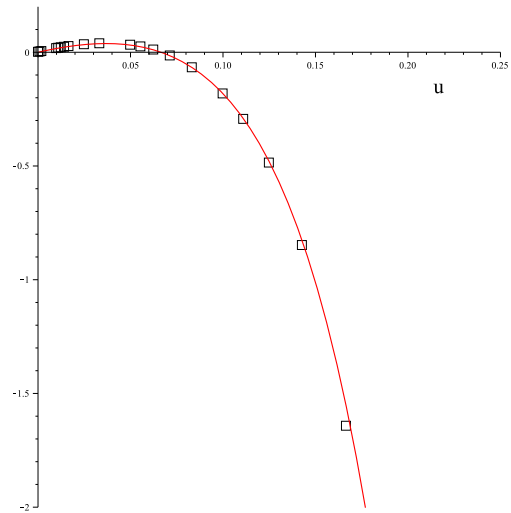}
\caption{The 7.5PN-accurate analytic expression for the $X_1-$linear piece $\hat A_{1}^{(2^-)1\rm SF}(u)$ in the quadrupolar-magnetic tidal factor is plotted as a function of the EOB variable $u$ (solid line). The numerical data (boxes) are obtained by combining  the $1$SF results of \cite{Dolan:2014pja} for $\Delta U$ and $\lambda^{\rm (B)1SF}$,  and model \#14 from \cite{Akcay:2012ea} for the EOB function $a_{1\rm SF}(u)$.  }
\end{figure} 

In Fig. 5 we plot the 7.5PN accurate result for $\hat A_{1}^{(2^-)1\rm SF}(u)$. As expected, from Eq. (\ref{eq:8.4}), it exhibits a maximum, it crosses the zero level around $u\simeq 0.066484$,   and then plunges towards large, negative values.
Using the same tools as in the electric case (i.e., combining numerical data from \cite{Dolan:2014pja} and model \#14 from \cite{Akcay:2012ea}), we also computed a strong-field numerical estimate of $\hat A_{1}^{(2^-)1\rm SF}(u)$. The numerical data points are indicated by boxes in Fig. 5. These data confirm the strong-field behavior inferred above from analytical arguments.

Clearly, one could extend the fitting technique we have used above in the electric case to the present magnetic one, by considering templates of the form
\beq
\label{eq:8.7}
\hat A_{1}^{(2^-)1\rm SF}(u)=\frac{11}{6}\frac{u(1-a_1u)(1-a_2u)}{(1-3u)^{7/2}} f_0(u)\,,
\eeq
with a suitably chosen fudge factor $f_0(u)$. However, as there is no pressing need to have in hand an accurate global representation of $\hat A_{1}^{(2^-)1\rm SF}(u)$ we leave such a task to future work. Let us only mention that the $X_1-$nonlinear contributions probably play also a very important role in the magnetic tidal factor. Indeed, contrary to the electric case, they start at the 1PN level with the contribution $+X_1^2u$.

To conclude,  let us only exhibit the current knowledge of the structure of $\hat A_{1}^{(2^-)}(u; X_1)$
\begin{eqnarray}
\label{eq:8.8}
\hat A_{1}^{(2^-)}(u;X_1)&=& \frac{1-2u}{1-3u}+X_1 \frac{\widetilde A_{1}^{(2^-)1\rm SF}(u)}{(1-3u)^{7/2}}\nonumber\\
&& +X_1^2 \frac{\widetilde A_{1}^{(2^-)2\rm SF}(u)}{(1-3u)^{p}}+O(X_1^3u^2)\,,
\end{eqnarray}
with $4\le p \le 6$ and
\beq
\label{eq:8.9}
\widetilde A_{1}^{(2^-)2\rm SF}(u)=u+\ldots
\eeq
Again we expect such a representation to be able to maintain the positivity of $\hat A_{1}^{(2^-)}(u;X_1)$ in the strong-field domain.

\section{Summary and concluding remarks}

We used a synergetic combination of the results of several approximation formalisms (post-Newtonian, analytic black hole perturbation, numerical self-force, effective one-body theory) to improve our analytic knowledge of tidal interactions in binary systems. Our results concern a gravitationally interacting two-body system  moving on circular orbits. The main new results of our work are:
\begin{itemize}
  \item The analytic computation, to linear order in the mass ratio $q=m_1/m_2$ and to the 7.5PN accuracy, i.e., up to $(v/c)^{15}$, of several tidal invariant functions
$J(y)$, where $y=(Gm_2\Omega/c^3)^{2/3}$, namely ${\rm Tr}{\mathcal E}^2(k)$, ${\rm Tr}{\mathcal E}^3(k)$, ${\rm Tr}{\mathcal B}^2(k)$, as well as some octupolar-level invariants discussed in Appendix D. 
  \item The comparison of our analytic results to the recent numerical self-force calculations by Dolan et al. \cite{Dolan:2014pja} of several invariant functions (in the form of tidal eigenvalues $\lambda_1(y)$, $\lambda_2(y)$, $\lambda_B(y)$, to first order in $q$).
  \item Our work is, to the best of our knowledge, the first where high-accuracy analytical results 
allow one  to inform numerical self-force studies about the presence of probable, hitherto undetected, systematic numerical errors (at a small, but significant level) affecting data points at large radii $r\ge 30 Gm_2/c^2$. Moreover, the high-accuracy analytic formulas
we give in Appendix D for several octupolar level invariants might serve as useful test beds of future numerical SF computations of invariants involving three derivatives of the metric.

  \item We provided an analytic understanding of the light ring asymptotic behavior of tidal eigenvalues found in Ref. \cite{Dolan:2014pja}.
  \item We transcribed the analytical and numerical results on tidal invariants in the more dynamically significant knowledge of certain tidal relativistic factors
$\hat A^{l^\epsilon}(u;X_1)$ (with $X_1=m_1/(m_1+m_2)$) entering the effective one-body description of tidal interactions. This knowledge is encoded both in the 7.5PN accurate expansion of the $X_1-$linear
 piece in $\hat A^{l^\epsilon}(u;X_1)$, and in numerical data for this $X_1-$linear
 piece.
\item We discussed the analytically expected light ring behaviors for several such 1SF pieces, $\hat A^{(l^\epsilon)1 \rm SF}(u)$, notably for $l=2^+$ (quadrupolar-electric) and $2^-$
(quadrupolar-magnetic). We then provided simple, accurate analytic representations of the global, strong-field behavior of $\hat A^{(2^+) 1 \rm SF}(u)$, see subsection VII C. 
\item Our most striking finding is that the $X_1-$linear piece of $\hat A^{(2^+) 1\rm SF}(u)$ in the quadrupolar-electric relativistic tidal factor $\hat A^{2^+}(u;X_1)$
for body 1 has a positive maximum in the weak field domain, and then, after crossing zero, plunges towards rather large negative values in the strong-field domain $u\gtrsim 0.12$. This unexpected behavior was found both in our 7.5PN-accurate analytical results and in the numerical results of \cite{Dolan:2014pja} (completed by model \#14 in \cite{Akcay:2012ea}), and was shown to be related to the analytically expected singular behavior of the function $\hat A^{(2^+) 1\rm SF}(u)$ as $u\to \frac13^-$, i.e. near the LR.
\item We, however, argued that the negative character of the $X_1-$linear piece   $X_1\hat A^{(2^+) 1\rm SF}(u)$ is more than compensated by a probable fast growth of the $X_1-$quadratic piece  $X_1^2\hat A^{(2^+) 2\rm SF}(u)$ in the strong-field regime, and we provided plausible parametrizations (and estimates) of the growth.
 
The latter result shows the importance of further improving the analytic knowledge of the relativistic tidal factors. This can be done either by using second-order SF theory, or by extending the 2PN-accurate results of \cite{Bini:2012gu} to higher PN accuracies (and to all orders in the mass  fraction $X_1$), or, finally, by comparing our analytical representation Eq. (\ref{eq:7.30}) to full numerical simulations of coalescing binary neutron stars (e.g., using the exponent $p$ as a fitting parameter).
All these research avenues should be pursued because they might all contribute to improving our ability at theoretically describing the late dynamics of inspiralling neutron star binaries.

\end{itemize}

\appendix

\begin{widetext}

\section{Details about the analytical computation of $m_2^4{\rm Tr}[{\mathcal E}^2(k)]$}

We start from the expression of $m_2^4{\rm Tr} {\mathcal E}(k)^2$,  Eqs. (\ref{eq:3.6}) and (\ref{eq:3.7}), where
the first-order term $\delta_{e^2}(y)$ is computed in the Regge-Wheeler gauge, by decomposing the perturbed metric in tensor spherical harmonics \cite{Zerilli:1970se,Zerilli:1971wd}. Following a  standard procedure, outlined and detailed in our previous works 
\cite{Bini:2013zaa,Bini:2013rfa,Bini:2014nfa,Bini:2014ica}, one decomposes $\delta$ into even-parity and odd-parity parts.
The even-parity part is given by:
\begin{eqnarray}
\delta_{\rm (even)}&=&   \left(A_0H_0+A_1H_1+A_2H_2 +A_3 K +A_4 H_0'+A_5 K'+A_6 H_0''+A_7 K''\right)Y_{lm}(\pi/2,0) 
\end{eqnarray}
where we used the standard Regge-Wheeler notation for the even metric perturbations $H_0,H_1\ldots $, and where the coefficients $A_k=A_k(y; l,m)$ are listed below (using the notation  $L\equiv l(l+1)$)
\begin{eqnarray}
\label{A-ele-odd-coeff}
A_0 &=& -  [(6 y^2-L) (1-2 y)+y m^2 (4-9 y)]y^5\nonumber\\
A_1&=&  2 i m (1-3y)y^{11/2}  \nonumber\\
A_2&=& -2 (18 y^2-18 y+5) y^6  \nonumber\\
A_3&=& -\frac{y^6}{(1-2y)} \big[y m^2 (4-9 y)-[L+2(24 y^2-21 y+5)](1-2 y) \big]  \nonumber\\  
A_4&=& 2m_2(1-y)(1-2y)(2-3y)y^4\nonumber\\
A_5&=& 2m_2(21y^2-20y+5)y^5\nonumber\\
A_6&=& m_2^2(2-3 y) (1-2 y)^2 y^3\nonumber\\
A_7&=& m_2^2 (1-2 y) (2-3y)y^4 \,. 
\end{eqnarray}
Similarly,  the odd part yields  the following expression
\begin{eqnarray}
\delta_{\rm (odd)}&=&   \left(B_0h_0+B_1h_1+B_2h_0' +B_3 h_0'' \right)\frac{dY_{lm}}{d\theta}\Bigg|_{(\pi/2,0) } \,,
\end{eqnarray}
where
\begin{eqnarray}
\label{B-ele-odd-coeff}
B_0&=&-\frac{2y^{13/2}}{m_2(1-2 y)}  [(-9 y+4 )y m^2
-(1-2 y)(L+2y)]\nonumber\\
B_1&=& \frac{2}{m_2}im (1-3 y) y^7  \nonumber\\
B_2&=&6 (1-3y)(1-2y)y^{11/2}  \nonumber\\
B_3&=& 2m_2  (1-2 y) (2-3 y) y^{9/2}\,.
\end{eqnarray}
One then re-expresses the metric functions $H_0(r),H_1(r),\ldots$ in terms of suitable radial factors $R^{\rm (even/odd)}_{lm\omega}(r)$ (for example, $h_1(r)=r^2/(r-2m_2)R^{\rm (odd)}_{lm\omega}(r)$, etc.). These auxiliary functions $R^{\rm (even/odd)}_{lm\omega}(r)$ are  chosen so as to be both solutions of the (odd-parity) Regge-Wheeler (RW) equation, with different (distributional) source terms
$S^{\rm (even/odd)}(r)$, namely
\beq
{\mathcal L}^{(r)}_{\rm (RW)}[R^{\rm (even/odd)}_{lm\omega}(r)]=S^{\rm (even/odd)}(r)\,,
\eeq
where
\beq
{\mathcal L}^{(r)}_{\rm (RW)}=\frac{d^2}{dr_*^2}+\omega^2 -\left(1-\frac{2m_2}{r}  \right)\left(\frac{l(l+1)}{r^2}-\frac{6m_2}{r^3} \right)\,,
\eeq
with $dr_*=dr/f(r)$, $f(r)=1-2m_2/r$.

The solution of the RW equation with source terms is written by using the (retarded) Green's function 
\begin{eqnarray}
\label{eq31}
G(r,r')&=&\frac{1}{W}\Bigl[X_{\rm (in)}(r)X_{\rm (up)}(r')H(r'-r) +X_{\rm (in)}(r')X_{\rm (up)}(r)H(r-r')  \Bigl] \,,
\end{eqnarray} 
with $H(x)$ denoting the Heaviside step function, $X_{\rm (in/up)}(r)$  being  solutions of the {\it homogeneous} RW equation,
and  $W$ denoting their (constant) Wronskian
\begin{eqnarray}
\label{eq32}
W&=&\left(1-\frac{2m_2}{r}  \right)\biggl[X_{\rm (in)}(r)\frac{d}{dr }X_{\rm (up)}(r )
-\frac{d}{dr}X_{\rm (in)}(r)X_{\rm (up)}(r) \biggl]={\rm constant} \,,
\end{eqnarray}
so that
\beq
R^{\rm (even/odd)}(r)=\int dr' G(r,r')f(r')^{-1}S^{\rm (even/odd)}(r')\,. 
\eeq
One can then compute  the (in general, discontinuous) limit when $r\to r_0^\pm$ of $R^{\rm (even/odd)}$ and their first derivatives $dR^{\rm (even/odd)}/dr$.

In this way, we get a new form of   $\delta^{\pm \rm (even/odd)}_{lm}$, which depends on the (left/right) direction of approach to the particle location $r_0$. E.g.,
in the odd case we have
\begin{eqnarray}
\delta^{-\rm (odd)}_{lm}(y)&=&-\frac{1}{\sqrt{1-3y}}
 \frac{96\pi y^6(m_2 X'_{\rm (up)}+y X_{\rm (up)})(C^{\rm (odd)} m_2 X'_{\rm (in)}+ D^{\rm (odd)} X_{\rm (in)} )}
{L(L-2)m_2(X_{\rm (in)}X'_{\rm (up)} -X_{\rm (up)} X'_{\rm (in)})}\,,
\end{eqnarray}
with 
\begin{eqnarray}
C^{\rm (odd)}&=& -(1-2 y) [L (y-1) -2+2 y m^2+4 y (3-4 y)]\nonumber\\
D^{\rm (odd)}&=&  2 y [L(1-2 y)^2 +y (8y-3) (1-2 y)+y m^2 (5 y-2)] \,.
\end{eqnarray}
Similar expressions hold for $\delta^{+\rm (odd)}_{lm}$.
In the even case we have instead
\begin{eqnarray}
\delta^{-\rm (even)}_{lm}(y)&=&\frac{24\pi y^5}{(1-2y)\sqrt{1-3y}}
 \frac{(A^{\rm (even)} m_2 X'_{\rm (up)}+B^{\rm (even)} X_{\rm (up)} )(C^{\rm (even)} m_2 X'_{\rm (in)}+ D^{\rm (even)} X_{\rm (in)} )}
{L(L-2)[L^2(L-2)^2+144 m^2y^3]m_2(X_{\rm (in)}X'_{\rm (up)} -X_{\rm (up)} X'_{\rm (in)})}\,.
\end{eqnarray}
with 
\begin{eqnarray}
A^{\rm (even)}&=&  2 ( 1-2 y) (-6 y^2 L+12 m^2 y^2-4 y L+2 y L^2+2 L-L^2)\\
B^{\rm (even)}&=&  y [2 y (-2 L+L^2-24 y^2+12 y) m^2+L (y L^2-12 y+4 y L-12 y^2 L+2 L+12 y^2+24 y^3-L^2)]  \nonumber\\
C^{\rm (even)}&=&  -2 ( 1-2 y) [-24 m^4 y^3-2 y (-96 y^3-12 y+5 y L^2+4 L-16 y L-2 L^2+72 y^2) m^2\nonumber\\
&& +L (-L^2-96 y^4+80 y^3+2 L+2 y L^2+2 y^3 L-16 y^2-6 y^2 L+y^2 L^2-4 y L)] \nonumber\\
D^{\rm (even)}&=&  y [4 y^2 (-2 L-60 y^2+L^2+24 y) m^4\nonumber\\
&+& 4 y (-168 y^3-18 y^3 L+192 y^4-L^3-14 y^2 L^2+64 y^2 L-30 y L+2 L+y L^3+7 y L^2+L^2+36 y^2) m^2\nonumber\\
&+& L (-100 y^3 L-2 y L^3-2 L^2+L^3+2 y^2 L^2+72 y^4 L+8 y L+384 y^4-96 y^3-384 y^5+8 y^3 L^2\nonumber\\
&& +16 y^2 L+y^2 L^3)]  \nonumber \,.
\end{eqnarray}
Similar expressions hold for  $\delta^{+\rm (even)}_{lm}$.

The functions $X_{\rm (in/up)}$ must be solutions of the homogeneous Regge-Wheeler equation.
These can be simple PN-type solutions for $l>N\equiv l_{\rm (max)}$ [the choice of $l_{\rm (max)}$ depending on the PN order accuracy required for the final result] or they can  incorporate all radiative and tail corrections  [for $2\le l\le  N$]. The low multipoles $l=0$ and $l=1$ (even and odd) are computed separately. In this work we use $l_{\rm (max)}=N=5$ so that we used PN solutions for $l\ge 6$ and radiative-corrected solutions for $l=2,3,4,5$ (the latter solutions of the homogeneous Regge-Wheeler equation are obtained  by using the technique developed by Mano, Suzuki and Tagasugi (MST), Refs. \cite{Mano:1996vt,Mano:1996mf,Mano:1996gn}).

After summing over $m$, the sum over $l$ of the regularized  values $\delta_{e^2}{}_l(y)$ comprises: i) low multipoles ($l=0$ and $l=1$), ii) radiatively-exact  contributions $l=2\ldots N$ and iii) (approximate) PN-type contributions ($l>N$).

The $l=0$ and $l=1$ multipoles are obtained by inserting, in the general expression (\ref{eq:3.7}), the corresponding Zerilli metric solutions (reformulated in a flat gauge) [see, e.g., Eqs. (4.14)--(4.18) of Ref. \cite{Bini:2014ica}]. 
The $l=0$ contribution (jump-regularized but still $B$-unsubtracted) yields
\beq
\delta_0^{\rm (unsub)}=-6y^7\frac{(1-2y)(1-y)}{\sqrt{1-3y}}\,,
\eeq
so that the final (subtracted) regularized value  $\delta_0^{\rm reg}=\delta_0^{\rm (unsub)}-B(y;0)$ is (in the following we omit the
superscript ``reg" indicating a regularized value)
\begin{eqnarray}
\delta_0^{}&=&\frac{135}{8}y^7-\frac{3759}{64}y^8 +\frac{40749}{512}y^9-\frac{169959}{8192}y^{10}-\frac{7148589}{131072}y^{11}\nonumber\\
&& 
-\frac{154173093}{1048576} y^{12}-\frac{3359407239}{8388608} y^{13} 
+O(y^{14})\,.
\end{eqnarray}

Similarly, in the case $l=1$ we find
\beq
\delta_1^{\rm (unsub)}=-6y^7\frac{14y^2-13y +5}{\sqrt{1-3y}}\,,
\eeq
and a corresponding (subtracted) regularized value $\delta_1^{}=\delta_1^{\rm (unsub)}-B(y;1)$ equal to
\begin{eqnarray}
\delta_1^{}&=&-12y^6+\frac{447}{8}y^7-\frac{7227}{64}y^8 +\frac{7653}{512}y^9-\frac{1085565}{8192}y^{10}-\frac{43461429}{131072}y^{11}\nonumber\\
&& -\frac{924074961}{1048576} y^{12}-\frac{20150969679}{8388608} y^{13} 
+O(y^{14})\,.
\end{eqnarray}
Let us also quote, as an example of radiatively-exact contribution to $\delta$ the $l=2$ contribution to the
regularized value of $\delta_{e^2}$ (i.e., with both the jump and the divergent part removed).
It reads
\begin{eqnarray}
\delta_{e^2,l=2}(y)&=& -\frac{135}{56} y^7 +\frac{31195}{448} y^8 -\frac{9757144099}{20697600} y^9 +
\left(-\frac{2304}{5}\ln(y)+\frac{7685857267007}{2739609600} -\frac{9216}{5}\ln(2)-\frac{4608}{5}\gamma  \right) y^{10}\nonumber\\
&& +\left(\frac{293824}{35}\gamma-\frac{43588443462350807}{4339541606400} +\frac{146912}{35}\ln(y)+\frac{588992}{35}\ln(2)\right) y^{11}\nonumber\\
&& -\frac{164352}{175}\pi y^{23/2}\nonumber\\
&&+\left(-\frac{737992797721187716241}{262175075205120000}-\frac{18037904}{735}\gamma-\frac{9018952}{735}\ln(y)-\frac{7245200}{147}\ln(2)\right) y^{12}\nonumber\\
&& +\frac{31511072}{3675}\pi y^{25/2}\nonumber\\
&&+\left(\frac{3741529451495079869012522279}{15714424441027952640000}-\frac{18443821276}{606375}\ln(y)-\frac{109568}{35}\pi^2+\frac{328704}{175}\ln^2(y)\right.\nonumber\\
&& -\frac{36887642552}{606375}\gamma-\frac{31525213816}{259875}\ln(2)-\frac{73728}{5}\zeta(3)+\frac{5259264}{175}\ln^2(2)+\frac{1314816}{175}\gamma^2\nonumber\\
&& \left. +\frac{1314816}{175}\gamma\ln(y)+\frac{2629632}{175}\ln(y)\ln(2)+\frac{5259264}{175}\ln(2)\gamma\right) y^{13}\nonumber\\
&&-\frac{77523640}{3087}\pi y^{27/2} +O_{\ln{}}(y^{14})\,.
\end{eqnarray}
Finally, let us exhibit the result of summing the regularized PN contributions from $l=6$ to $\infty$:
\begin{eqnarray}
S_{\rm PN}^{(6)}&\equiv &\sum_{l=6}^\infty \delta^{}_l(y)= -\frac{1215}{572} y^7+\frac{138305}{4576} y^8+\left(-\frac{580910755745041}{2943478137600}+\frac{1779}{128}\pi^2\right) y^9\nonumber\\
&& +\left(\frac{52597086939524482189}{63979440798873600}-\frac{38949}{512}\pi^2\right) y^{10}
+\left(\frac{1393795}{4096}\pi^2-\frac{131193085408413930569}{40143962854195200}\right) y^{11}\nonumber\\
&& +\left(-\frac{4725416287}{1179648}\pi^2-\frac{6580119}{524288}\pi^4+\frac{10521244939882371215189680568987}{254534169400554279075840000}  \right) y^{12}\nonumber\\
&& +\left(-\frac{31931321740721}{3303014400}\pi^2+\frac{16267066167}{33554432}\pi^4+\frac{1165940046898463658979566580099506079}{28624956957598403886482718720000}\right) y^{13}\nonumber\\
&&+O(y^{14})\,.\nonumber
\end{eqnarray}
Combining the various terms one gets the final result quoted in text,   Eq. (\ref{eq:3.13}).

\section{Details about the analytical computation of $m_2^6{\rm Tr}[{\mathcal E}^3(k)]$}

A straightforward calculation shows that the rescaled cubic invariant $m_2^6 {\rm Tr}[{\mathcal E}(k)^3]$ has the following expression
\beq
\label{trecube_1}
m_2^6 {\rm Tr}[{\mathcal E}(k)^3]
= - 3 (1-3y) (2-3 y) y^9 (1+q \widehat \delta_{e^3}(y))+O(q^2)\,,
\eeq
where
\begin{eqnarray}
\widehat \delta_{e^3}(y)&=& \frac{(-1+2y)(27 y^2-27 y+7)}{(3 y-1) (3 y-2)} h_{rr}
-m_2\frac{ (3 y-2)}{y^2}\partial_r h_{kk}-\frac{(-1+2 y) (3 y-2)}{2 y^3 (3 y-1)}m_2^2 \partial_{rr} h_{kk}\nonumber\\
&&
 -\frac{y (18 y^2-18 y+5)}{ (3 y-1) (3 y-2) (-1+2 y) }h_{kk} -\frac{(y-1) (18 y^2-18 y+5) y^2}{ (-1+2 y) (3 y-2)m_2^2} h_{\phi\phi}
-\frac{3 y^2 }{m_2^2 (-9 y+3) (3 y-2)}h_{\theta \theta}\nonumber\\
&& +\frac{2 (18 y^2-18 y+5) y^{3/2}}{m_2 (-1+2 y)  (3 y-2)} h_{t\phi}
-\frac{3}{ 2 (-9 y+3) (3 y-2) y }\partial_{\theta\theta} h_{kk}
+\frac{ (3 y-1)^2}{ 2 (-1+2 y) y (3 y-2)}\partial_{\bar \phi\bar \phi}h_{kk}\nonumber\\
&&-\frac{(18 y^2-18 y+5)}{ y^{1/2} (3 y-2)}\partial_{\bar \phi}h_{tr}
-\frac{(18 y^2-18 y+5) y}{m_2(3 y-2) } \partial_{\bar \phi} h_{r\phi}
+\frac{(18 y^2-18 y+5) y}{m_2 (3 y-2) }\partial_r  h_{\phi\phi}\nonumber\\
&& +\frac{(18 y^2-18 y+5)}{ y^{1/2}(3 y-2)} \partial_r h_{t\phi}\,.
\end{eqnarray}

We proceed along the lines explained above. Let us only quote the most relevant new features. The singular part of $\widehat \delta_{l}(y)$ (which needs to be subtracted 
to regularize $\widehat \delta_{e^3}(y)$) is found to be
\beq
\widehat B(y; l)= L\widehat  b_0(y)+\widehat b_1 (y)\,,
\eeq
where
\begin{eqnarray}
\widehat b_0(y)&=&  \frac{3}{2}-\frac{27}{8} y -\frac{297}{128} y^2-\frac{1623}{512} y^3-\frac{131805}{32768} y^4  \nonumber\\
&&   -\frac{562131}{131072} y^5-\frac{4688361}{2097152} y^6+\frac{57419361}{8388608} y^7+O(y^8)\nonumber\\
\widehat b_1 (y)&=& -\frac{183}{32} y+\frac{717}{256} y^2+\frac{13155}{2048} y^3+\frac{403797}{32768} y^4\nonumber\\
&& +\frac{13249725}{524288} y^5+\frac{231713151}{4194304} y^6\nonumber\\
&& +\frac{4282692327}{33554432} y^7 +O(y^8)\,.
\end{eqnarray}
The unsubtracted contributions of the  low multipoles $l=0,1$  are
\begin{eqnarray}
\widehat \delta_{l=0}^{\rm (unsub)} &=& -3 \frac{ (1-2 y)  (1-y)y}{(2-3 y)(1-3 y)^{3/2}}\,, \nonumber\\
\widehat\delta_{l=1}^{\rm (unsub)} &=& -3\frac{ y (26 y^2-25 y+5)}{(2-3 y)(1-3 y)^{3/2} }
\end{eqnarray}
with associated  subtracted  regularized values $\widehat \delta_0^{}=\widehat \delta_0^{\rm (unsub)}-\widehat B(y;0)$ and $\widehat \delta_1^{}=\widehat \delta_1^{\rm (unsub)}-\widehat B(y;1)$.
As examples of intermediate results let us quote the radiatively-correct regularized solution for $l=2$,  
\begin{eqnarray}
\widehat \delta_2^{}&=&
-\frac{135}{224} y+\frac{29305}{1792} y^2-\frac{5248670899}{82790400} y^3\nonumber\\
&& +\left(-\frac{2304}{5}\ln 2 -\frac{576}{5}\ln y +\frac{53165214824923}{120542822400} -\frac{1152}{5}\gamma \right) y^4\nonumber\\
&& +\left( \frac{22616}{35}\ln y +\frac{18160}{7}\ln 2 +\frac{45232}{35}\gamma-\frac{14464275420235367}{17358166425600} \right) y^5\nonumber\\
&& -\frac{41088}{175}\pi y^{11/2} \nonumber\\
&& +\left(-\frac{373726}{735}\ln(y)-\frac{747452}{735}\gamma-\frac{63948592610334337491983}{13633103910666240000}-\frac{302492}{147}\ln(2)\right) y^6\nonumber\\
&& +\frac{194312}{147}\pi y^{13/2}\nonumber\\
&& +\left(\frac{2808274681091617913731512647}{62857697764111810560000}+\frac{82176}{175}\ln^2(y)-\frac{6611028394}{606375}\ln(y)+\frac{657408}{175}\ln(y)\ln(2)\right. \nonumber\\
&& +\frac{328704}{175}\ln(y)\gamma-\frac{13222056788}{606375}\gamma+\frac{1314816}{175}\ln^2(2)+\frac{1314816}{175}\ln(2)\gamma+\frac{328704}{175}\gamma^2\nonumber\\
&& \left.-\frac{18432}{5}\zeta(3)-\frac{27392}{35}\pi^2-\frac{11333715604}{259875}\ln(2)\right) y^7\nonumber\\
&& -\frac{16183322}{15435}\pi y^{15/2}  +O_{\ln{}}(y^8)\,,
\end{eqnarray}
and the summation of the PN contributions for $l\ge 6$
\begin{eqnarray}
\widehat S_{\rm PN}^{(6)}&=&\sum_{l=6}^\infty \widehat \delta^{}_l(y)= 
-\frac{1215}{2288} y+\frac{121295}{18304} y^2+\left(\frac{1779}{512}\pi^2-\frac{310927685187841}{11773912550400}  \right) y^3\nonumber\\
&& +\left(-\frac{2271}{256}\pi^2+\frac{7526860183399951567}{63979440798873600}  \right) y^4\nonumber\\
&& +\left(-\frac{1213007580302673923737}{2729789474085273600}+\frac{907147}{16384}\pi^2  \right) y^5\nonumber\\
&& +\left(\frac{156773216742226524002873899261}{59890392800130418606080000}-\frac{6580119}{2097152}\pi^4-\frac{900450163}{4718592}\pi^2\right) y^6\nonumber\\
&& +\left(\frac{5955078711}{134217728}\pi^4+\frac{17913668814079}{13212057600}\pi^2-\frac{9611524527185390804403803435123566091}{526699208019810631511282024448000}\right) y^7+O(y^8)\,.
\end{eqnarray}

\section{Details about the analytical computation of $m_2^4{\rm Tr}[{\mathcal B}^2(k)]$}

Our starting expression for $1$SF expansion of  $m_2^4{\rm Tr}[{\mathcal B}^2(k)]$ is
\beq
\tilde {\mathcal J}_{2m}=
m_2^4{\rm Tr} {\mathcal B}(k)^2=18 (1-2 y) y^7+q \delta_{b^2}(y)+O(q^2)\,.
\eeq
where 
\begin{eqnarray}
\delta_{b^2}(y)
&=&  3(3y-1)(-1+2y)^2y^5 m_2 \partial_r h_{rr} -3\frac{(3 y-1) (-1+2 y) y^7}{m_2}\partial_r h_{\theta\theta}+\frac{6(3y-1)(-1+2y)y^7}{m_2}\partial_\theta h_{r\theta}\nonumber\\
&&- 3(3 y-1) (-1+2 y) y^{9/2}m_2\partial_{r\phi}  h_{tr}-3 (3y-1)(-1+2y)y^6 \partial_{r\phi}   h_{r\phi}-\frac{3(3y-1)y^8}{m_2^2}\partial_{\theta\phi} h_{\theta\phi} \nonumber\\
&&-\frac{3(3y-1) y^{13/2}}{m_2}\partial_{\theta\phi}  h_{t\theta} - 3(-1+2y)y^4m_2^2 \partial_{rr}  h_{kk}-\frac{6y^8}{(-1+2 y)} h_{kk} - 6(-1+2y)(15y^2-9y+1)y^6  h_{rr}\nonumber\\
&&- 9(3y-1)y^5 m_2\partial_r h_{kk}+\frac{6(3y-1)(3y^2-3y+1)y^8}{(-1+2y)m_2^2} h_{\phi\phi}+\frac{18(-1+2y)y^9}{m_2^2} h_{\theta\theta}-\frac{12(3y-1)(3y-2)y^{15/2}}{m_2(-1+2y)} h_{t\phi}\nonumber\\
&&- 3y^6 \partial_{\theta\theta}h_{kk}
+6(3y-1) (-1+2 y) y^{11/2} \partial_\phi   h_{tr}+\frac{6(3 y-1) (-1+2 y) y^7}{m_2}\partial_\phi h_{r\phi}   \nonumber\\
&&-\frac{3(3 y-1) (5 y-3) y^7 }{m_2}\partial_r h_{\phi\phi} - 6 (3 y-1) (y-1) y^{11/2} \partial_r h_{t\phi}+ 3(3y-1) (-1+2 y) y^6  \partial_{rr}   h_{\phi\phi}\nonumber\\
&&+ 3(3y-1)(-1+2 y) y^{9/2}m_2 \partial_{rr}  h_{t\phi}+\frac{3(3 y-1) y^8}{m_2^2}\partial_{\theta\theta}  h_{\phi\phi} +\frac{3(3y-1) y^{13/2}}{m_2}\partial_{\theta\theta} h_{t\phi} 
\end{eqnarray}
Following the same procedure outlined above,  
one identifies an even part and an odd part for $\tilde \delta^R(y)$:
\begin{eqnarray}
\tilde \delta_{\rm (even)}&=&   \left(\tilde A_0H_0+\tilde A_1H_1+\tilde A_2H_2 +\tilde A_3 K +\tilde A_4 H_0'+\tilde A_5 K'+\tilde A_6 H_0''+\tilde A_7 K''+\tilde A_8 H_1'+\tilde A_9 H_2'\right)Y_{lm}(\pi/2,0) 
\end{eqnarray}
where 
\begin{eqnarray}
\label{A-magn-odd-coeff}
\tilde A_0 &=& -3(8y^2-2m^2y-2y+2Ly+m^2-L)y^6\nonumber\\  
\tilde A_1&=&   6i m (1-2 y) (1-3 y) y^{11/2} \nonumber\\   
\tilde A_2&=&  6 (1-6 y) (1-2 y) y^6\nonumber\\   
\tilde A_3&=&  -3y^6(32y^2-2m^2y+2Ly-18y-L+2+m^2)\nonumber\\  
\tilde A_4&=&  3m_2  (1-2 y) (3-5 y)y^5 \nonumber\\  
\tilde A_5&=&  6m_2 (3-7 y)y^6 \nonumber\\   
\tilde A_6&=&  3m_2 (1-2 y)^2y^4 \nonumber\\   
\tilde A_7&=&  3m_2^2(1-2 y)^2 y^4 \nonumber\\   
\tilde A_8&=&   -3im m_2(1-2 y) (1-3 y) y^{9/2}\nonumber\\  
\tilde A_9&=& -3m_2(1-3 y) (1-2 y) y^5\,.  
\end{eqnarray}

Similarly, for the odd part we have
\begin{eqnarray}
\tilde \delta_{\rm (odd)}&=&   \left(\tilde B_0h_0+\tilde B_1h_1'+\tilde B_2h_0' +\tilde B_3 h_0'' \right)\frac{dY_{lm}}{d\theta}\Bigg|_{(\pi/2,0) } \,,
\end{eqnarray}
where
\begin{eqnarray}
\label{B-magn-odd-coeff}
\tilde B_0&=& -\frac{3}{m_2} (20 y^2-8 y+L y+2 y m^2-L) y^{13/2} \nonumber\\  
\tilde B_1&=&  -3im (1-2 y) (1-3 y) y^6\nonumber\\   
\tilde B_2&=&  -6 (1-4y)(1-3y)y^{11/2}\nonumber\\   
\tilde B_3&=&   3m_2(1-2y)(1-y)y^{9/2}\,.   
\end{eqnarray}
[Note that the  above (odd) coefficients (\ref{A-magn-odd-coeff}) and  (\ref{B-magn-odd-coeff})  do not correspond to the analogous electric coefficients (\ref{A-ele-odd-coeff}) and  (\ref{B-ele-odd-coeff}).]
Re-expressing the metric functions $H_0(r),H_1(r),\ldots$ in terms of the Regge-Wheeler-Zerilli radial functions $R^{\rm (even/odd)}_{lm\omega}(r)$, as for the electric-type quantity explained above, we get a new form of $\tilde \delta^{\pm \rm (even/odd)}_{lm}$ which also depends on the (left/right) direction of approach  to the particle location $r_0$.
We have
\begin{eqnarray}
\tilde \delta^{-\rm (odd)}_{lm}&=&-\frac{1}{(1-2y)\sqrt{1-3y}}
 \frac{96\pi y^6(m_2 X'_{\rm (up)}+yX_{\rm (up)})(\tilde C^{\rm (odd)} m_2 X'_{\rm (in)}+ \tilde D^{\rm (odd)} X_{\rm (in)} )}
{L(L-2)m_2(X_{\rm (in)}X'_{\rm (up)} -X_{\rm (up)} X'_{\rm (in)})}\,,
\end{eqnarray}
with 
\begin{eqnarray}
\tilde C^{\rm (odd)}&=& -( 1-2 y)^2 (-L+2+2m^2y+Ly-16y+32y^2)  \nonumber\\   
\tilde D^{\rm (odd)}&=&  2y^2(-64y^3+72y^2+3+3m^2y-26y-m^2y^2+2L-9Ly-m^2+10y^2L)\,.  
\end{eqnarray}
Similar expressions hold for $\tilde \delta^{+\rm (odd)}_{lm}$.
In the even case we have instead
\begin{eqnarray}
\tilde \delta^{-\rm (even)}_{lm}(y)&=&\frac{48\pi y^6}{(1-2y)^2\sqrt{1-3y}}
 \frac{(\tilde A^{\rm (even)} m_2 X'_{\rm (up)}+\tilde B^{\rm (even)} X_{\rm (up)} )(\tilde C^{\rm (even)} m_2 X'_{\rm (in)}+ \tilde D^{\rm (even)} X_{\rm (in)} )}
{L(L-2)[L^2(L-2)^2+144 m^2y^3]m_2(X_{\rm (in)}X'_{\rm (up)} -X_{\rm (up)} X'_{\rm (in)})}\,.
\end{eqnarray}
with 
\begin{eqnarray}
\tilde A^{\rm (even)}&=&  2 (1-2y)((12m^2-6L)y^2+(-4L+2L^2)y+2L-L^2) \\   
\tilde B^{\rm (even)}&=& (-48m^2+24L)y^3+(12L-12L^2+24m^2)y^2+(2m^2L^2+L^3-12L+4L^2-4m^2L)y+2L^2-L^3 
\nonumber\\
\tilde C^{\rm (even)}&=&   (1-2y) \left[(-768m^2+384L)y^4+(-416L+768m^2+28L^2-48m^4)y^3\right.\nonumber\\
&& +(-2m^2L^2+11L^3+24m^4+144L+28m^2L-52L^2-240m^2)y^2\nonumber\\
&& \left.+(24m^2-24m^2L-12L^3+32L^2+6m^2L^2-16L)y-6L^2-2m^2L^2+4m^2L+3L^3\right] \nonumber\\
\tilde D^{\rm (even)}&=&   y \left[(1536m^2-768L)y^5+(960L-1728m^2+24m^4-252m^2L+72L^2)y^4  \right. \nonumber\\
&& +(-72m^4-68L^2-2L^3-384L-40m^2L^2+332m^2L+624m^2)y^3\nonumber\\
&& +(48m^2L^2-4m^2L^3+8Lm^4+48L+16L^2+8L^4-140m^2L+24m^4-15L^3-72m^2-4L^2m^4)y^2 \nonumber\\
&& +(2L^2m^4-4L^2+28m^2L+6m^2L^3-26m^2L^2-4Lm^4-8L^4+18L^3)y\nonumber\\
&& \left.+2L^2-2m^2L^3+6m^2L^2-4m^2L+2L^4-5L^3  \right]\nonumber \,.
\end{eqnarray}
Similar expressions hold for  $\tilde \delta^{+\rm (even)}_{lm}$.

The calculation proceeds as in the cases discussed above.
The subtraction term  for $\delta_{b^2}(y)$ is found to be
\beq
\tilde B(y; l)=L \tilde b_0(y)+\tilde b_1(y)
\eeq
with
\begin{eqnarray}
\tilde b_0(y) &=&  \frac{22863555}{268435456}y^{15}+O(y^{16}) \nonumber\\
\tilde b_1(y) &=&  6y^7-\frac{243}{2}y^8+\frac{7215}{32}y^9+\frac{267}{32}y^{10} +\frac{196803}{8192}y^{11}+\frac{2224713}{32768}y^{12}  +\frac{99490095}{524288}y^{13}
+\frac{2210742531}{4194304}y^{14}\nonumber\\
&&   +\frac{381877833969}{268435456}y^{15}+O(y^{16}) 
\end{eqnarray}

The $l=0$ contribution (jump-regularized but still $B$-unsubtracted by the $B$-term) results
\beq
\tilde \delta_0^{\rm (unsub)}= -6y^7\frac{ (1-4 y)  (y-1)}{\sqrt{1-3y}}\,,
\eeq
so that the final (subtracted) regularized value  $\tilde \delta_0^{}=\tilde \delta_0^{\rm (unsub)}-\tilde B(y;0)$ is
\begin{eqnarray}
\tilde \delta_0^{}&=&\frac{201}{2}y^8-\frac{7239}{32}y^9-\frac{735}{32}y^{10}-\frac{518211}{8192}y^{11} -\frac{5604681}{32768}y^{12}-\frac{244061487}{524288}y^{13}\nonumber\\
&&-\frac{5354983683}{4194304}y^{14}
+O(y^{15})\,. 
\end{eqnarray}
Similarly, in the case $l=1$ we find
\beq
\tilde \delta_1^{\rm (unsub)}=  6y^7\frac{28 y^2-7y -3}{\sqrt{1-3y}}\,,
\eeq
with a corresponding  (subtracted) regularized value $\tilde \delta_1^{}=\tilde \delta_1^{\rm (unsub)}-\tilde B(y;1)$ equal to
\begin{eqnarray}
\tilde \delta_1^{}&=&-24y^7+\frac{105}{2}y^8-\frac{5799}{32}y^9-\frac{1599}{32}y^{10}-\frac{1720899}{8192}y^{11}
-\frac{21529929}{32768}y^{12}-\frac{1017431343}{524288}y^{13}\nonumber\\
&& -\frac{23629205763}{4194304}y^{14}
+O(y^{15})\,.
\end{eqnarray}
As an example of radiatively-exact term, we give the regularized value of $\tilde \delta$ for $l=2$,
\begin{eqnarray}
\tilde \delta^{}_2(y)&=&  -\frac{15}{2} y^8+\frac{73901}{672} y^9-\frac{218530317}{431200} y^{10}\nonumber\\
&& +\left(-\frac{3616}{5} \ln(y)+\frac{302133200808277}{90407116800} -2880\ln(2)-\frac{7232}{5}\gamma\right) y^{11}\nonumber\\
&& +\left(\frac{110632}{21}\ln(y)+\frac{221264}{21}\gamma+\frac{63280}{3}\ln(2)-\frac{31059866029753481}{3254656204800} \right) y^{12}\nonumber\\
&& -\frac{10272}{7}\pi y^{25/2}\nonumber\\
&& +\left(-\frac{469856}{49}\ln(y)-\frac{38106213603422952309467}{1704137988833280000} -\frac{939712}{49}\gamma-\frac{5217472}{135}\ln(2)\right) y^{13}\nonumber\\
&& +\frac{96728}{9}\pi y^{27/2}\nonumber\\
&& +\left(\frac{4715300156233844764198612711}{14142981996925157376000}+\frac{1537376}{525}\ln^2(y)-\frac{17032098407}{259875}\ln(y)+\frac{6149504}{525}\ln(y)\gamma\right.\nonumber\\
&& +\frac{4095104}{175}\ln(y)\ln(2)-\frac{1537376}{315}\pi^2-\frac{67951632142}{259875}\ln(2)-\frac{34064196814}{259875}\gamma+\frac{24556928}{525}\ln^2(2)\nonumber\\
&& \left. +\frac{6149504}{525}\gamma^2-\frac{114944}{5}\zeta(3)+\frac{8190208}{175}\ln(2)\gamma\right) y^{14}
\nonumber\\
&& -\frac{50274592}{5145}\pi y^{29/2}\nonumber\\
&& +O_{\ln {}}(y^{15})\,.
\end{eqnarray}
The sum of the PN  contributions  for $l\le 6$ is found to be
\begin{eqnarray}
\tilde S_{\rm (PN)}^{(6)}&=&\sum_{l=6}^\infty  \tilde \delta_{l }^{}(y)=-\frac{945}{143} y^8+\frac{22357}{528} y^9+\left(-\frac{53778169487729}{183967383600} +\frac{123}{4}\pi^2\right) y^{10}  \nonumber\\
&& +\left(\frac{57815}{256}\pi^2-\frac{10284424419054240163}{4739217836953600}\right)y^{11}\nonumber\\
&& +\left(-\frac{651042905735117257129}{90323916421939200} +\frac{191973}{256}\pi^2\right) y^{12}\nonumber\\
&& +\left(\frac{80074047}{131072}\pi^4-\frac{7934674343}{294912}\pi^2+\frac{3133543922680342689253334319888697}{15144783079332979605012480000} \right) y^{13}\nonumber\\
&& +\left(\frac{50187001206383676002431993886829037909}{25322077308644741899580866560000} -\frac{116483667391}{8388608}\pi^4-\frac{4420569077921}{68812800}\pi^2\right) y^{14}\nonumber\\
&& 
+O(y^{15})\,.
\end{eqnarray}
The final result was given in Eq. (\ref{eq:3.18}) of the text.

\section{Octupolar-type tidal invariants}

In this section, we discuss the computation of the {\it irreducible} octupolar invariant (linked to $l^\epsilon=3^+$ tidal effects) 
\beq
J_{3^+}= G_{\alpha\beta\gamma}G^{\alpha\beta\gamma} \equiv \Gamma^4 {\mathcal J}_{3^+} \nonumber\\
\eeq
where
\beq
G_{\alpha\beta\gamma}=-\left[{\rm Sym}_{\alpha\beta\gamma} \left(\nabla_\gamma^\perp  R_{ \alpha \mu \beta \nu} \right)\right] U^\mu U^\nu\,.
\eeq
Here $\nabla_\gamma^\perp=P(U)^\mu{}_\gamma \nabla_\mu$, with $P(U)=g+U\otimes U$, denoting the projection operator orthogonal to $U$.
We found convenient to compute the octupolar invariant $J_{3^+}$ in terms of two other invariants (see \cite{Bini:2012gu}), namely
\beq
\label{from_faye}
J_{3^+}=K_{3^+}+\frac13 J_{\dot 2^+}\,,
\eeq
where
\begin{eqnarray}
K_{3^+}&=&C_{\alpha\beta\gamma}C^{\alpha\beta\gamma}\equiv \Gamma^4 {\mathcal K}_{3^+}\nonumber\\
J_{\dot 2^+} &=& \dot G_{\alpha\beta} \dot   G^{\alpha\beta}\equiv \Gamma^6 {\mathcal J}_{\dot 2\rm e}\,,
\end{eqnarray}
with 
\begin{eqnarray}
C_{\alpha\beta\gamma}&=&\left[{\rm Sym}_{\alpha\beta\gamma} \left(\nabla_\gamma R_{ \alpha \mu \beta \nu}\right)\right]U^\mu U^\nu \nonumber\\
\dot G_{\alpha\beta}&=& -\nabla_U {\mathcal E}(U)_{\alpha\beta}\,.
\end{eqnarray}
Let us introduce their suitably $\Gamma$-rescaled counterparts, namely 
\begin{eqnarray}
{\mathcal C}_{\alpha\beta\gamma} 
&=&\left[{\rm Sym}_{\alpha\beta\gamma} \left(\nabla_\gamma R_{ \alpha \mu \beta \nu}\right)\right]k^\mu k^\nu \equiv \Gamma^{-2}C_{\alpha\beta\gamma}\nonumber\\
{\mathcal G}_{\alpha\beta\gamma}
&=&-\left[{\rm Sym}_{\alpha\beta\gamma} \left(\nabla_\gamma^\perp  R_{ \alpha \mu \beta \nu} \right)\right]k^\mu k^\nu\equiv \Gamma^{-2}G_{\alpha\beta\gamma}\nonumber\\
{\dot {\mathcal G}}_{\alpha\beta}&=& - \nabla_k {\mathcal E}(k)_{\alpha\beta}\equiv \Gamma^{-3}\dot   G_{\alpha\beta}\,.
\end{eqnarray}

We work  with the $\Gamma$-rescaled quantities with the  following results
\begin{eqnarray}
m_2^6 {\mathcal K}_{3^+}&=&6 (1-2 y) (42 y^2-46 y+15) y^8  +m_2^6 {\mathcal K}_{3^+}^{\rm 1SF}=6 (1-2 y) (42 y^2-46 y+15) y^8  (1+q \,\widehat \delta_{K3^+}) +O(q^2)
\end{eqnarray}
where, for example, 
\begin{eqnarray}
{\mathcal K}_{3^+}^{\rm 1SF}&=&
-\frac{4}{m_2}(7y-4)(2y-3)y^{17/2} \partial_{\theta\theta}(h_{t\phi}) -2(-1+2y)(40y^2-53y+15)y^8 \partial_{rr}(h_{\phi\phi})\nonumber\\
&&  -(42y^3+3y^2-44y+18)y^6 m_2\partial_{r}(h_{tt})-8(-1+2y)(17y^2-21y+6)y^{13/2}m_2\partial_{rr}(h_{t\phi})\nonumber\\
&&-\frac{2}{(-1+2y)}y^9 (154y^2+60-189y)h_{tt} +\frac{4}{m_2^2}(3y-1)(2y-3)y^{10}\partial_{\phi\theta}(h_{\theta\phi})\nonumber\\
&& +(138y^2-145y+42)(-1+2y)^2y^7m_2\partial_{r}(h_{rr}) +\frac{2}{m_2}(6y-5)(-1+2y)(2y-3)y^9 \partial_\theta(h_{r\theta})\nonumber\\
&& +3(3y-2)(-1+2y)^2y^7m_2\partial_{rrr}(h_{\phi\phi})+6(3y-2)(-1+2y)^2y^{11/2} m_2^2\partial_{rrr}(h_{t\phi})\nonumber\\
&&+6(-1+2y)(2y-3)y^{15/2}\partial_{r\theta\theta}(h_{t\phi}) +4(3y-1)(-1+2y)(2y-3)y^8\partial_{\phi r}(h_{r\phi}) \nonumber\\
&&+\frac{4}{m_2}(3y-1)(2y-3)y^{17/2}\partial_{\phi\theta}(h_{t\theta}) +\frac{3}{m_2}(3y-1)(7y-3)y^9\partial_{\phi\phi r}(h_{\phi\phi}) \nonumber\\
&&  -\frac{1}{m_2}(6y-11)(-1+2y)(2y-3)y^9\partial_{r}(h_{\theta\theta}) +3(3y-1)(7y-3)y^6m_2 \partial_{\phi\phi r}(h_{tt})\nonumber\\
&&+6(3y-1)(7y-3)y^{15/2}\partial_{\phi\phi r}(h_{t\phi}) +3(3y-2)(-1+2y)^2y^4 m_2^3\partial_{rrr}(h_{tt})+3(-1+2y)(2y-3)y^6 m_2\partial_{r\theta\theta}(h_{tt}) \nonumber\\
&&+4(3y-1)(-1+2y)(2y-3)y^{13/2} m_2\partial_{\phi r}(h_{tr}) -2(486y^3-724y^2+363y-60)y^6 m_2\partial_r (h_{kk})\nonumber\\
&& -2(4y-3)(2y-3)y^7\partial_{\theta\theta}(h_{tt}) -\frac{4}{m_2  (-1+2y)}y^{17/2}(3y-1)(23y^2-31y+9)\partial_{\phi\phi }(h_{t\phi}) \nonumber\\
&&+\frac{6}{m_2^2}(2y-5)(-1+2y)(2y-3)y^{10}h_{\theta\theta} -\frac{2}{m_2^2 (-1+2y)}y^{10}(3y-1)(23y^2-31y+9)\partial_{\phi\phi }(h_{\phi\phi})\nonumber\\
&& +\frac{2}{m_2}(-1+2y)(84y^2-40y+15)y^9 \partial_{\phi }(h_{r\phi}) +\frac{3}{m_2}(-1+2y)(2y-3)y^9\partial_{r\theta\theta}(h_{\phi\phi})\nonumber\\
&&  +\frac{2}{m_2^2}(-1+2y)(92y^2-52y+15)y^{10}h_{\phi\phi} -2(-1+2y)(28y^2-31y+9)y^5 m_2^2\partial_{rr}(h_{tt})\nonumber\\
&& +\frac{1}{ m_2}(-1+2y)(12y^2-104y+27)y^9\partial_{r}(h_{\phi\phi}) +\frac{16}{m_2}(5y-2)(2y-3)y^{19/2}h_{t\phi}\nonumber\\
&& -2(-1+2y)(7y-6)(15y-2)y^{15/2}\partial_{\phi }(h_{tr}) -\frac{2}{(-1+2y)}y^7(3y-1)(23y^2-31y+9)\partial_{\phi\phi }(h_{tt}) \nonumber\\
&&-\frac{10}{m_2^2}(-1+2y)(2y-3)y^{10}\partial_{\theta\theta}(h_{\phi\phi})-2(-1+2y)(894y^3-1335y^2+694y-123)y^8h_{rr}\nonumber\\
&&  +4(-1+2y)(36y^2-50y+15)y^{15/2}\partial_{r}(h_{t\phi})\,.
\end{eqnarray}
We have for the (jump-regularized but still unsubtracted) low multipoles (and for the fractional 1SF correction to  $K_{3^+}$)  
\begin{eqnarray}
\widehat \delta_{K3^+}^{l=0, \rm (unsub)}&=& \frac{-100y^3+84y^4+94y^2+15-60y}{3(-1+2y)(42 y^2-46 y+15) \sqrt{1-3 y}}\nonumber\\
&=&   -\frac13-\frac{77}{90} y-\frac{13883}{5400} y^2-\frac{1048681}{162000} y^3-\frac{317621843}{19440000} y^4-\frac{24889884451}{583200000} y^5-\frac{4039031134639}{34992000000} y^6\nonumber\\
&& -\frac{335874959831873}{1049760000000} y^7-\frac{226643666008108163}{251942400000000} y^8+O(y^9) 
\end{eqnarray}
\begin{eqnarray}
\widehat \delta_{K3^+}^{l=1, \rm (unsub)}&=& -\frac{4}{3} y  \frac{(357 y^3-566 y^2+345 y-75)}{\sqrt{1-3 y}(-1+2y)(42 y^2-46 y+15)} \nonumber\\
&=&  -\frac{32}{3} y -\frac{416}{45}y^2-\frac{10436}{675}y^3 -\frac{372776}{10125}y^4 -\frac{62113439}{607500}y^5 -\frac{2686057049}{9112500}y^6\nonumber\\
&& -\frac{929114577097}{1093500000} y^7 -\frac{9989205434563}{4100625000} y^8 +O(y^9) \,.
\end{eqnarray}
The final (regularized) result is found to be
\begin{eqnarray}
\label{final_K3}
\widehat \delta_{K3^+}&=&-\frac{8}3+\frac{358}{45} y +\frac{11848}{675}y^2\nonumber\\
&&+\left(-\frac{3581903}{40500}+\frac{4681}{1536}\pi^2\right)y^3\nonumber\\
&&+\left(\frac{614794483}{2430000}-\frac{4096}{15}\ln(2)-\frac{1024}{15}\ln(y)-\frac{2048}{15}\gamma-\frac{790931}{92160}\pi^2\right)y^4\nonumber\\
&&+\left(\frac{431520437}{11059200}\pi^2-\frac{759123028241}{1020600000}+\frac{535352}{1575}\ln(y)+\frac{354064}{225}\ln(2)+\frac{1070704}{1575}\gamma-\frac{1458}{7}\ln(3)\right)y^5\nonumber\\
&&-\frac{219136}{1575}\pi y^{11/2}\nonumber\\
&&+\left(\frac{12569905047667}{2187000000}+\frac{181080056}{212625}\gamma+\frac{1024}{75}\pi-\frac{123628168}{212625}\ln(2)+\frac{90540028}{212625}\ln(y)+\frac{73953}{35}\ln(3)\right.\nonumber\\
&& \left. -\frac{1903269674027}{1769472000}\pi^2-\frac{42147341}{6291456}\pi^4\right)y^6\nonumber\\
&&+\frac{118163398}{165375}\pi y^{13/2}\nonumber\\
&&+\left( \frac{52369829422440012073}{990186120000000}+\frac{351206984461}{6039797760}\pi^4-\frac{4143716714678}{245581875}\gamma-\frac{2071858357339}{245581875}\ln(y)\right. \nonumber\\
&& +\frac{9765625}{14256}\ln(5)+\frac{438272}{1575}\ln^2(y)-\frac{214350489}{30800}\ln(3)-\frac{6124042466966}{245581875}\ln(2)-\frac{4176344893416403}{990904320000}\pi^2\nonumber\\
&& +\frac{1753088}{1575}\gamma^2-\frac{32768}{15}\zeta(3)+\frac{7012352}{1575}\ln^2(2)+\frac{1753088}{1575}\ln(y)\gamma+\frac{7012352}{1575}\gamma\ln(2)\nonumber\\
&& \left. +\frac{3506176}{1575}\ln(2)\ln(y)\right)y^7\nonumber\\
&&+ \frac{169822838237}{245581875}\pi y^{15/2}\,.
\end{eqnarray}

The complete 7.5PN computation of the $k-$normalized, $m_2-$rescaled octupolar tidal invariant ${\mathcal J}_{\dot 2^+}$ is given by
\begin{eqnarray}
m_2^6 {\mathcal J}_{\dot 2^+}&=& 
18 (1-3y)(1-2y)^2 y^9 (1+\nu \widehat \delta_{\dot 2 ^+})\,,
\end{eqnarray}
where 
\begin{eqnarray}
\label{final_J2dot}
\widehat \delta_{\dot 2^+}&= &-2+\frac{13}{3} y+\frac{167}{12} y^2+\left(-\frac{1327}{24}+\frac{1165}{512}\pi^2\right) y^3\nonumber\\
&& +\left(-\frac{512}{3}\gamma-\frac{256}{3}\ln(y)-\frac{1024}{3}\ln(2)+\frac{4309069}{14400}-\frac{919}{192}\pi^2  \right) y^4 \nonumber\\
&& +\left(-\frac{266364907}{604800}+\frac{25855117}{294912}\pi^2+\frac{79712}{315}\ln(y)+\frac{159424}{315}\gamma+\frac{94912}{63}\ln(2)-486\ln(3)\right) y^5\nonumber\\
&& -\frac{54784}{315}\pi y^{11/2}\nonumber\\
&& +\left(\frac{640244}{567}\ln(y)+\frac{30132}{7}\ln(3)+\frac{1280488}{567}\gamma-\frac{3388984}{2835}\ln(2)-\frac{12986651897}{7257600}-\frac{1807468975}{7077888}\pi^2-\frac{12666295}{1048576}\pi^4\right) y^6\nonumber\\
&& +\frac{18653182}{33075}\pi y^{13/2}\nonumber\\
&& +\left(-\frac{8192}{3}\zeta(3)+\frac{1194070016}{3274425}\ln(2)+\frac{109568}{315}\ln^2(y)-\frac{49548974752}{3274425}\gamma-\frac{132126957}{12320}\ln(3)-\frac{30510962277409}{23781703680}\pi^2\right.\nonumber\\
&& +\frac{142382205611}{805306368}\pi^4+\frac{2182965632437}{938843136}-\frac{24774487376}{3274425}\ln(y)-\frac{9765625}{2592}\ln(5)+\frac{1753088}{315}\ln^2(2)+\frac{438272}{315}\gamma^2\nonumber\\
&& \left. +\frac{1753088}{315}\ln(2)\gamma+\frac{438272}{315}\ln(y)\gamma+\frac{876544}{315}\ln(y)\ln(2)\right) y^7\nonumber\\
&& +\frac{20370348976}{9823275}\pi y^{15/2}+O_{\ln{}}(y^8)\,.
\end{eqnarray}
In this case, the contribution of the (jump-regularized but still unsubtracted) low multipoles turns out to be
\beq
\widehat \delta_{\dot 2^+}^{l=0, \rm (unsub)}= -\frac13 \frac{(18y-7)y(-1+y)}{ (1-2 y) (1-3 y)^{3/2}}
\eeq
and
\beq
\widehat \delta_{\dot 2^+}^{l=1, \rm (unsub)}= -\frac13  \frac{y(126 y^2-91 y+15)}{ (1-2 y) (1-3 y)^{3/2}}\,.
\eeq
The subtraction term reads $\widehat B(y;l)=L \widehat b_0(y)+ \widehat b_1(y)$, with
\begin{eqnarray}
\widehat b_0(y)&=& 1-\frac{15}{8} y-\frac{185}{64} y^2-\frac{4155}{1024} y^3-\frac{76275}{16384} y^4-\frac{309213}{131072} y^5+\frac{9719235}{1048576} y^6+\frac{1568005965}{33554432} y^7 \nonumber\\
\widehat b_1(y)&=& -\frac{419}{96} y+\frac{285}{128} y^2+\frac{6401}{4096} y^3-\frac{104915}{49152} y^4-\frac{7021983}{524288} y^5-\frac{86553773}{2097152} y^6-\frac{13597215787}{134217728} y^7  \,.
\end{eqnarray}
Finally, the PN-solution terms contribute 
\begin{eqnarray}
S_{\rm (PN)}^{(6)} &=&   -\frac{45}{176} y+\frac{1120709}{247104} y^2+(-\frac{17267970589193}{1086822696960}+\frac{1165}{512}\pi^2) y^3\nonumber\\
&& +\left(-\frac{919}{192}\pi^2+\frac{38043781607578480627}{575814967189862400}\right) y^4\nonumber\\
&& +\left(-\frac{178102591076445413767429}{221112947400907161600}+\frac{25855117}{294912}\pi^2  \right) y^5\nonumber\\
&& +\left(-\frac{1807468975}{7077888}\pi^2+\frac{143778662972046270001175803568729}{36347479390399151052029952000}-\frac{12666295}{1048576}\pi^4  \right) y^6\nonumber\\
&& +\left(\frac{142382205611}{805306368}\pi^4-\frac{81105348347393296735582762457425451401}{7292758264889685667079289569280000}-\frac{16724117257249}{23781703680}\pi^2\right) y^7\,.
\end{eqnarray}

The final result for the $1$SF contribution to the irreducible octupolar invariant ${\mathcal J}_{3^+}$ is obtained by inserting Eqs. (\ref{final_J2dot})  (\ref{final_K3}) in Eq. (\ref{from_faye}).

The octupolar-level invariants presented in this appendix have not yet been numerically computed. The high-accuracy analytic formulas that we give for them
might serve as useful test beds of future numerical SF computations.
\end{widetext}

\noindent {\bf Acknowledgments}  
We thank Sam Dolan and Niels Warburton for informative email exchanges.
D.B. thanks the Italian INFN (Naples) for partial support and IHES for hospitality during the development of this project.
Both  authors are grateful to ICRANet for partial support.


\begin{thebibliography}{99}

\bibitem{Dolan:2014pja} 
  S.~R.~Dolan, P.~Nolan, A.~C.~Ottewill, N.~Warburton and B.~Wardell,
  ``Tidal invariants for compact binaries on quasi-circular orbits,''
  arXiv:1406.4890 [gr-qc].
 
\bibitem{Flanagan:2007ix} 
  E.~E.~Flanagan and T.~Hinderer,
  ``Constraining neutron star tidal Love numbers with gravitational wave detectors,''
  Phys.\ Rev.\ D {\bf 77}, 021502 (2008)
  [arXiv:0709.1915 [astro-ph]].

\bibitem{Read:2009yp} 
  J.~S.~Read, C.~Markakis, M.~Shibata, K.~Uryu, J.~D.~E.~Creighton and J.~L.~Friedman,
  ``Measuring the neutron star equation of state with gravitational wave observations,''
  Phys.\ Rev.\ D {\bf 79}, 124033 (2009)
  [arXiv:0901.3258 [gr-qc]].

\bibitem{Baiotti:2010xh} 
  L.~Baiotti, T.~Damour, B.~Giacomazzo, A.~Nagar and L.~Rezzolla,
  ``Analytic modelling of tidal effects in the relativistic inspiral of binary neutron stars,''
  Phys.\ Rev.\ Lett.\  {\bf 105}, 261101 (2010)
  [arXiv:1009.0521 [gr-qc]].

\bibitem{Bernuzzi:2012ci} 
  S.~Bernuzzi, A.~Nagar, M.~Thierfelder and B.~Brugmann,
  ``Tidal effects in binary neutron star coalescence,''
  Phys.\ Rev.\ D {\bf 86}, 044030 (2012)
  [arXiv:1205.3403 [gr-qc]].

\bibitem{Damour:2012yf} 
  T.~Damour, A.~Nagar and L.~Villain,
  ``Measurability of the tidal polarizability of neutron stars in late-inspiral gravitational-wave signals,''
  Phys.\ Rev.\ D {\bf 85}, 123007 (2012)
  [arXiv:1203.4352 [gr-qc]].

\bibitem{Bernuzzi:2011aq} 
  S.~Bernuzzi, M.~Thierfelder and B.~Bruegmann,
  ``Accuracy of numerical relativity waveforms from binary neutron star mergers and their comparison with post-Newtonian waveforms,''
  Phys.\ Rev.\ D {\bf 85}, 104030 (2012)
  [arXiv:1109.3611 [gr-qc]].

\bibitem{Read:2013zra} 
  J.~S.~Read, L.~Baiotti, J.~D.~E.~Creighton, J.~L.~Friedman, B.~Giacomazzo, K.~Kyutoku, C.~Markakis and L.~Rezzolla {\it et al.},
  ``Matter effects on binary neutron star waveforms,''
  Phys.\ Rev.\ D {\bf 88}, 044042 (2013)
  [arXiv:1306.4065 [gr-qc]].

\bibitem{DelPozzo:2013ala} 
  W.~Del Pozzo, T.~G.~F.~Li, M.~Agathos, C.~Van Den Broeck and S.~Vitale,
  ``Demonstrating the feasibility of probing the neutron star equation of state with second-generation gravitational wave detectors,''
  Phys.\ Rev.\ Lett.\  {\bf 111}, no. 7, 071101 (2013)
  [arXiv:1307.8338 [gr-qc]].

\bibitem{Hotokezaka:2013mm} 
  K.~Hotokezaka, K.~Kyutoku and M.~Shibata,
  ``Exploring tidal effects of coalescing binary neutron stars in numerical relativity,''
  Phys.\ Rev.\ D {\bf 87}, no. 4, 044001 (2013)
  [arXiv:1301.3555 [gr-qc]].

\bibitem{Radice:2013hxh} 
  D.~Radice, L.~Rezzolla and F.~Galeazzi,
  ``Beyond second-order convergence in simulations of binary neutron stars in full general-relativity,''
  Mon.\ Not.\ Roy.\ Astron.\ Soc.\  {\bf 437}, L46 (2014)
  [arXiv:1306.6052 [gr-qc]].

\bibitem{Bernuzzi:2014kca} 
  S.~Bernuzzi, A.~Nagar, S.~Balmelli, T.~Dietrich and M.~Ujevic,
  ``Quasiuniversal properties of neutron star mergers,''
  Phys.\ Rev.\ Lett.\  {\bf 112}, 201101 (2014)
  [arXiv:1402.6244 [gr-qc]].

\bibitem{Buonanno:1998gg} 
  A.~Buonanno and T.~Damour,
   ``Effective one-body approach to general relativistic two-body dynamics,''
  Phys.\ Rev.\ D {\bf 59}, 084006 (1999)
  [gr-qc/9811091].

\bibitem{Buonanno:2000ef} A.~Buonanno and T.~Damour,
``Transition from inspiral to plunge in binary black hole coalescences,''
Phys.\ Rev.\ D {\bf 62}, 064015 (2000)
[gr-qc/0001013].

\bibitem{Damour:2000we} 
  T.~Damour, P.~Jaranowski and G.~Schaefer,
  ``On the determination of the last stable orbit for circular general relativistic binaries at the third postNewtonian approximation,''
  Phys.\ Rev.\ D {\bf 62}, 084011 (2000)
  [gr-qc/0005034].

\bibitem{Damour:2001tu} 
  T.~Damour,
  ``Coalescence of two spinning black holes: an effective one-body approach,''
  Phys.\ Rev.\ D {\bf 64}, 124013 (2001)
  [gr-qc/0103018].

\bibitem{Damour:2009sm} 
  T.~Damour,
  ``Gravitational Self Force in a Schwarzschild Background and the Effective One Body Formalism,''
  Phys.\ Rev.\ D {\bf 81}, 024017 (2010)
  [arXiv:0910.5533 [gr-qc]].

\bibitem{Barack:2010ny} 
  L.~Barack, T.~Damour and N.~Sago,
  ``Precession effect of the gravitational self-force in a Schwarzschild spacetime and the effective one-body formalism,''
  Phys.\ Rev.\ D {\bf 82}, 084036 (2010)
  [arXiv:1008.0935 [gr-qc]].

\bibitem{Akcay:2012ea} 
S.~Akcay, L.~Barack, T.~Damour and N.~Sago,
``Gravitational self-force and the effective-one-body formalism between the innermost stable circular orbit and the light ring,''
Phys.\ Rev.\ D {\bf 86}, 104041 (2012)
[arXiv:1209.0964 [gr-qc]].

\bibitem{Taracchini:2013rva} 
  A.~Taracchini, A.~Buonanno, Y.~Pan, T.~Hinderer, M.~Boyle, D.~A.~Hemberger, L.~E.~Kidder and G.~Lovelace {\it et al.},
  ``Effective-one-body model for black-hole binaries with generic mass ratios and spins,''
  Phys.\ Rev.\ D {\bf 89}, 061502 (2014)
  [arXiv:1311.2544 [gr-qc]].

\bibitem{Damour:2014sva} 
  T.~Damour and A.~Nagar,
 ``A new effective-one-body description of coalescing nonprecessing spinning black-hole binaries,''
  arXiv:1406.6913 [gr-qc].

\bibitem{Damour:2014afa} 
  T.~Damour, F.~Guercilena, I.~Hinder, S.~Hopper, A.~Nagar and L.~Rezzolla,
  ``Strong-Field Scattering of Two Black Holes: Numerics Versus Analytics,''
  Phys.\ Rev.\ D {\bf 89}, 081503 (2014)
  [arXiv:1402.7307 [gr-qc]].

\bibitem{Bini:2014ica} 
  D.~Bini and T.~Damour,
  ``Two-body gravitational spin-orbit interaction at linear order in the mass ratio,''
  Phys.\ Rev.\ D {\bf 90}, 024039 (2014)
  [arXiv:1404.2747 [gr-qc]].

\bibitem{Bini:2012gu} 
  D.~Bini, T.~Damour and G.~Faye,
  ``Effective action approach to higher-order relativistic tidal interactions in binary systems and their effective one body description,''
  Phys.\ Rev.\ D {\bf 85}, 124034 (2012)
  [arXiv:1202.3565 [gr-qc]].

\bibitem{Bini:2013zaa} 
  D.~Bini and T.~Damour,
  ``Analytical determination of the two-body gravitational interaction potential at the fourth post-Newtonian approximation,''
  Phys.\ Rev.\ D {\bf 87}, no. 12, 121501 (2013)
  [arXiv:1305.4884 [gr-qc]].

\bibitem{Bini:2013rfa} 
  D.~Bini and T.~Damour,
  ``High-order post-Newtonian contributions to the two-body gravitational interaction potential from analytical gravitational self-force calculations,''
  arXiv:1312.2503 [gr-qc].

\bibitem{Bini:2014nfa} 
  D.~Bini and T.~Damour,
  ``Analytic determination of the eight-and-a-half post-Newtonian self-force contributions to the two-body gravitational interaction potential,''
  arXiv:1403.2366 [gr-qc].

\bibitem{Damour:2009wj} 
  T.~Damour and A.~Nagar,
  ``Effective One Body description of tidal effects in inspiralling compact binaries,''
  Phys.\ Rev.\ D {\bf 81}, 084016 (2010)
  [arXiv:0911.5041 [gr-qc]].

\bibitem{Vines:2010ca} 
  J.~E.~Vines and E.~E.~Flanagan,
  ``Post-1-Newtonian quadrupole tidal interactions in binary systems,''
  Phys.\ Rev.\ D {\bf 88}, 024046 (2013)
  [arXiv:1009.4919 [gr-qc]].

\bibitem{Damour:1982wm} T.~Damour, Gravitational Radiation And The Motion Of
  Compact Bodies, in {\it Gravitational Radiation}, edited by N.~Deruelle and
  T.~Piran (North-Holland, Amsterdam, 1983), pp.~59-144.

\bibitem{Damour:1995kt} T.~Damour and G.~Esposito-Far\`ese, Testing gravity to
  second post-Newtonian order: A Field theory approach, {\it Phys. Rev. D}
  {\bf 53}, 5541 (1996) [gr-qc/9506063].  

\bibitem{Damour:1998jk} T.~Damour and G.~Esposito-Far\`ese, Gravitational wave
  versus binary-pulsar tests of strong field gravity, {\it Phys. Rev. D} {\bf
    58}, 042001 (1998) [gr-qc/9803031].

\bibitem{Goldberger:2004jt} W.D.~Goldberger and I.Z.~Rothstein, An Effective
  field theory of gravity for extended objects, {\it Phys. Rev. D} {\bf 73},
  104029 (2006) [hep-th/0409156].

\bibitem{Zhang86} Xiao-He~Zhang, Multipole expansions of the
  general-relativistic gravitational field of the external universe, {\it
    Phys. Rev. D} {\bf 34}, 991 (1986).

\bibitem{Damour:1990pi} T.~Damour, M.~Soffel and C.-m.~Xu, General
  relativistic celestial mechanics. 1. Method and definition of reference
  systems, {\it Phys. Rev. D} {\bf 43}, 3272 (1991).

\bibitem{Damour:1991yw} T.~Damour, M.~Soffel and C.-m.~Xu, General
  relativistic celestial mechanics. 2. Translational equations of motion, {\it
    Phys. Rev. D} {\bf 45}, 1017 (1992).

\bibitem{Damour:1992qi} T.~Damour, M.~Soffel and C.-m.~Xu, General
  relativistic celestial mechanics. 3. Rotational equations of motion, {\it
    Phys. Rev. D} {\bf 47}, 3124 (1993).

\bibitem{Damour:1993zn} T.~Damour, M.~Soffel and C.-m.~Xu, General
  relativistic celestial mechanics. 4: Theory of satellite motion, {\it Phys.
    Rev. D} {\bf 49}, 618 (1994).

\bibitem{Taylor:2008xy} 
  S.~Taylor and E.~Poisson,
  ``Nonrotating black hole in a post-Newtonian tidal environment,''
  Phys.\ Rev.\ D {\bf 78}, 084016 (2008)
  [arXiv:0806.3052 [gr-qc]].

\bibitem{JohnsonMcDaniel:2009dq} 
  N.~K.~Johnson-McDaniel, N.~Yunes, W.~Tichy and B.~J.~Owen,
  ``Conformally curved binary black hole initial data including tidal deformations and outgoing radiation,''
  Phys.\ Rev.\ D {\bf 80}, 124039 (2009)
  [arXiv:0907.0891 [gr-qc]].

\bibitem{Detweiler:2008ft} 
  S.~L.~Detweiler,
  ``A Consequence of the gravitational self-force for circular orbits of the Schwarzschild geometry,''
  Phys.\ Rev.\ D {\bf 77}, 124026 (2008)
  [arXiv:0804.3529 [gr-qc]].

\bibitem{Blanchet:2009sd} 
  L.~Blanchet, S.~L.~Detweiler, A.~Le Tiec and B.~F.~Whiting,
  ``Post-Newtonian and Numerical Calculations of the Gravitational Self-Force for Circular Orbits in the Schwarzschild Geometry,''
  Phys.\ Rev.\ D {\bf 81}, 064004 (2010)
  [arXiv:0910.0207 [gr-qc]].

\bibitem{Blanchet:2010zd} 
  L.~Blanchet, S.~L.~Detweiler, A.~Le Tiec and B.~F.~Whiting,
  ``High-Order Post-Newtonian Fit of the Gravitational Self-Force for Circular Orbits in the Schwarzschild Geometry,''
  Phys.\ Rev.\ D {\bf 81}, 084033 (2010)
  [arXiv:1002.0726 [gr-qc]].

\bibitem{LeTiec:2011ab} 
  A.~Le Tiec, L.~Blanchet and B.~F.~Whiting,
  ``The First Law of Binary Black Hole Mechanics in General Relativity and Post-Newtonian Theory,''
  Phys.\ Rev.\ D {\bf 85}, 064039 (2012)
  [arXiv:1111.5378 [gr-qc]].

\bibitem{Blanchet:2014bz} 
  L.~Blanchet, G.~Faye and B.~F.~Whiting,
  ``High-order half-integral conservative post-Newtonian coefficients in the redshift factor of black hole binaries,''
  Phys.\ Rev.\ D {\bf 90}, 044017 (2014)
  [arXiv:1405.5151 [gr-qc]].

\bibitem{Blanchet:2012at} 
  L.~Blanchet, A.~Buonanno and A.~Le Tiec,
  ``First Law of Mechanics for Black Hole Binaries with Spins,''
  Phys.\ Rev.\ D {\bf 87}, 024030 (2013)
  [arXiv:1211.1060 [gr-qc]].

\bibitem{Zerilli:1970se} 
  F.~J.~Zerilli,
  ``Effective potential for even parity Regge-Wheeler gravitational perturbation equations,''
  Phys.\ Rev.\ Lett.\  {\bf 24}, 737 (1970).

\bibitem{Zerilli:1971wd} 
  F.~J.~Zerilli,
   ``Gravitational field of a particle falling in a schwarzschild geometry analyzed in tensor harmonics,''
  Phys.\ Rev.\ D {\bf 2}, 2141 (1970).

\bibitem{Mano:1996vt} S.~Mano, H.~Suzuki and E.~Takasugi,
``Analytic solutions of the Teukolsky equation and their low frequency expansions,''
Prog.\ Theor.\ Phys.\  {\bf 95}, 1079 (1996)
[gr-qc/9603020].

\bibitem{Mano:1996mf} S.~Mano, H.~Suzuki and E.~Takasugi,
``Analytic solutions of the Regge-Wheeler equation and the postMinkowskian expansion,''
Prog.\ Theor.\ Phys.\  {\bf 96}, 549 (1996)
[gr-qc/9605057].

\bibitem{Mano:1996gn} S.~Mano and E.~Takasugi,
``Analytic solutions of the Teukolsky equation and their properties,''
Prog.\ Theor.\ Phys.\  {\bf 97}, 213 (1997)
[gr-qc/9611014].

\bibitem{Nakano:2003he} H.~Nakano, N.~Sago and M.~Sasaki,
``Gauge problem in the gravitational self force. 2. First postNewtonian force under Regge-Wheeler gauge,''
Phys.\ Rev.\ D {\bf 68}, 124003 (2003)
[gr-qc/0308027].

\bibitem{Shah:2013uya} 
  A.~G.~Shah, J.~L.~Friedman and B.~F.~Whiting,
  ``Finding high-order analytic post-Newtonian parameters from a high-precision numerical self-force calculation,''
  Phys.\ Rev.\ D {\bf 89}, 064042 (2014)
  [arXiv:1312.1952 [gr-qc]].
 
\bibitem{LeTiec:2011dp} 
  A.~Le Tiec, E.~Barausse and A.~Buonanno,
   ``Gravitational Self-Force Correction to the Binding Energy of Compact Binary Systems,''
  Phys.\ Rev.\ Lett.\  {\bf 108}, 131103 (2012)
  [arXiv:1111.5609 [gr-qc]].
 
\bibitem{Dolan:2013roa} 
  S.~R.~Dolan, N.~Warburton, A.~I.~Harte, A.~L.~Tiec, B.~Wardell and L.~Barack,
  ``Gravitational self-torque and spin precession in compact binaries,''
  Phys.\ Rev.\ D {\bf 89}, 064011 (2014)
  [arXiv:1312.0775 [gr-qc]].
  
\bibitem{Buonanno:2007pf} 
  A.~Buonanno, Y.~Pan, J.~G.~Baker, J.~Centrella, B.~J.~Kelly, S.~T.~McWilliams and J.~R.~van Meter,
  ``Toward faithful templates for non-spinning binary black holes using the effective-one-body approach,''
  Phys.\ Rev.\ D {\bf 76}, 104049 (2007)
  [arXiv:0706.3732 [gr-qc]].
  
\bibitem{Damour:2009kr} 
  T.~Damour and A.~Nagar,
  ``An Improved analytical description of inspiralling and coalescing black-hole binaries,''
  Phys.\ Rev.\ D {\bf 79}, 081503 (2009)
  [arXiv:0902.0136 [gr-qc]].
  
\bibitem{Buonanno:2009qa} 
  A.~Buonanno, Y.~Pan, H.~P.~Pfeiffer, M.~A.~Scheel, L.~T.~Buchman and L.~E.~Kidder,
  ``Effective-one-body waveforms calibrated to numerical relativity simulations: Coalescence of non-spinning, equal-mass black holes,''
  Phys.\ Rev.\ D {\bf 79}, 124028 (2009)
  [arXiv:0902.0790 [gr-qc]].
  
\bibitem{Damour:2012ky} 
  T.~Damour, A.~Nagar and S.~Bernuzzi,
  ``Improved effective-one-body description of coalescing nonspinning black-hole binaries and its numerical-relativity completion,''
  Phys.\ Rev.\ D {\bf 87}, no. 8, 084035 (2013)
  [arXiv:1212.4357 [gr-qc]].
  

\end{thebibliography}
\end{document}